\newcommand{\goesto}{\rightarrow}

\newcommand{\VanD}{\Delta\!^{1/2}}

\newcommand{\lC}{\left[}
\newcommand{\rC}{\right]}

\newcommand{\w}{\omega}
\newcommand{\symmeq}{\stackrel{.}{=}}

\documentclass[prd,aps,eqsecnum,showpacs]{revtex4}
\usepackage{amsmath}
\usepackage{amssymb}

\begin{document}

\title{Noise Kernel and Stress Energy Bi-Tensor of Quantum Fields in
Conformally-Optical Metrics: Schwarzschild Black Holes}

\author{Nicholas G. Phillips}
\email{Nicholas.G.Phillips@gsfc.nasa.gov}
\affiliation{ SSAI, Laboratory for Astronomy and Solar Physics, Code 685, 
             NASA/GSFC, Greenbelt, Maryland 20771}

\author{B. L. Hu}
\email{hub@physics.umd.edu}
\affiliation{Department of Physics, University of Maryland, College
             Park, Maryland 20742-4111}

\date{umdpp 03-017, Submitted to Phys. Rev. D, Sept 12, 2002}
\begin{abstract}
In Paper II [N. G. Phillips and B. L. Hu, this journal]
we presented the details for the
regularization of the noise kernel of a quantum scalar field in
optical spacetimes by the modified point separation scheme, and a
Gaussian approximation for the Green function. We worked out the
regularized noise kernel for two examples: hot flat space and
optical Schwarzschild metric. In this paper we consider noise
kernels for a scalar field in the Schwarzschild black hole.
Much of the work in the point separation approach is to determine
how the divergent piece conformally transforms. For the
Schwarzschild metric we find that the fluctuations of the stress
tensor of the Hawking flux in the far field region checks with
the analytic results given by Campos and Hu earlier
[A. Campos and B. L. Hu, Phys. Rev. D {\bf 58} (1998) 125021;
Int. J. Theor. Phys. {\bf 38} (1999) 1253].
We also verify Page's result 
[D. N. Page, Phys. Rev. {\bf D25}, 1499 (1982)]
for the stress tensor,
which, though used often, still lacks a rigorous proof, as in his
original work the direct use of the conformal transformation was
circumvented.
However, as in the optical case, we show that the Gaussian
approximation applied to the Green function produces significant
error in the noise kernel on the Schwarzschild horizon. As before
we identify the failure as occurring at the fourth covariant
derivative order.

\end{abstract}

\maketitle

\section{Introduction}
\label{sec-intro}

Many physically interesting metrics are conformally related to an
optical metric. When the metric has an optical form, the Gaussian
approximation \cite{BekPark81} for the Green function
\cite{Page82} yields  a closed analytic form. For a given noise
kernel in an optical metric, one can take advantage of the simple
conformal transformation property of the scalar field's Green
function to compute the noise kernel for the corresponding
conformally-optical metrics. In Paper II we have derived the
regularized noise kernel in the optical metric by the modified
point separation method and worked out two examples: hot flat
space and optical Schwarzschild metrics. Hot flat space is
related to thermal fields in the (spatially-flat)
Robertson-Walker universe by a time dependent scale factor, and
the optical Schwarzschild metric is of course conformally related
to the Schwarzschild black hole metric. In this paper we present
the details for computing the regularized noise kernel.


According to the procedure outlined in Paper II, the main obstacle
here is the subtraction of the Hadamard ansatz. The divergent
Green function is defined in terms of the optical metric while
the Hadamard ansatz in terms of the physical metric. We need to
re-express the transformed optical metric in terms of the
physical metric. The defining equations for the geometric objects
on the optical metric are transformed to the physical metric and
solved recursively.

Now the Green function series expansion can be written solely in
terms of the physical metric. While for the ultrastatic
spacetimes considered in Paper II the Hadamard subtraction was
straightforward, for spacetimes only conformally ultrastatic this
subtraction is non-trivial. Before the subtraction can be carried
out and the Green function regularized, we need to study the
conformal transformation properties of the point separation
objects used to define the Green function. As we will show,
regularizing the divergent structure to sufficient order for the
noise kernel is no small task, when we include the conformal
transformation between the physical and the optical metrics.
(e.g., the fourth order term in the expansion of the renormalized
Green function has over 1100 terms). It is at this point that the
symbolic computation environment takes over.

Since all the series expansions used are recursively derived, as
a check of the full expression it is sufficient that we get the
correct results for the lower order terms. Indeed our general
expression contains and confirms the Page \cite{Page82} result for
the vacuum expectation value of the stress tensor in the
Schwarzschild black hole.  With the general expansion of the
renormalized Green function at hand, one can choose a metric and
compute the coincident limit of the noise kernel as outlined in
Paper II. In Section \ref{sec-ch4-physnoise}  we compute the
noise kernel for spacetimes conformally related to ultrastatic
spacetimes. In Section \ref{sec-Schw} we specialize to the
Schwarzschild black hole and discuss the significance of our
results in Sec. IV. The Appendices contain formulas whose
usefulness goes beyond the specific approximations adopted here.

\section{Noise Kernel in Conformally-Optical Spacetimes }
\label{sec-ch4-physnoise}

For a static physical metric $g_{ab}$, we consider the
conformally related optical metric
\begin{equation}
   \bar g_{ab} = e^{-2\w} g_ab
\end{equation}
where the conformal factor $e^{-2\w}$ is the
space-dependent function such that $\bar g_{\tau\tau} = 1$, {\it
i.e.}, the metric $\bar g_{ab}$ is for an ultrastatic
spacetime. In general, we use the overbar to denote objects
defined in terms of the optical metric $\bar g_{ab}$ and without for those
in terms of the physical metric $g_{ab}$. For a conformally invariant
field, the Green functions on the two spacetimes are related via
\begin{equation}
   G(x,y) = e^{-\w(x)} \bar G(x,y) e^{-\w(y)}.
\end{equation}
This is our starting
point. In terms of the Gaussian approximation
\begin{equation}
 \bar G_{\rm Gauss}(x,y) =
      \frac{\kappa\bar\VanD}{8\pi^2 \bar r}
       \frac{\sinh\kappa \bar r}
            {\left(\cosh\kappa \bar r-\cos\kappa\tau\right)},
\end{equation}
where $r=\left(2\,{}^{(3)}\!\bar\sigma\right)^\frac{1}{2}$ and
$\kappa$ is the period of the imaginary time dimension, we have
the expansion of the physical Green function as
\begin{eqnarray}
G_{\rm Gauss } &=&
{\frac{{\bar\Delta^{{\frac{1}{2}}}}\,{e^{-\w - {\w'}}}}
   {8\,{\bar\sigma}\,{{\pi }^2}}}
+ {\frac{{\bar\Delta^{{\frac{1}{2}}}}\,{e^{-\w -
{\w'}}}}{8\,{{\pi }^2}}} \left\{ {\frac{{{\kappa}^2}}{6}}
  + {\frac{{{\kappa}^4}}{180}} \left( -{\bar\sigma} + 2\,{{{\Delta\tau}}^2}
\right) \right. \cr && \left. \hspace{20mm} +
{\frac{{{\kappa}^6}}{3780}} \left( {{{\bar\sigma}}^2} -
6\,{\bar\sigma}\,{{{\Delta\tau}}^2} +
  4\,{{{\Delta\tau}}^4} \right)
\right\} +O\left(\sigma^\frac{5}{2},\Delta\tau^5\right)
\label{Phys-Gseries1}
\end{eqnarray} where $\w=\w(x)$ and $\w'=\w(y)$.

We proceed as in Paper II and subtract from ${G_{\rm Gauss}}$ the
Hadamard ansatz \begin{equation} S(x,y) = {\frac{1}{16\,{{\pi }^2}}} \left(
{\frac{2\,{\Delta^{{\frac{1}{2}}}}}{\sigma}} + \sigma\,w_{1} +
  {{\sigma}^2}\,w_{2} \right)
 + O\left(\sigma^3\right)
\label{Phys-Had1}
\end{equation}
to get the renormalized Green function
\begin{equation}
  G_{\rm ren} = G_{\rm gauss} - S.
\label{Phys-Gren1}
\end{equation}
In the Hadamard ansatz, the $V(x,x')$ term is not included since there
is no $\log\sigma$ divergences present in the expansion of the
Gaussian approximation to the Green function. This can also be
viewed as an extension of the Gaussian approximation.

Now the situation here is different from the optical case as the
divergent terms present (\ref{Phys-Gseries1}) and
(\ref{Phys-Gren1}) do not directly cancel. Much of the work for
regularizing the Green function will entail showing that indeed
the difference between these two divergent terms is finite and to
develop this finite difference to sufficient order to compute the
noise kernel. To this end, we write the renormalized Green
function as \begin{equation} G_{\rm ren} = G_{\rm div,ren} + G_{\rm fin} -
\frac{1}{(4\pi)^2} W \end{equation} with 
\begin{subequations}
\begin{eqnarray}
 G_{\rm div,ren}
&=&\frac{1}{8\pi^2}\left(
 {\frac{{\bar\Delta^{{\frac{1}{2}}}}\,{e^{-\w - {\w'}}}}{{\bar\sigma}}} -
  {\frac{{\Delta^{{\frac{1}{2}}}}}{\sigma}} \right) \\
G_{\rm fin} &=&
{\frac{{\bar\Delta^{{\frac{1}{2}}}}\,{e^{-\w - {\w'}}}}{8\,{{\pi
}^2}}} \left\{ {\frac{{{\kappa}^2}}{6}}
  + {\frac{{{\kappa}^4}}{180}} \left( -{\bar\sigma} + 2\,{{{\Delta\tau}}^2}
\right) \right. \cr && \left. \hspace{20mm} +
{\frac{{{\kappa}^6}}{3780}} \left( {{{\bar\sigma}}^2} -
6\,{\bar\sigma}\,{{{\Delta\tau}}^2} +
  4\,{{{\Delta\tau}}^4} \right)
\right\}
\label{Phys-Gfin1} \\
W &=&  \sigma\,w_{1} + {{\sigma}^2}\,w_{2} 
\end{eqnarray}
\end{subequations}
Both $G_{\rm
fin}$ and $W$ have well behaved coincident limits. As these
functions stand, it is only the last one, $W$, that we can
readily handle. Appendix E of Paper II gives the series expansion
in terms of the world function $\sigma$ defined with respect to
the physical metric $g_{ab}$. Since we want the noise kernel in
the physical metric, it is the covariant derivative commensurate
with this metric we must use when computing the noise kernel. On
the other hand, the function $G_{\rm fin}$ is defined in terms of
the world function and the VanVleck-Morette determinant of the
optical metric $\bar g_{ab}$. Thus the coincident limit
expressions for $\sigma$ and the series expansion for
${\Delta^{{\frac{1}{2}}}}$ derived in the Appendices of Paper II
cannot be used to determine the contribution to a series
expansion of the renormalized Green function $G_{\rm ren}$.

In Appendix \ref{apx-conformal} here we show how to get around
this problem. Both $\bar\sigma$ and
${\bar\Delta^{{\frac{1}{2}}}}$ are defined in terms of covariant
differential equations with respect to the optical metric. The
conformal transformation properties of the covariant derivative
are used to re-express these equations in terms of the covariant
derivative commensurate with the physical metric. Then end
point series solutions built from the physical metric world
function $\sigma$ are found. The details, along with the found
solution are collected in the Appendices of this paper. Using these
results, we can determine the contribution from $G_{\rm fin}$ to
the noise kernel.

This leaves the first function defined above to deal with. The
key to unlocking this term is to introduce the symmetric function
\begin{equation}
  \Sigma(x,y) = e^{\w(x)}\bar\sigma(x,y)e^{\w(y)} - \sigma(x,y)
\end{equation}
The important property of this function as
shown in Appendix \ref{apx-conformal} is $\Sigma
\sim\sigma^2$ as $\sigma\goesto 0$. With this function, we have
\begin{equation}
G_{\rm div,ren} = {\frac{1}{8\,{{\pi }^2}}}
{\frac{\sigma\,\left( {\bar\Delta^{{\frac{1}{2}}}} -
        {\Delta^{{\frac{1}{2}}}} \right)  -
\Sigma\,{\Delta^{{\frac{1}{2}}}}}
     {\sigma\,\left( \sigma + \Sigma \right) }}
\label{Phys-Gdiv-Sigma-form} \end{equation} To see that this is  indeed
finite, we use the expansions of
\begin{subequations}\begin{eqnarray}
\Sigma &\sim& {{\sigma}^2}\,S^{(4)} \\
\label{Phys-Sigma-series}
{\Delta^{{\frac{1}{2}}}} &\sim& 1 + \sigma\,\Delta^{(2)} \\
\label{Phys-VanD-series} {\bar\Delta^{{\frac{1}{2}}}} &\sim& 1 +
\sigma\,\bar\Delta^{(2)} \label{Phys-bVanD-series} 
\end{eqnarray}\end{subequations} to obtain
the leading order behavior 
\begin{equation} G_{\rm div,ren} =
-{\frac{1}{8\,{{\pi }^2}}} {\frac{-\bar\Delta^{(2)} +
\Delta^{(2)} + S^{(4)} +
      \sigma\,\Delta^{(2)}\,S^{(4)}}{1 + \sigma\,S^{(4)}}}
\end{equation} which is finite as $\sigma\goesto 0$ and has the value \begin{equation}
{\left[G_{\rm div,ren}\right]}  = {\frac{1}{8\,{{\pi }^2}}}
 \left(   \bar\Delta^{(2)} - \Delta^{(2)} - S^{(4)} \right)
\label{Phys-Gdiv-order0}
\end{equation}

Using Eqns (\ref{Apx-Sigma4}) and (\ref{Apx-bVan2}),
along with (C8) of Paper II, we get the explicit form of the
leading order expansion tensors used above as
\begin{subequations}\begin{eqnarray}
\Sigma^{(4)}_{abcd} &=&
{\frac{\w{}_{;}{}_{c}\,\w{}_{;}{}_{d}\,g{}_{a}{}_{b}}{12}} +
  {\frac{\w{}_{;}{}_{c}{}_{d}\,g{}_{a}{}_{b}}{12}} -
  {\frac{\w{}_{;}{}_{p}\,\w{}^{;}{}^{p}\,g{}_{a}{}_{b}\,g{}_{c}{}_{d}}{24}}
\\
\Delta^{(2)}_{ab} &=& {\frac{R{}_{a}{}_{b}}{12}}
\\
\bar\Delta^{(2)}_{ab} &=&
{\frac{\w{}_{;}{}_{a}\,\w{}_{;}{}_{b}}{6}} +
  {\frac{\w{}_{;}{}_{a}{}_{b}}{6}} -
  {\frac{\w{}_{;}{}_{p}\,\w{}^{;}{}^{p}\,g{}_{a}{}_{b}}{6}} +
  {\frac{\w{}_{;}{}_{p}{}^{p}\,g{}_{a}{}_{b}}{12}} +
  {\frac{R{}_{a}{}_{b}}{12}}
\end{eqnarray}\end{subequations}
From these we can get the expansion scalars we need for
(\ref{Phys-Gdiv-order0}): 
\begin{subequations}\begin{eqnarray}
S^{(4)} &=& 4 p^p p^q p^r p^s
\Sigma^{(4)}_{pqrs}
  = -{\frac{\left( \w{}_{;}{}_{p}\,\w{}^{;}{}^{p} \right) }{6}} +
  {\frac{\w{}_{;}{}_{p}\,\w{}_{;}{}_{q}\,p{}^{p}\,p{}^{q}}{3}} +
  {\frac{\w{}_{;}{}_{p}{}_{q}\,p{}^{p}\,p{}^{q}}{3}} \\
\Delta^{(2)} &=& 2 p^p p^q \Delta^{(2)}_{pq}
  = {\frac{p{}^{p}\,p{}^{q}\,R{}_{p}{}_{q}}{6}} \\
\bar\Delta^{(2)} &=& 2 p^p p^q \bar\Delta^{(2)}_{pq}
  = -{\frac{\left( \w{}_{;}{}_{p}\,\w{}^{;}{}^{p} \right) }{3}} +
  {\frac{\w{}_{;}{}_{p}{}^{p}}{6}} +
  {\frac{\w{}_{;}{}_{p}\,\w{}_{;}{}_{q}\,p{}^{p}\,p{}^{q}}{3}} +
  {\frac{\w{}_{;}{}_{p}{}_{q}\,p{}^{p}\,p{}^{q}}{3}} +
  {\frac{p{}^{p}\,p{}^{q}\,R{}_{p}{}_{q}}{6}}
\end{eqnarray}\end{subequations}
 where $p^a$ is a unit vector. From these expressions, we
might expect there to be residual direction dependence for
(\ref{Phys-Gdiv-order0}). But when we substitute the expansion
scalars into the $\sigma\goesto 0$ value of $G_{\rm div,ren}$,
the direction dependences of the three expansion scalars cancel
and we are left with \begin{equation} {\left[G_{\rm div,ren}\right]} =
{\frac{1}{48\,{{\pi }^2}}}
  \left( \w{}_{;}{}_{p}{}^{p} - \w{}_{;}{}_{p}\,\w{}^{;}{}^{p} \right)
\end{equation}

Using ${\left[{\bar\Delta^{{\frac{1}{2}}}}\right]}=1$ and
${\left[{\bar\sigma}\right]} = 0$, we can immediately get
the coincident limit of (\ref{Phys-Gfin1})
\begin{equation}
\left[ G_{\rm fin} \right] =
 {\frac{{{\kappa}^2}}{48\,{e^{2\,\w}}\,{{\pi }^2}}}
\end{equation} Since $\left[ W \right] = 0$, we readily obtain the
coincident limit of the renormalized Green function \begin{equation}
{\left[G_{{\rm ren}}\right]} = {\frac{1}{48\,{{\pi }^2}}} \left(
  {\frac{{{\kappa}^2}}{{e^{2\,\w}}}} - \w{}_{;}{}_{p}\,\w{}^{;}{}^{p} +
  \w{}_{;}{}_{p}{}^{p} \right),
\label{Phys-CiGren} \end{equation} which is the result derived by Page (Eq.
(29) of \cite{Page82}).

Now that we know we can regularize $G_{\rm div,ren}$, we turn to
developing the series expansion of $G_{\rm div,ren}$ to a
sufficient order to compute the coincident limit of the noise
kernel. We need the expansion \begin{equation} G_{\rm div,ren} =
{\frac{1}{8\,{{\pi }^2}}} \left( G_{\rm div,ren}^{(0)} +
{\sqrt{\sigma}}\,G_{\rm div,ren}^{(1)} +
  \sigma\,G_{\rm div,ren}^{(2)} +
  {{\sigma}^{{\frac{3}{2}}}}\,G_{\rm div,ren}^{(3)} +
  {{\sigma}^2}\,G_{\rm div,ren}^{(4)} \right)
\label{Phys-Gdiv-series1} \end{equation} The computation of ${\left[G_{\rm
div,ren}\right]}$, i.e. $G_{\rm div,ren}^{(0)}$, involves
$\Sigma$ to order $\sigma^2$ and both ${\Delta^{{\frac{1}{2}}}}$
and ${\bar\Delta^{{\frac{1}{2}}}}$ to order $\sigma$. Thus to
carry out the expansion (\ref{Phys-Gdiv-series1}) we will need
these functions expanded to order $\sigma^4$ for $\Sigma$ and
$\sigma^3$ for ${\Delta^{{\frac{1}{2}}}}$ and
${\bar\Delta^{{\frac{1}{2}}}}$. These are done in Appendix
\ref{apx-conformal}. For this section we will use expansions in
terms of the scalar coefficients: 
\begin{subequations}\begin{eqnarray}
 \Sigma &\sim&
{{\sigma}^2}\,S^{(4)} + {{\sigma}^{{\frac{5}{2}}}}\,S^{(5)} +
  {{\sigma}^3}\,S^{(6)} + {{\sigma}^{{\frac{7}{2}}}}\,S^{(7)} +
  {{\sigma}^4}\,S^{(8)} \\
{\Delta^{{\frac{1}{2}}}} &\sim& 1 + \sigma\,\Delta^{(2)} +
{{\sigma}^{{\frac{3}{2}}}}\,\Delta^{(3)} +
  {{\sigma}^2}\,\Delta^{(4)} + {{\sigma}^{{\frac{5}{2}}}}\,\Delta^{(5)} +
  {{\sigma}^3}\,\Delta^{(6)} \\
{\bar\Delta^{{\frac{1}{2}}}} &\sim& 1 + \sigma\,\bar\Delta^{(2)} +
{{\sigma}^{{\frac{3}{2}}}}\,\bar\Delta^{(3)} +
  {{\sigma}^2}\,\bar\Delta^{(4)} +
  {{\sigma}^{{\frac{5}{2}}}}\,\bar\Delta^{(5)} +
  {{\sigma}^3}\,\bar\Delta^{(6)}.
\end{eqnarray}\end{subequations}
(Recall the scalar expansion coefficients are related to the
tensor expansion cofficients via
$A^{(n)} = 2^\frac{n}{2} p^{p_1}\cdots p^{p_n}
A^{(n)}_{p_1\cdots p_n}$, where in general we take
$\sqrt{\sigma}\, p^a = \sigma^{;a}$).

Putting these in (\ref{Phys-Gdiv-Sigma-form}), the expansion
coefficients $G_{\rm div,ren}^{(n)}$ are
\begin{subequations}\begin{eqnarray}
 G_{\rm div,ren}^{(0)} &=&  \bar\Delta^{(2)} - \Delta^{(2)}
- S^{(4)}
\\
G_{\rm div,ren}^{(1)} &=&  \bar\Delta^{(3)} - \Delta^{(3)} -
S^{(5)}
\\
G_{\rm div,ren}^{(2)} &=&  \bar\Delta^{(4)} - \Delta^{(4)} -
\Delta^{(2)}\,S^{(4)} +
  S^{(4)}\,\left( -\bar\Delta^{(2)} + \Delta^{(2)} + S^{(4)} \right)  -
  S^{(6)}
\\
G_{\rm div,ren}^{(3)} &=&  \bar\Delta^{(5)} - \Delta^{(5)} -
\Delta^{(3)}\,S^{(4)} -
  \Delta^{(2)}\,S^{(5)} + \left( -\bar\Delta^{(2)} + \Delta^{(2)} +
     S^{(4)} \right) \,S^{(5)} \cr
                & & + S^{(4)}\,\left( -\bar\Delta^{(3)} + \Delta^{(3)} +
S^{(5)} \right)  - S^{(7)}
\\
G_{\rm div,ren}^{(4)} &=&  \bar\Delta^{(6)} - \Delta^{(6)} -
\Delta^{(4)}\,S^{(4)} -
  \Delta^{(3)}\,S^{(5)} + S^{(5)}\,
   \left( -\bar\Delta^{(3)} + \Delta^{(3)} + S^{(5)} \right)  \cr
                & &  -\left( \left( -\bar\Delta^{(2)} + \Delta^{(2)} +
S^{(4)}
\right) \,
     \left( {{S^{(4)}}^2} - S^{(6)} \right)  \right)  -
\Delta^{(2)}\,S^{(6)}
\cr
                & & + S^{(4)}\,\left( -\bar\Delta^{(4)} + \Delta^{(4)} +
\Delta^{(2)}\,S^{(4)} +
     S^{(6)} \right)  - S^{(8)}
\end{eqnarray}\end{subequations}
 We have shown that for $G_{\rm div,ren}^{(0)}$ a direction
dependence could arise, but when we use the values of the
expansion tensors for $\Sigma$, ${\Delta^{{\frac{1}{2}}}}$ and
${\bar\Delta^{{\frac{1}{2}}}}$, this direction dependence
cancels. We expect this to happen because we see the coefficients
that contribute to any given coefficient of $G_{\rm div,ren}$ are
of a higher power than the $G_{\rm div,ren}^{(n)}$ coefficient in
question. When going from the scalar expansion coefficient form
to the tensor expansion form, the rank of the tensor is the same
as the order of the coefficient. And each of the $G_{\rm
div,ren}^{(n)}$ is made up of higher order coefficients. But as
we find by direct substitution from the expansion tensors listed
in Appendix \ref{apx-conformal}, for each case the direction
dependence always cancel. In Appendix \ref{apx-Gdiv}, we give the
complete form of the coefficients $G_{\rm div,ren}^{(n)}$, in
their corresponding tensorial form. This is one of our main
results: The regularization of the leading order divergence of the
Green function, when Green function has been computed in an
optical metric conformal to the physical metric in which the
problem is given. What is new here is that this regularization
has been carried out to the order needed for the computation of
the noise kernel. This analysis has been developed so as to take
advantage of the symbolic computing potential of current
workstations.

The last remaining obstacle in  computing the noise kernel comes
from $G_{\rm fin}$. As it stands, it is defined in terms of the
optical metric,  while it is the physical metric with which we
need to take the covariant derivatives that determine the noise
kernel. By using the function $\Sigma$ and the series expansions
(\ref{Phys-Sigma-series}), (\ref{Phys-VanD-series}) and
(\ref{Phys-bVanD-series}), along with the series expansions 
\begin{eqnarray}
\left(\delta\tau\right)^2 &=&
        \sigma\,\delta\tau^{(2)} +
{{\sigma}^{{\frac{3}{2}}}}\,\delta\tau^{(3)} +
  {{\sigma}^2}\,\delta\tau^{(4)} \\
e^{-a(\w+\w')} &=&
        \w^{(0,a)} + {\sqrt{\sigma}}\,\w^{(1,a)} + \sigma\,\w^{(2,a)} +
  {{\sigma}^{{\frac{3}{2}}}}\,\w^{(3,a)} + {{\sigma}^2}\,\w^{(4,a)}
\end{eqnarray}
$G_{\rm fin}$ can be expanded in terms of the physical
$\sigma$ as 
\begin{equation}
G_{\rm fin} = {\frac{1}{8\,{{\pi }^2}}} \left(
G_{\rm fin}^{(0)} + {\sqrt{\sigma}}\,G_{\rm fin}^{(1)} +
  \sigma\,G_{\rm fin}^{(2)} + {{\sigma}^{{\frac{3}{2}}}}\,G_{\rm fin}^{(3)}
+
  {{\sigma}^2}\,G_{\rm fin}^{(4)} \right)
\end{equation}
where these expansion coefficients are 
\begin{subequations}\begin{eqnarray}
G_{\rm fin}^{(0)} &=& {\frac{{{\kappa}^2}}{6}}
                      \w^{(0,1)}
\\
G_{\rm fin}^{(1)} &=&  {\frac{{{\kappa}^2}}{6}}
                      \w^{(1,1)}
\\
G_{\rm fin}^{(2)} &=&
  {\frac{{{\kappa}^2}}{6}}
  \left( \bar\Delta^{(2)}\,\w^{(0,1)} + \w^{(2,1)} \right)
+  {\frac{{{\kappa}^4}}{180}}
  \left( 2\,\delta\tau^{(2)}\,\w^{(0,1)} - \w^{(0,2)} \right)
\\
G_{\rm fin}^{(3)} &=&
  {\frac{{{\kappa}^2}}{6}}
  \left( \bar\Delta^{(3)}\,\w^{(0,1)} + \bar\Delta^{(2)}\,\w^{(1,1)} +
\w^{(3,1)} \right) \cr && +  {\frac{{{\kappa}^4}}{180}}
  \left( 2\,\delta\tau^{(3)}\,\w^{(0,1)} + 2\,\delta\tau^{(2)}\,\w^{(1,1)} -
\w^{(1,2)} \right)
\\
G_{\rm fin}^{(4)} &=&
  {\frac{{{\kappa}^2}}{6}}
  \left( \bar\Delta^{(4)}\,\w^{(0,1)} + \bar\Delta^{(3)}\,\w^{(1,1)} +
  \bar\Delta^{(2)}\,\w^{(2,1)} + \w^{(4,1)} \right) \cr
&&-  {\frac{{{\kappa}^4}}{180}}
  \left( -2\,\delta\tau^{(4)}\,\w^{(0,1)} + S^{(4)}\,\w^{(0,2)} +
  \bar\Delta^{(2)}\,\left( -2\,\delta\tau^{(2)}\,\w^{(0,1)} + \w^{(0,2)}
      \right)  \right. \cr
&&\hspace{20mm}
  \left. -2\,\delta\tau^{(3)}\,\w^{(1,1)} - 2\,\delta\tau^{(2)}\,\w^{(2,1)}
+
  \w^{(2,2)} \right) \cr
&&+  {\frac{{{\kappa}^6}}{3780}}
  \left( 4\,{{\delta\tau^{(2)}}^2}\,\w^{(0,1)} -
6\,\delta\tau^{(2)}\,\w^{(0,2)} +
  \w^{(0,3)} \right)
\end{eqnarray}\end{subequations}
 The explicit expressions for these expansion tensors are
given in Appendix \ref{apx-Gfin}.

With this we have regularized and expanded $G_{\rm div,ren}$, and
expanded in the physical metric $G_{\rm fin}$, both to fourth
order. The Hadamard ansatz function $W$'s series expansion is
derived in Appendix E of Paper II, also to fourth order. From
here we proceed as we did in Paper II for the optical metrics. To
review, once a metric is selected, the component values of the
expansions of $G_{\rm div,ren}$, $G_{\rm fin}$ and $W$ are
computed symbolically on the computer. Then Eqn (3.10) of Paper
II is used to determine the
component values of the coincident limit of up to four covariant
derivatives of $G_{{\rm ren}}$, along with the needed covariant
derivatives of the coincident limits. From here the component
values of the coincident limit of the noise kernel are computed.
This procedure is adopted owing to the large size of the
expansions. In fact, initial attempts to delay the specification
of the metric until a full determination of the noise kernel ended
up yielding a general tensorial expression of nearly 60,000
terms. By working out from the expansion tensors in terms of
their component values, we found this had the added advantage of
enabling one to study each of the separate terms that go into
computing the noise kernel.

\section{Schwarzschild Black Hole} \label{sec-Schw}

We can now turn our attention to a specific example. Consider a
massless, conformally coupled scalar field on a Schwarzschild
black hole with mass $M$. The line element is given in the usual
coordinates $x^a=(r,\theta,\phi,\tau)$
\begin{equation} ds^2 =
\left(1-\frac{2M}{r}\right) d\tau^2 +
       \left(1-\frac{2M}{r}\right)^{-1} dr^2 +
        r^2\left(d\theta^2 + \sin^2\!\theta\,d\phi^2\right)
\end{equation} where $\tau$ is the imaginary time in the Euclidean sector of
the spacetime. With the conformal factor \begin{equation} e^{2w} = 1 -
\frac{2M}{r} \end{equation} this metric is the physical metric
corresponding to the optical metric considered in Paper II. As in
that case, the imaginary time dimension has periodicity $\kappa =
1/4M$, corresponding to a temperature associated with the
Hartle-Hawking state. We use the scaled spatial coordinate $x
\equiv 2M/r \equiv 1/2\kappa r$. With this choice, the black hole
horizon $r=2M$ is at $x=1$ while spatial infinity is at $x=0$.

We now use the results for the expansion tensors above to compute
the noise kernel coincident limit, along with the stress tensor
itself. For the stress tensor: \begin{eqnarray} \left<T_a{}^b \right> =
\left( \rho_{\infty} \right) {\rm diag}\left\{\right.&
 &  \left( 1 + 2\,x + 3\,{x^2} + 4\,{x^3} + 5\,{x^4} + 6\,{x^5} + 15\,{x^6}
\right)/3, \cr &
 &         \left( 1 + 2\,x + 3\,{x^2} \right) \,\left( 1 + 4\,{x^3} -
3\,{x^4}
\right) /3        , \cr &
 &         \left( 1 + 2\,x + 3\,{x^2} \right) \,\left( 1 + 4\,{x^3} -
3\,{x^4}
\right) /3        , \cr &
 &                    -1 - 2\,x - 3\,{x^2} - 4\,{x^3} - 5\,{x^4} - 6\,{x^5}
+ 33\,{x^6} \left. \right\} \label{Phys-emt-full} \end{eqnarray} where \begin{equation}
\rho_{\infty} = {\frac{{{\kappa}^4}}{480\,{{\pi }^2}}} = -\left.
\left<T_\tau{}^\tau \right> \right|_{r\goesto\infty} \end{equation} This
result agrees with Page's Eq. (83) \cite{Page82}. In his work,
Page showed the stress tensor must satisfy a
functional-differential scale equation under conformal
transformations. He then found, via trial and error,
 a general solution to this equation. This became the
basis for his computation of the stress tensor in the black hole
metric.

In contrast, we have worked directly with the Green function and
the point separation definition of the stress tensor.  The
conformal transformation properties of the geometric objects that
go into the Green function are studied and the corresponding
series expansions are developed. As can be seen from Appendices
\ref{apx-Gdiv} and \ref{apx-Gfin}, these expansions can get quite
involved even for just up to the second order terms needed for
the stress tensor. Thus our agreement with Page on the stress
tensor serves as an anchor for this work. Moreover, the methods
used for determining the series expansions are fundamentally
recursive. The zero order result (\ref{Phys-CiGren}) and the
second order result (\ref{Phys-emt-full}) are in agreement with
known results. Since the terms needed for the noise kernel are
generated recursively from these the symbolically computed
results for the Schwarzschild noise kernel coincident limit
should be accurate, up to the validity of the Gaussian
approximation to the Green function. These results are 
\begin{subequations}\begin{eqnarray}
N_\tau{}^\tau{}_\tau{}^\tau &=&
    {\frac{{{\rho_{\infty}}^2}}{756}} \left(
    219 + 876\,x + 2190\,{x^2}
    + 4380\,{x^3} + 7215\,{x^4} + 10464\,{x^5} +
\right. \cr  && \hspace{10mm}\left.
    + 16920\,{x^6}
    + 21424\,{x^7}
    -2984943\,{x^8} + 219140\,{x^9} + 197314\,{x^{10}}
\right. \cr  && \hspace{10mm}\left.
    + 180260\,{x^{11}}
    + 3292493\,{x^{12}}
\right)
\\
N_r{}^r{}_r{}^r &=&
    {\frac{{{\rho_{\infty}}^2}}{2268}} \left(
    137 + 548\,x + 1370\,{x^2}
    + 3860\,{x^3} + 9005\,{x^4} + 98432\,{x^5}
\right. \cr  && \hspace{10mm}\left.
    + 225080\,{x^6}
    + 408752\,{x^7}
    + 2553371\,{x^8} + 1725900\,{x^9} + 1206822\,{x^{10}}
\right. \cr  && \hspace{10mm}\left.
    + 761100\,{x^{11}}
    - 4090521\,{x^{12}}
\right)
\\
N_\theta{}^\theta{}_\theta{}^\theta &=&
    {\frac{{{\rho_{\infty}}^2}}{2268}} \left(
    137 + 548\,x + 1370\,{x^2} +
    + 2180\,{x^3} + 3965\,{x^4} + 37952\,{x^5}
\right. \cr  && \hspace{10mm}\left.
    + 77240\,{x^6}
    + 102992\,{x^7} -11973349\,{x^8} + 1817100\,{x^9} + 1721382\,{x^{10}}
\right. \cr  && \hspace{10mm}\left.
    + 1630140\,{x^{11}}
    + 12756759\,{x^{12}}
\right)
\\
N_\tau{}^\tau{}_r{}^r &=&
    {\frac{{{\rho_{\infty}}^2}}{2268}} \left(
    -219 - 876\,x - 2190\,{x^2}
    - 2140\,{x^3} - 495\,{x^4} + 2976\,{x^5}
\right. \cr  && \hspace{10mm}\left.
    - 10984\,{x^6}
    - 49872\,{x^7} + 1327551\,{x^8} - 230916\,{x^9} - 50834\,{x^{10}}
\right. \cr  && \hspace{10mm}\left.
    + 95356\,{x^{11}}
    - 774845\,{x^{12}}
\right)
\\
N_\tau{}^\tau{}_\theta{}^\theta &=&
    {\frac{{{\rho_{\infty}}^2}}{2268}} \left(
    -219 - 876\,x - 2190\,{x^2}
    - 5500\,{x^3} - 10575\,{x^4} - 17184\,{x^5}
\right. \cr  && \hspace{10mm}\left.
    - 19888\,{x^6}
    - 7200\,{x^7} -10917561\,{x^8} + 1056828\,{x^9} + 999526\,{x^{10}}
\right. \cr  && \hspace{10mm}\left.
    + 952012\,{x^{11}}
    + 11087083\,{x^{12}}
\right)
\\
N_r{}^r{}_\theta{}^\theta &=&
    {\frac{{{\rho_{\infty}}^2}}{2268}} \left(
    41 + 164\,x + 410\,{x^2}
    - 860\,{x^3} - 4255\,{x^4} - 50704\,{x^5}
\right. \cr  && \hspace{10mm}\left.
    - 107048\,{x^6}
    - 179440\,{x^7} + 1023059\,{x^8} + 159708\,{x^9} + 329206\,{x^{10}}
\right. \cr  && \hspace{10mm}\left.
    + 478972\,{x^{11}}
    + 1465003\,{x^{12}}
\right) 
\label{Phys-noise-results} 
\end{eqnarray}\end{subequations} 
Knowing the component
values, we determine the trace to be \begin{eqnarray} N_p{}^p{}_q{}^q =
{\frac{32000\,{{\rho_{\infty}}^2}\,{x^8}}{21}} \left( 1 + x
\right) \,\left( -27 + 31\,x \right) \,\left( 1 + {x^2} \right)
\label{Phys-noise-trace} 
\end{eqnarray}
As with the optical Schwarzschild
case, the trace under the Gaussian approximation fails to vanish,
which it should, for the massless conformal coupling case we are
considering. By taking the trace of the coincident limit
expressions for the noise kernel above, term by term, we find
that this arises from the non-vanishing of the fourth order
derivative terms such as $\left[ G_{\rm ren}{}_{;p}{}^p{}_q{}^q
\right]$. This is what we had expected, based on our analysis of
the optical Schwarzschild metric undertaken in Paper II. It is
not from our implementation of  the conformal transformation.

Our result yields a zero trace at spatial infinity $x\goesto 0$,
or $r\goesto\infty$. Using the fluctuation measure \begin{equation}
\Delta_{abcd} = \left|
  \frac{ \left< T _{ab} T_{cd} \right> -
     \left< T_{ab}\right> \left< T_{cd} \right>}
       { \left< T_{ab} T_{cd} \right> }
                \right|
 = \left| \frac{ 4 N_{abcd} }
    {4N_{abcd} + \left< T_{ab}\right> \left< T_{cd} \right> }
   \right|,
\label{def-Delta-measure} \end{equation} at $r\goesto\infty$, we get \begin{equation}
\begin{array}{ccccccc}
abcd: & \tau\tau\tau\tau & rrrr & \theta\theta\theta\theta &
\tau\tau rr & \tau\tau\theta\theta & rr\theta\theta \\
\Delta_{abcd}: &
  {\frac{73}{136}} &
  {\frac{137}{200}} &
  {\frac{137}{200}} &
  {\frac{73}{136}} &
  {\frac{73}{136}} &
  {\frac{41}{104}}
\end{array}
\end{equation} As we can expect for a Hartle-Hawking state, this matches
the thermal results of hot flat space we obtained in Paper II.

We also report the magnitude of the error at the horizon:
\begin{equation}
\begin{array}{ccccccc}
abcd: & \tau\tau\tau\tau & rrrr & \theta\theta\theta\theta &
\tau\tau rr & \tau\tau\theta\theta & rr\theta\theta \\
\frac{N_p{}^p{}_q{}^q}{N_a{}^b{}_c{}^d}: &
  1904 \% &
  1904 \% &
  894 \% &
  18278 \% &
  1775 \% &
  1775 \%
\end{array}
\end{equation}
These results show the Gaussian approximation has completely
broken down at the horizon.

This is not to say we can draw no conclusions other than to
discover the inadequacies of the Gaussian approximation. What is
important is the finiteness at the horizon of both the noise
kernel components and the error as expressed by the failure of
the trace of the noise kernel to vanish. In our computation of
both the stress tensor and the noise kernel, we have discovered
finiteness at the horizon is a ``fragile'' property. By this we
mean any small error in the symbolic code would result in a noise
kernel that diverges as $x\goesto 1$. This has lead us to develop
more than one way to determine the series for the optical metric
geometric objects, just to test the symbolic code. We arrive at
the results (\ref{Phys-noise-trace}) using more than one
computational path. In contrast if there were one single error in
the code, the resulting noise kernel cannot be finite on the
horizon.

It is the noise kernel itself, via its trace, that provides a
measure of the error of the Gaussian approximation, {\it i.e.},
we self-consistently compute both the noise kernel and its error.
Statements that address correcting the error of the Gaussian
approximation also apply to the noise kernel. Correcting the
Gaussian approximation will amount to finding the terms that need
to be included such that it satisfies the field equation to
fourth order in $\sigma^a$. When this corrected approximation is
then in turn used to compute the noise kernel, it will have to
correct the current noise kernel results
(\ref{Phys-noise-results}) in such a manner as to exactly cancel
the current trace (\ref{Phys-noise-trace}). Hence the correction
to the noise kernel will itself be finite at the horizon. With
this in mind, we can conclude the fluctuations of the stress
tensor, as measured by the coincident limit of the noise kernel,
are finite at the black-hole's horizon. This is one of the main
lessons learned from the analysis of the noise kernel, as derived
via point separation.

\section{Discussions}

Let us summarize our findings pertaining to two sets of issues:
the range of validity of the Gaussian approximation and the
results and usefulness of our program in spite of this
approximation.

Despite its success for the stress tensor calculation there is no
compelling reason to expect the Gaussian approximated Green
function to produce reasonable results for expressions involving
higher order covariant derivatives, as in the noise kernel.
Nonetheless the Gaussian approximation is very relevant because it
contains the leading order divergence. This structure will remain
even with a better approximation, while this leading order
divergence must be regularized. This step is needed regardless of
what form of the Green function one adopts -- future improvements
to the Gaussian approximation remains desirable, or, if the exact
Green function is derived in the optical metric only. Our work
lays down the structure and provides the details for its
implementation.

Now for the successes and failures of the Gaussian approximation
as applied to the computation of the noise kernel. On the positive
side our results for the fluctuations of the stress tensor of the
Hawking flux in the far field region checks with the analytic
results of Campos and Hu \cite{CamHu}. A fringe benefit is that
we can verify our procedure by explicitly re-deriving the Page
result \cite{Page82} for the stress tensor. We note that in
Page's original work, the direct use of the conformal
transformation was circumvented by ``guessing'' the solution to a
functional differential equation. Our result is the first we know
where the methodology of point separation was carried all the way
through to the final result. That we get the known results is a
check on our method and its correct implementation. On the
negative side, our calculation shows that the fluctuations of the
stress tensor based on the Gaussian approximation is unreliable
in regions close to the event horizon. We show this by checking
that the trace anomaly fails to vanish there. This result is not
unexpected, as can be inferred from our findings in Paper II.
Corrections to the Gaussian approximation need be introduced to
improve the accuracy. One important result which may be
under-appreciated is that we find a {\it finite} expression for
the noise kernel. Even though the error is large, as long as it
is finite, we know corrections to the approximations will
themselves be finite. Hence, where the approximate noise kernel
is finite, we can expect the full noise kernel to be finite. That
the noise kernel on the horizon of a Schwarzschild black hole is
finite is in itself a qualitatively significant result. This
dispels claims to the contrary based on intuitive arguments or
less rigorous calculations \cite{ForSva,BFP,Sorkin,CEIMP}.

Along the way to regularizing the Green function to fourth order,
necessary for the coincident limit of the noise kernel, we have
developed the series expansions of the various geometric objects
that make up the Green function to a high order. Once the
additional terms necessary to correct the Gaussian approximation
are determined, it is a simple matter to use the work contained
herein to compute these correction terms to a high enough order.
In this sense, this work not only lays down the tracks and defines
the steps, but also provides all the details necessary for
implementing the point separation program to calculate the
regularized noise kernel for quantum fields in curved spacetime.
To end with a more practical note, for black hole fluctuations
and backreaction calculations \cite{Vishu,HRS} which requires
results from investigations like ours here, we can only reiterate
what other researchers have said (lamented) before:
A better approximation to the Green function is to be desired.

\begin{acknowledgments}
NGP thanks Professors R. Wald for
correspondences and S. Christensen and L. Parker for the use of
the MathTensor program. BLH thanks Professors Paul Anderson, Larry
Ford and E. Verdaguer for discussions. This work is supported in
part by NSF grant PHY98-00967.
\end{acknowledgments}


\appendix


\section{Conformal Transformation}
\label{apx-conformal}

Since we compute the Green function in the optical metric
conformally related to the physical metric we are interested in,
we need to know how the geometric objects we use
conformally transform.
We denote objects in the optical metric
with an overbar and the covariant derivative commensurate with
the optical metric by a vertical stroke.
The two metrics are related via
\begin{equation}
  \bar g_{ab} = e^{-2\w} g_{ab}
\end{equation}
We develop the series expansions of
the world function $\bar\sigma$, to eighth order, and $\bar\VanD$
to sixth order. We also  need the function
\begin{equation}
  \Sigma(x,y) = e^{\w(x)+\w(y)}\bar\sigma  - \sigma,
\end{equation}
whose series expansion
readily follows once the series for $\bar\sigma$ is determined.
The most straightforward method for determining
these series expansions is to start by considering the conformal
transformation of the differential equations each must satisfy.

Considering first $\bar\sigma$, it satisfies, in terms of the optical
metric, the equation
\begin{equation}
\bar\sigma = \frac{1}{2} \bar\sigma_{|p} \bar\sigma^{|p}
  = \frac{1}{2} \bar g^{pq} \bar\sigma_{|p} \bar\sigma_{|q}.
\end{equation}
Using $\bar\sigma_{|a}=\bar\sigma_{,a}=\bar\sigma_{;a}$,
this equation transforms to
\begin{equation} \bar\sigma
  = \frac{e^{2\w}}{2}  g^{pq} \bar\sigma_{;p} \bar\sigma_{;q}
 = \frac{e^{2\w}}{2} \bar\sigma_{;p} \bar\sigma^{;p}
\label{Apx-bsigma1}
\end{equation}
and we now have an equation purely in
terms of the physical metric. We still have
\begin{equation}
\lC \bar\sigma \rC = 0,
\hspace{2cm}
\lC \bar\sigma_{;a} \rC = 0.
\end{equation}
and we assume the series expansion
\begin{eqnarray}
\bar\sigma &=&
  \Upsilon^{(2)}_{pq}    \sigma^p \sigma^q
 + \Upsilon^{(3)}_{pqr}   \sigma^p \sigma^q \sigma^r
 + \Upsilon^{(4)}_{pqrs}  \sigma^p \sigma^q \sigma^r \sigma^s
 + \Upsilon^{(5)}_{pqrst} \sigma^p \sigma^q \sigma^r \sigma^s \sigma^t \cr
&&\hspace{2cm}
 + \Upsilon^{(6)}_{pqrstu}\sigma^p \sigma^q \sigma^r \sigma^s
        \sigma^t \sigma^u
 + \Upsilon^{(7)}_{pqrstuv}
   \sigma^p \sigma^q \sigma^r \sigma^s \sigma^t \sigma^u \sigma^v \cr
&&\hspace{2cm}
 + \Upsilon^{(8)}_{pqrstuvw}
   \sigma^p \sigma^q \sigma^r \sigma^s \sigma^t \sigma^u \sigma^v \sigma^w
\label{Apx-bsigma2}
\end{eqnarray}
To determine the expansion tensors, we
proceed as we did in Paper II for $\VanD$ and substitute the expansion into
the differential equation (\ref{Apx-bsigma1})
and collect terms by their order in $\sigma^a$.

The order $\sigma$ term must satisfy
\begin{equation}
2\,\Upsilon^{(2)}{}_{a}{}_{b} -
  {e^{2\,\omega}}\,\left( \Upsilon^{(2)}{}_{a}{}_{p}\,
      \Upsilon^{(2)}{}_{b}{}^{p} +
     \Upsilon^{(2)}{}_{p}{}_{a}\,\Upsilon^{(2)}{}^{p}{}_{b} +
     2\,\Upsilon^{(2)}{}_{a}{}_{p}\,\Upsilon^{(2)}{}^{p}{}_{b} \right)  = 0.
\end{equation}
This has the solution
\begin{subequations}
\begin{equation}
\Upsilon^{(2)}{}_{a}{}_{b} =
  {\frac{g{}_{a}{}_{b}}{2\,{e^{2\,\omega}}}},
\label{Apx-bsigma2rule}
\end{equation}
which can be seen via substitution.
Since $\bar\sigma$ is a symmetric function, we use
the results of Paper II's Appendix B, which give the odd order expansion
coefficients in terms of the even order one. Thus the third order
coefficient is
\begin{equation} \Upsilon^{(3)}{}_{a}{}_{b}{}_{c} = -
{\frac{\Upsilon^{(2)}{}_{a}{}_{b}{}_{;}{}_{c}}{2}}
 = {\frac{\omega{}_{;}{}_{c}\,g{}_{a}{}_{b}}{2\,{e^{2\,\omega}}}}
\end{equation}
With this, we have set up the recursion. Now it is just a
matter to proceed as with the computation of the series expansion
Van Vleck-Morette determinant as presented in Paper II.
The results for the rest of the
expansion tensors for $\bar\sigma$ are:
\begin{eqnarray}
%
{e^{2\,\omega}}\,\Upsilon^{(4)}{}_{a}{}_{b}{}_{c}{}_{d} &\symmeq&
{\frac{g{}_{c}{}_{d}}{6}} \left(
2\,\omega{}_{;}{}_{a}\,\omega{}_{;}{}_{b} -
\omega{}_{;}{}_{a}{}_{b} \right)
  - {\frac{g{}_{a}{}_{b}\,g{}_{c}{}_{d}}{24}}
\omega{}_{;}{}_{p}\,\omega{}^{;}{}^{p}
\\ \cr
%
%
{e^{2\,\omega}}\,\Upsilon^{(5)}{}_{a}{}_{b}{}_{c}{}_{d}{}_{e}
&\symmeq& {\frac{g{}_{a}{}_{b}}{24}} \left(
4\,\omega{}_{;}{}_{c}\,\omega{}_{;}{}_{d}\,\omega{}_{;}{}_{e} -
  6\,\omega{}_{;}{}_{c}\,\omega{}_{;}{}_{d}{}_{e} +
  \omega{}_{;}{}_{c}{}_{d}{}_{e} \right)
\cr &&
 - {\frac{g{}_{a}{}_{b}\,g{}_{c}{}_{d}}{24}} \omega{}_{;}{}_{p}\,\left(
\omega{}_{;}{}_{e}\,\omega{}^{;}{}^{p} -
    \omega{}_{;}{}_{e}{}^{p} \right)
\\ \cr
%
%
{e^{2\,\omega}}\,\Upsilon^{(6)}{}_{a}{}_{b}{}_{c}{}_{d}{}_{e}{}_{f}
&\symmeq& {\frac{g{}_{a}{}_{b}}{120}}\left(
   8\,\omega{}_{;}{}_{c}\,\omega{}_{;}{}_{d}\,\omega{}_{;}{}_{e}\,
   \omega{}_{;}{}_{f} - 24\,\omega{}_{;}{}_{c}\,\omega{}_{;}{}_{d}\,
   \omega{}_{;}{}_{e}{}_{f} + 6\,\omega{}_{;}{}_{c}{}_{d}\,
   \omega{}_{;}{}_{e}{}_{f}
\right. \cr && \hspace{10mm} \left.
   + 8\,\omega{}_{;}{}_{c}\,\omega{}_{;}{}_{d}{}_{e}{}_{f} -
  \omega{}_{;}{}_{c}{}_{d}{}_{e}{}_{f}
\right)\cr
&&-
  {\frac{g{}_{a}{}_{b}\,g{}_{c}{}_{d}}{720}}\left(
   18\,\omega{}_{;}{}_{p}\,\omega{}_{;}{}_{e}\,\omega{}_{;}{}_{f}\,
   \omega{}^{;}{}^{p} - 8\,\omega{}_{;}{}_{p}\,\omega{}^{;}{}^{p}\,
   \omega{}_{;}{}_{e}{}_{f}
\right. \cr && \hspace{10mm} \left.
   + 9\,\omega{}_{;}{}_{p}\,\omega{}_{;}{}_{e}{}_{f}{}^{p} +
  12\,\omega{}_{;}{}_{p}\,\omega{}_{;}{}_{q}\,R{}_{e}{}^{q}{}_{f}{}^{p}
\right)\cr
&&+ \frac{1}{720}
  g{}_{a}{}_{b}\,g{}_{c}{}_{d}\,g{}_{e}{}_{f}
    \omega{}_{;}{}_{p}\,\omega{}_{;}{}_{q}\,
  \left( \omega{}^{;}{}^{p}\,\omega{}^{;}{}^{q} -
    3\,\omega{}^{;}{}^{p}{}^{q} \right)
\\ \cr
%
%
{e^{2\,\omega}}\,\Upsilon^{(7)}{}_{a}{}_{b}{}_{c}{}_{d}{}_{e}{}_{f}{}_{g}
&\symmeq& {\frac{g{}_{a}{}_{b}}{720}} \left(
16\,\omega{}_{;}{}_{c}\,\omega{}_{;}{}_{d}\,\omega{}_{;}{}_{e}\,
   \omega{}_{;}{}_{f}\,\omega{}_{;}{}_{g} -
  80\,\omega{}_{;}{}_{c}\,\omega{}_{;}{}_{d}\,\omega{}_{;}{}_{e}\,
   \omega{}_{;}{}_{f}{}_{g}
\right. \cr && \hspace{6mm} \left.
    +
60\,\omega{}_{;}{}_{c}\,\omega{}_{;}{}_{d}{}_{e}\,\omega{}_{;}{}_{f}{}_{g}
+
  40\,\omega{}_{;}{}_{c}\,\omega{}_{;}{}_{d}\,
   \omega{}_{;}{}_{e}{}_{f}{}_{g} -
  20\,\omega{}_{;}{}_{c}{}_{d}\,\omega{}_{;}{}_{e}{}_{f}{}_{g}
\right. \cr && \hspace{10mm} \left.
      -10\,\omega{}_{;}{}_{c}\,\omega{}_{;}{}_{d}{}_{e}{}_{f}{}_{g} +
  \omega{}_{;}{}_{c}{}_{d}{}_{e}{}_{f}{}_{g}
\right)\cr
&&- {\frac{g{}_{a}{}_{b}\,g{}_{c}{}_{d}}{4320}} \left(
48\,\omega{}_{;}{}_{p}\,\omega{}_{;}{}_{e}\,\omega{}_{;}{}_{f}\,
   \omega{}_{;}{}_{g}\,\omega{}^{;}{}^{p} -
  27\,\omega{}_{;}{}_{p}\,\omega{}_{;}{}_{q}\,
   R{}_{e}{}^{q}{}_{f}{}^{p}{}_{;}{}_{g}
\right. \cr && \hspace{6mm} \left.
 -66\,\omega{}_{;}{}_{p}\,\omega{}_{;}{}_{e}\,\omega{}^{;}{}^{p}\,
   \omega{}_{;}{}_{f}{}_{g} + 50\,\omega{}_{;}{}_{p}\,
   \omega{}_{;}{}_{e}{}^{p}\,\omega{}_{;}{}_{f}{}_{g}
\right. \cr && \hspace{6mm} \left.
 -156\,\omega{}_{;}{}_{p}\,\omega{}_{;}{}_{e}\,\omega{}_{;}{}_{f}\,
   \omega{}_{;}{}_{g}{}^{p} + 72\,\omega{}_{;}{}_{e}\,
   \omega{}_{;}{}_{p}{}_{f}\,\omega{}_{;}{}_{g}{}^{p} +
  22\,\omega{}_{;}{}_{p}\,\omega{}_{;}{}_{e}{}_{f}\,\omega{}_{;}{}_{g}{}^{p}
\right. \cr && \hspace{6mm} \left.
 +9\,\omega{}_{;}{}_{p}\,\omega{}^{;}{}^{p}\,\omega{}_{;}{}_{e}{}_{f}{}_{g}
+
  78\,\omega{}_{;}{}_{p}\,\omega{}_{;}{}_{e}\,
   \omega{}_{;}{}_{f}{}_{g}{}^{p} -
  24\,\omega{}_{;}{}_{p}{}_{e}\,\omega{}_{;}{}_{f}{}_{g}{}^{p}
\right. \cr && \hspace{6mm} \left.
 -6\,\omega{}_{;}{}_{p}{}_{e}\,\omega{}_{;}{}_{f}{}^{p}{}_{g} -
  12\,\omega{}_{;}{}_{p}\,\omega{}_{;}{}_{e}{}_{f}{}_{g}{}^{p} -
  6\,\omega{}_{;}{}_{p}\,\omega{}_{;}{}_{e}{}_{f}{}^{p}{}_{g}
\right. \cr && \hspace{6mm}
 6\,\omega{}_{;}{}_{p}\,\omega{}_{;}{}_{e}{}^{p}{}_{f}{}_{g} -
  32\,\omega{}_{;}{}_{p}\,\omega{}_{;}{}_{q}{}_{e}\,
   R{}_{f}{}^{p}{}_{g}{}^{q} + 96\,\omega{}_{;}{}_{p}\,\omega{}_{;}{}_{q}\,
   \omega{}_{;}{}_{e}\,R{}_{f}{}^{q}{}_{g}{}^{p}
\cr && \hspace{6mm} \left.
-52\,\omega{}_{;}{}_{p}\,\omega{}_{;}{}_{q}{}_{e}\,R{}_{f}{}^{q}{}_{g}{}^{p}
\right)\cr
&&+\frac{1}{8640} g{}_{a}{}_{b}\,g{}_{c}{}_{d}\,g{}_{e}{}_{f}
\omega{}_{;}{}_{p} \left(
12\,\omega{}_{;}{}_{q}\,\omega{}_{;}{}_{g}\,\omega{}^{;}{}^{p}\,
   \omega{}^{;}{}^{q} - 14\,\omega{}_{;}{}_{q}\,\omega{}^{;}{}^{q}\,
   \omega{}_{;}{}_{g}{}^{p}
\right. \cr && \hspace{6mm} \left.
 -10\,\omega{}_{;}{}_{q}\,\omega{}^{;}{}^{p}\,\omega{}_{;}{}_{g}{}^{q} +
  9\,\omega{}_{;}{}_{q}{}^{p}\,\omega{}_{;}{}_{g}{}^{q} -
  36\,\omega{}_{;}{}_{q}\,\omega{}_{;}{}_{g}\,\omega{}^{;}{}^{p}{}^{q}
\right. \cr && \hspace{6mm} \left.
 +27\,\omega{}_{;}{}_{q}{}_{g}\,\omega{}^{;}{}^{p}{}^{q} -
  6\,\omega{}_{;}{}_{q}\,\omega{}_{;}{}_{g}{}^{p}{}^{q} +
  18\,\omega{}_{;}{}_{q}\,\omega{}_{;}{}_{g}{}^{q}{}^{p}
\right. \cr && \hspace{6mm} \left.
 +6\,\omega{}_{;}{}_{q}\,\omega{}^{;}{}^{p}{}^{q}{}_{g} -
  6\,\omega{}_{;}{}_{q}\,\omega{}_{;}{}_{r}\,R{}_{g}{}^{p}{}^{q}{}^{r}
\right) \end{eqnarray} \begin{eqnarray}
%
%
{e^{2\,\omega}}\,\Upsilon^{(8)}{}_{a}{}_{b}{}_{c}{}_{d}{}_{e}{}_{f}{}_{g}
   {}_{h} &\symmeq&
{\frac{g{}_{a}{}_{b}}{5040}} \left(
32\,\omega{}_{;}{}_{c}\,\omega{}_{;}{}_{d}\,\omega{}_{;}{}_{e}\,
   \omega{}_{;}{}_{f}\,\omega{}_{;}{}_{g}\,\omega{}_{;}{}_{h} -
  240\,\omega{}_{;}{}_{c}\,\omega{}_{;}{}_{d}\,\omega{}_{;}{}_{e}\,
   \omega{}_{;}{}_{f}\,\omega{}_{;}{}_{g}{}_{h}
\right. \cr && \hspace{6mm} \left.
 +360\,\omega{}_{;}{}_{c}\,\omega{}_{;}{}_{d}\,\omega{}_{;}{}_{e}{}_{f}\,
   \omega{}_{;}{}_{g}{}_{h} - 60\,\omega{}_{;}{}_{c}{}_{d}\,
   \omega{}_{;}{}_{e}{}_{f}\,\omega{}_{;}{}_{g}{}_{h}
\right. \cr && \hspace{6mm} \left.
 +160\,\omega{}_{;}{}_{c}\,\omega{}_{;}{}_{d}\,\omega{}_{;}{}_{e}\,
   \omega{}_{;}{}_{f}{}_{g}{}_{h} -
  240\,\omega{}_{;}{}_{c}\,\omega{}_{;}{}_{d}{}_{e}\,
   \omega{}_{;}{}_{f}{}_{g}{}_{h}
\right. \cr && \hspace{6mm} \left.
 +20\,\omega{}_{;}{}_{c}{}_{d}{}_{e}\,\omega{}_{;}{}_{f}{}_{g}{}_{h} -
  60\,\omega{}_{;}{}_{c}\,\omega{}_{;}{}_{d}\,
   \omega{}_{;}{}_{e}{}_{f}{}_{g}{}_{h} +
  30\,\omega{}_{;}{}_{c}{}_{d}\,\omega{}_{;}{}_{e}{}_{f}{}_{g}{}_{h}
\right. \cr && \hspace{6mm} \left.
 +12\,\omega{}_{;}{}_{c}\,\omega{}_{;}{}_{d}{}_{e}{}_{f}{}_{g}{}_{h} -
  \omega{}_{;}{}_{c}{}_{d}{}_{e}{}_{f}{}_{g}{}_{h}
\right)\cr
%
&&-{\frac{g{}_{a}{}_{b}\,g{}_{c}{}_{d}}{120960}} \left(
480\,\omega{}_{;}{}_{p}\,\omega{}_{;}{}_{e}\,\omega{}_{;}{}_{f}\,
  \omega{}_{;}{}_{g}\,\omega{}_{;}{}_{h}\,\omega{}^{;}{}^{p}
\right. \cr && \hspace{6mm} \left.
  -1360\,\omega{}_{;}{}_{p}\,\omega{}_{;}{}_{e}\,\omega{}_{;}{}_{f}\,
   \omega{}^{;}{}^{p}\,\omega{}_{;}{}_{g}{}_{h} +
  308\,\omega{}_{;}{}_{p}\,\omega{}^{;}{}^{p}\,\omega{}_{;}{}_{e}{}_{f}\,
   \omega{}_{;}{}_{g}{}_{h}
\right. \cr && \hspace{6mm} \left.
 +1936\,\omega{}_{;}{}_{p}\,\omega{}_{;}{}_{e}\,\omega{}_{;}{}_{f}{}^{p}\,
   \omega{}_{;}{}_{g}{}_{h} - 200\,\omega{}_{;}{}_{p}{}_{e}\,
   \omega{}_{;}{}_{f}{}^{p}\,\omega{}_{;}{}_{g}{}_{h}
\right. \cr && \hspace{6mm} \left.
  -2080\,\omega{}_{;}{}_{p}\,\omega{}_{;}{}_{e}\,\omega{}_{;}{}_{f}\,
   \omega{}_{;}{}_{g}\,\omega{}_{;}{}_{h}{}^{p} +
  1552\,\omega{}_{;}{}_{e}\,\omega{}_{;}{}_{f}\,\omega{}_{;}{}_{p}{}_{g}\,
   \omega{}_{;}{}_{h}{}^{p}
\right. \cr && \hspace{6mm} \left.
 +1048\,\omega{}_{;}{}_{p}\,\omega{}_{;}{}_{e}\,\omega{}_{;}{}_{f}{}_{g}\,
   \omega{}_{;}{}_{h}{}^{p} - 504\,\omega{}_{;}{}_{p}{}_{e}\,
   \omega{}_{;}{}_{f}{}_{g}\,\omega{}_{;}{}_{h}{}^{p}
\right. \cr && \hspace{6mm} \left.
 +384\,\omega{}_{;}{}_{p}\,\omega{}_{;}{}_{e}\,\omega{}^{;}{}^{p}\,
   \omega{}_{;}{}_{f}{}_{g}{}_{h} -
  444\,\omega{}_{;}{}_{p}\,\omega{}_{;}{}_{e}{}^{p}\,
   \omega{}_{;}{}_{f}{}_{g}{}_{h}
\right. \cr && \hspace{6mm} \left.
 +1584\,\omega{}_{;}{}_{p}\,\omega{}_{;}{}_{e}\,\omega{}_{;}{}_{f}\,
   \omega{}_{;}{}_{g}{}_{h}{}^{p} -
  1128\,\omega{}_{;}{}_{e}\,\omega{}_{;}{}_{p}{}_{f}\,
   \omega{}_{;}{}_{g}{}_{h}{}^{p}
\right. \cr && \hspace{6mm} \left.
  -828\,\omega{}_{;}{}_{p}\,\omega{}_{;}{}_{e}{}_{f}\,
   \omega{}_{;}{}_{g}{}_{h}{}^{p} +
  6\,\omega{}_{;}{}_{p}{}_{e}{}_{f}\,\omega{}_{;}{}_{g}{}_{h}{}^{p}
\right. \cr && \hspace{6mm} \left.
 +135\,\omega{}_{;}{}_{e}{}_{f}{}_{p}\,\omega{}_{;}{}_{g}{}_{h}{}^{p} +
  48\,\omega{}_{;}{}_{p}\,\omega{}_{;}{}_{e}\,\omega{}_{;}{}_{f}\,
   \omega{}_{;}{}_{g}{}^{p}{}_{h}
\right. \cr && \hspace{6mm} \left.
  -264\,\omega{}_{;}{}_{e}\,\omega{}_{;}{}_{p}{}_{f}\,
   \omega{}_{;}{}_{g}{}^{p}{}_{h} +
  72\,\omega{}_{;}{}_{p}\,\omega{}_{;}{}_{e}{}_{f}\,
   \omega{}_{;}{}_{g}{}^{p}{}_{h}
\right. \cr && \hspace{6mm} \left.
 +84\,\omega{}_{;}{}_{p}{}_{e}{}_{f}\,\omega{}_{;}{}_{g}{}^{p}{}_{h} -
  90\,\omega{}_{;}{}_{e}{}_{f}{}_{p}\,\omega{}_{;}{}_{g}{}^{p}{}_{h} -
  36\,\omega{}_{;}{}_{p}\,\omega{}^{;}{}^{p}\,
   \omega{}_{;}{}_{e}{}_{f}{}_{g}{}_{h}
\right. \cr && \hspace{6mm} \left.
  -504\,\omega{}_{;}{}_{p}\,\omega{}_{;}{}_{e}\,
   \omega{}_{;}{}_{f}{}_{g}{}_{h}{}^{p} +
  144\,\omega{}_{;}{}_{p}{}_{e}\,\omega{}_{;}{}_{f}{}_{g}{}_{h}{}^{p}
\right. \cr && \hspace{6mm} \left.
  -288\,\omega{}_{;}{}_{p}\,\omega{}_{;}{}_{e}\,
   \omega{}_{;}{}_{f}{}_{g}{}^{p}{}_{h} +
  96\,\omega{}_{;}{}_{p}{}_{e}\,\omega{}_{;}{}_{f}{}_{g}{}^{p}{}_{h}
\right. \cr && \hspace{6mm} \left.
  +264\,\omega{}_{;}{}_{p}\,\omega{}_{;}{}_{e}\,
   \omega{}_{;}{}_{f}{}^{p}{}_{g}{}_{h} -
  48\,\omega{}_{;}{}_{p}{}_{e}\,\omega{}_{;}{}_{f}{}^{p}{}_{g}{}_{h} +
  60\,\omega{}_{;}{}_{p}\,\omega{}_{;}{}_{e}{}_{f}{}_{g}{}_{h}{}^{p}
\right. \cr && \hspace{6mm} \left.
  +36\,\omega{}_{;}{}_{p}\,\omega{}_{;}{}_{e}{}_{f}{}_{g}{}^{p}{}_{h} +
  12\,\omega{}_{;}{}_{p}\,\omega{}_{;}{}_{e}{}_{f}{}^{p}{}_{g}{}_{h} -
  48\,\omega{}_{;}{}_{p}\,\omega{}_{;}{}_{e}{}^{p}{}_{f}{}_{g}{}_{h}
\right. \cr && \hspace{6mm} \left.
  +36\,\omega{}_{;}{}_{p}\,\omega{}_{;}{}_{e}{}_{f}{}_{g}{}^{p}{}_{h} +
  12\,\omega{}_{;}{}_{p}\,\omega{}_{;}{}_{e}{}_{f}{}^{p}{}_{g}{}_{h} -
  48\,\omega{}_{;}{}_{p}\,\omega{}_{;}{}_{e}{}^{p}{}_{f}{}_{g}{}_{h}
\right. \cr && \hspace{6mm} \left.
     -1068\,\omega{}_{;}{}_{p}\,\omega{}_{;}{}_{q}\,\omega{}_{;}{}_{e}\,
   R{}_{f}{}^{q}{}_{g}{}^{p}{}_{;}{}_{h} -
  912\,\omega{}_{;}{}_{p}\,R{}_{q}{}_{e}{}_{f}{}^{p}{}_{;}{}_{g}\,
   \omega{}_{;}{}_{h}{}^{q}
\right. \cr && \hspace{6mm} \left.
   + 180\,\omega{}_{;}{}_{p}\,\omega{}_{;}{}_{q}\,
   R{}_{e}{}^{q}{}_{f}{}^{p}{}_{;}{}_{g}{}_{h} -
  336\,\omega{}_{;}{}_{p}\,\omega{}_{;}{}_{q}\,R{}_{r}{}_{e}{}_{f}{}^{p}\,
   R{}_{g}{}^{q}{}_{h}{}^{r}
\right. \cr && \hspace{6mm} \left.
  -8\,\omega{}_{;}{}_{p}\,R{}_{g}{}^{p}{}_{h}{}^{q} \left(
171\,\omega{}_{;}{}_{e}\,\omega{}_{;}{}_{q}{}_{f} +
  14\,\omega{}_{;}{}_{q}{}_{e}{}_{f} - 45\,\omega{}_{;}{}_{e}{}_{f}{}_{q}
\right) \right. \cr && \hspace{6mm} \left.
 + 8\,R{}_{g}{}^{q}{}_{h}{}^{p} \left(
    232\,\omega{}_{;}{}_{p}\,\omega{}_{;}{}_{q}\,\omega{}_{;}{}_{e}\,
   \omega{}_{;}{}_{f} - 271\,\omega{}_{;}{}_{p}\,\omega{}_{;}{}_{e}\,
   \omega{}_{;}{}_{q}{}_{f}
\right. \right. \cr && \hspace{21mm} \left. \left.
    +82\,\omega{}_{;}{}_{p}{}_{e}\,\omega{}_{;}{}_{q}{}_{f} -
  120\,\omega{}_{;}{}_{p}\,\omega{}_{;}{}_{q}\,\omega{}_{;}{}_{e}{}_{f}
\right. \right. \cr && \hspace{21mm} \left. \left.
    +56\,\omega{}_{;}{}_{p}\,\omega{}_{;}{}_{q}{}_{e}{}_{f} +
  15\,\omega{}_{;}{}_{p}\,\omega{}_{;}{}_{e}{}_{f}{}_{q}
\right) \right)\cr
%
&&{\frac{g{}_{a}{}_{b}\,g{}_{c}{}_{d}\,g{}_{e}{}_{f}}{241920}}
\left(
192\,\omega{}_{;}{}_{p}\,\omega{}_{;}{}_{q}\,\omega{}_{;}{}_{g}\,
  \omega{}_{;}{}_{h}\,\omega{}^{;}{}^{p}\,\omega{}^{;}{}^{q}
\right. \cr && \hspace{6mm} \left.
  -96\,\omega{}_{;}{}_{p}\,\omega{}_{;}{}_{q}\,\omega{}^{;}{}^{p}\,
   \omega{}^{;}{}^{q}\,\omega{}_{;}{}_{g}{}_{h} -
  360\,\omega{}_{;}{}_{p}\,\omega{}_{;}{}_{q}\,\omega{}_{;}{}_{g}\,
   \omega{}^{;}{}^{q}\,\omega{}_{;}{}_{h}{}^{p}
\right. \cr && \hspace{6mm} \left.
  +464\,\omega{}_{;}{}_{p}\,\omega{}_{;}{}_{q}\,\omega{}_{;}{}_{g}{}^{q}\,
   \omega{}_{;}{}_{h}{}^{p} - 408\,\omega{}_{;}{}_{p}\,\omega{}_{;}{}_{q}\,
   \omega{}_{;}{}_{g}\,\omega{}^{;}{}^{p}\,\omega{}_{;}{}_{h}{}^{q}
\right. \cr && \hspace{6mm} \left.
  160\,\omega{}_{;}{}_{p}\,\omega{}^{;}{}^{p}\,\omega{}_{;}{}_{q}{}_{g}\,
   \omega{}_{;}{}_{h}{}^{q} + 316\,\omega{}_{;}{}_{p}\,\omega{}_{;}{}_{g}\,
   \omega{}_{;}{}_{q}{}^{p}\,\omega{}_{;}{}_{h}{}^{q}
\right. \cr && \hspace{6mm} \left.
  +-64\,\omega{}_{;}{}_{p}{}_{g}\,\omega{}_{;}{}_{q}{}^{p}\,
   \omega{}_{;}{}_{h}{}^{q} - 64\,\omega{}_{;}{}_{p}{}_{q}\,
   \omega{}_{;}{}_{g}{}^{p}\,\omega{}_{;}{}_{h}{}^{q}
\right. \cr && \hspace{6mm} \left.
  -544\,\omega{}_{;}{}_{p}\,\omega{}_{;}{}_{q}\,\omega{}_{;}{}_{g}\,
   \omega{}_{;}{}_{h}\,\omega{}^{;}{}^{p}{}^{q} +
  948\,\omega{}_{;}{}_{p}\,\omega{}_{;}{}_{g}\,\omega{}_{;}{}_{q}{}_{h}\,
   \omega{}^{;}{}^{p}{}^{q}
\right. \cr && \hspace{6mm} \left.
  -128\,\omega{}_{;}{}_{p}{}_{g}\,\omega{}_{;}{}_{q}{}_{h}\,
   \omega{}^{;}{}^{p}{}^{q} + 216\,\omega{}_{;}{}_{p}\,\omega{}_{;}{}_{q}\,
   \omega{}_{;}{}_{g}{}_{h}\,\omega{}^{;}{}^{p}{}^{q}
\right. \cr && \hspace{6mm} \left.
  +168\,\omega{}_{;}{}_{p}\,\omega{}_{;}{}_{q}\,\omega{}^{;}{}^{q}\,
   \omega{}_{;}{}_{g}{}_{h}{}^{p} -
  144\,\omega{}_{;}{}_{p}\,\omega{}_{;}{}_{q}\,\omega{}^{;}{}^{p}\,
   \omega{}_{;}{}_{g}{}_{h}{}^{q}
\right. \cr && \hspace{6mm} \left.
  -300\,\omega{}_{;}{}_{p}\,\omega{}_{;}{}_{q}{}^{p}\,
   \omega{}_{;}{}_{g}{}_{h}{}^{q} -
  40\,\omega{}_{;}{}_{p}\,\omega{}_{;}{}_{q}\,\omega{}^{;}{}^{q}\,
   \omega{}_{;}{}_{g}{}^{p}{}_{h}
\right. \cr && \hspace{6mm} \left.
  +184\,\omega{}_{;}{}_{p}\,\omega{}_{;}{}_{q}\,\omega{}^{;}{}^{p}\,
   \omega{}_{;}{}_{g}{}^{q}{}_{h} -
  288\,\omega{}_{;}{}_{p}\,\omega{}_{;}{}_{q}\,\omega{}_{;}{}_{g}\,
   \omega{}_{;}{}_{h}{}^{p}{}^{q}
\right. \cr && \hspace{6mm} \left.
  -96\,\omega{}_{;}{}_{p}\,\omega{}_{;}{}_{q}{}_{g}\,
   \omega{}_{;}{}_{h}{}^{p}{}^{q} +
  672\,\omega{}_{;}{}_{p}\,\omega{}_{;}{}_{q}\,\omega{}_{;}{}_{g}\,
   \omega{}_{;}{}_{h}{}^{q}{}^{p}
\right. \cr && \hspace{6mm} \left.
  -264\,\omega{}_{;}{}_{p}\,\omega{}_{;}{}_{q}{}_{g}\,
   \omega{}_{;}{}_{h}{}^{q}{}^{p} +
  240\,\omega{}_{;}{}_{p}\,\omega{}_{;}{}_{q}\,\omega{}_{;}{}_{g}\,
   \omega{}^{;}{}^{p}{}^{q}{}_{h}
\right. \cr && \hspace{6mm} \left.
  -168\,\omega{}_{;}{}_{p}\,\omega{}_{;}{}_{q}{}_{g}\,
   \omega{}^{;}{}^{p}{}^{q}{}_{h} -
  12\,\omega{}_{;}{}_{p}\,\omega{}_{;}{}_{q}\,
   \omega{}_{;}{}_{g}{}_{h}{}^{p}{}^{q}
\right. \cr && \hspace{6mm} \left.
  -132\,\omega{}_{;}{}_{p}\,\omega{}_{;}{}_{q}\,
   \omega{}_{;}{}_{g}{}_{h}{}^{q}{}^{p} +
  66\,\omega{}_{;}{}_{p}\,\omega{}_{;}{}_{q}\,
   \omega{}_{;}{}_{g}{}^{p}{}_{h}{}^{q}
\right. \cr && \hspace{6mm} \left.
  +54\,\omega{}_{;}{}_{p}\,\omega{}_{;}{}_{q}\,
   \omega{}_{;}{}_{g}{}^{p}{}^{q}{}_{h} +
  6\,\omega{}_{;}{}_{p}\,\omega{}_{;}{}_{q}\,
   \omega{}_{;}{}_{g}{}^{q}{}_{h}{}^{p}
\right. \cr && \hspace{6mm} \left.
  -78\,\omega{}_{;}{}_{p}\,\omega{}_{;}{}_{q}\,
   \omega{}_{;}{}_{g}{}^{q}{}^{p}{}_{h} -
  48\,\omega{}_{;}{}_{p}\,\omega{}_{;}{}_{q}\,
   \omega{}^{;}{}^{p}{}^{q}{}_{g}{}_{h}
\right. \cr && \hspace{6mm} \left.
  -3\,\omega{}_{;}{}_{p}\,\omega{}_{;}{}_{q}\,\omega{}_{;}{}_{r} \left(
    15\,R{}_{g}{}^{p}{}_{h}{}^{q}{}^{;}{}^{r} -
  13\,R{}_{g}{}^{p}{}^{q}{}^{r}{}_{;}{}_{h} \right.
\right. \cr && \hspace{21mm} \left. \left.
   + 30\,R{}_{g}{}^{r}{}_{h}{}^{p}{}^{;}{}^{q} +
  15\,R{}_{g}{}^{r}{}_{h}{}^{q}{}^{;}{}^{p} +
  13\,R{}_{g}{}^{r}{}^{p}{}^{q}{}_{;}{}_{h} \right)
\right. \cr && \hspace{6mm} \left.
 +-80\,\omega{}_{;}{}_{p}\,\omega{}_{;}{}_{q}\,\omega{}_{;}{}_{r}{}^{q}\,
  \left( 5\,R{}_{g}{}^{p}{}_{h}{}^{r} + 4\,R{}_{g}{}^{r}{}_{h}{}^{p} \right)
\right. \cr && \hspace{6mm} \left.
 +48\,\omega{}_{;}{}_{p}\,\omega{}_{;}{}_{q}\,
  \left( 3\,\omega{}_{;}{}_{r}{}_{g}\,R{}_{h}{}^{p}{}^{q}{}^{r} +
    5\,\omega{}_{;}{}_{r}\,\omega{}_{;}{}_{g}\,R{}_{h}{}^{q}{}^{p}{}^{r}
     \right)
\right. \cr && \hspace{6mm} \left.
  32\,\omega{}_{;}{}_{p}\,\omega{}_{;}{}_{q}\,
  \left( 3\,\omega{}_{;}{}_{r}\,\omega{}^{;}{}^{r}\,
     R{}_{g}{}^{p}{}_{h}{}^{q} +
    4\,\omega{}_{;}{}_{r}{}_{g}\,R{}_{h}{}^{r}{}^{p}{}^{q} \right)
\right) \cr
%
&&-
{\frac{g{}_{a}{}_{b}\,g{}_{c}{}_{d}\,g{}_{e}{}_{f}\,g{}_{g}{}_{h}}{241920}}
    \omega{}_{;}{}_{p}\,\omega{}_{;}{}_{q} \left(
   6\,\omega{}_{;}{}_{r}\,\omega{}^{;}{}^{p}\,\omega{}^{;}{}^{q}\,
   \omega{}^{;}{}^{r} - 67\,\omega{}_{;}{}_{r}\,\omega{}^{;}{}^{r}\,
   \omega{}^{;}{}^{p}{}^{q}
\right. \cr && \hspace{6mm} \left.
  +31\,\omega{}_{;}{}_{r}\,\omega{}^{;}{}^{q}\,\omega{}^{;}{}^{p}{}^{r} +
  102\,\omega{}_{;}{}_{r}{}^{q}\,\omega{}^{;}{}^{p}{}^{r} -
  6\,\omega{}_{;}{}_{r}\,\omega{}^{;}{}^{p}{}^{q}{}^{r} +
  27\,\omega{}_{;}{}_{r}\,\omega{}^{;}{}^{p}{}^{r}{}^{q}
\right. \cr && \hspace{6mm} \left.
  +15\,\omega{}_{;}{}_{r}\,\omega{}^{;}{}^{q}{}^{r}{}^{p}
\right) \end{eqnarray}
\end{subequations}
where $\symmeq$ denotes equality upon symmetrization.

By using the expansion
\begin{eqnarray} e^{\omega+\omega'} &=& e^{2\omega}\left( 1
- {\sqrt{2}}\,{\sqrt{\sigma}}\,p{}^{p} \omega{}_{;}{}_{p}
+ \sigma\,p{}^{p}\,p{}^{q}\left(
\omega{}_{;}{}_{p}\,\omega{}_{;}{}_{q} + \omega{}_{;}{}_{p}{}_{q}
\right) \right. \cr && \hspace{5mm}
- {\frac{{\sqrt{2}}}{3}}
{{\sigma}^{{\frac{3}{2}}}}\,p{}^{p}\,p{}^{q}\,p{}^{r}\left(
\omega{}_{;}{}_{p}\,\omega{}_{;}{}_{q}\,\omega{}_{;}{}_{r} +
  3\,\omega{}_{;}{}_{p}\,\omega{}_{;}{}_{q}{}_{r} +
  \omega{}_{;}{}_{p}{}_{q}{}_{r} \right)
\cr && \hspace{5mm}
+ {\frac{1}{6}}
{{\sigma}^2}\,p{}^{p}\,p{}^{q}\,p{}^{r}\,p{}^{s}\left(
 \omega{}_{;}{}_{p}\,\omega{}_{;}{}_{q}\,\omega{}_{;}{}_{r}\,
   \omega{}_{;}{}_{s} + 6\,\omega{}_{;}{}_{p}\,\omega{}_{;}{}_{q}\,
   \omega{}_{;}{}_{r}{}_{s} + 3\,\omega{}_{;}{}_{p}{}_{q}\,
   \omega{}_{;}{}_{r}{}_{s}
 \right. \cr && \hspace{10mm} \left.
 +4\,\omega{}_{;}{}_{p}\,\omega{}_{;}{}_{q}{}_{r}{}_{s} +
  \omega{}_{;}{}_{p}{}_{q}{}_{r}{}_{s} \right)
\cr && \hspace{5mm}
- {\frac{1}{15\,{\sqrt{2}}}}
{{\sigma}^{{\frac{5}{2}}}}\,p{}^{p}\,p{}^{q}\,p{}^{r}\,p{}^{s}\,p{}^{t}\left(
 \omega{}_{;}{}_{p}\,\omega{}_{;}{}_{q}\,\omega{}_{;}{}_{r}\,
   \omega{}_{;}{}_{s}\,\omega{}_{;}{}_{t} +
  10\,\omega{}_{;}{}_{p}\,\omega{}_{;}{}_{q}\,\omega{}_{;}{}_{r}\,
   \omega{}_{;}{}_{s}{}_{t}
 \right. \cr && \hspace{10mm} \left.
 +15\,\omega{}_{;}{}_{p}\,\omega{}_{;}{}_{q}{}_{r}\,\omega{}_{;}{}_{s}{}_{t}
+
  10\,\omega{}_{;}{}_{p}\,\omega{}_{;}{}_{q}\,
   \omega{}_{;}{}_{r}{}_{s}{}_{t} +
  10\,\omega{}_{;}{}_{p}{}_{q}\,\omega{}_{;}{}_{r}{}_{s}{}_{t}
 \right. \cr && \hspace{10mm} \left.
 +5\,\omega{}_{;}{}_{p}\,\omega{}_{;}{}_{q}{}_{r}{}_{s}{}_{t} +
  \omega{}_{;}{}_{p}{}_{q}{}_{r}{}_{s}{}_{t} \right)
\cr && \hspace{5mm}
+ {\frac{1}{90}}
{{\sigma}^3}\,p{}^{p}\,p{}^{q}\,p{}^{r}\,p{}^{s}\,p{}^{t}\,p{}^{u}\left(
 \omega{}_{;}{}_{p}\,\omega{}_{;}{}_{q}\,\omega{}_{;}{}_{r}\,
  \omega{}_{;}{}_{s}\,\omega{}_{;}{}_{t}\,\omega{}_{;}{}_{u}
 \right. \cr && \hspace{10mm} \left.
 +15\,\omega{}_{;}{}_{p}\,\omega{}_{;}{}_{q}\,\omega{}_{;}{}_{r}\,
   \omega{}_{;}{}_{s}\,\omega{}_{;}{}_{t}{}_{u} +
  45\,\omega{}_{;}{}_{p}\,\omega{}_{;}{}_{q}\,\omega{}_{;}{}_{r}{}_{s}\,
   \omega{}_{;}{}_{t}{}_{u} + 15\,\omega{}_{;}{}_{p}{}_{q}\,
   \omega{}_{;}{}_{r}{}_{s}\,\omega{}_{;}{}_{t}{}_{u}
 \right. \cr && \hspace{10mm} \left.
 +20\,\omega{}_{;}{}_{p}\,\omega{}_{;}{}_{q}\,\omega{}_{;}{}_{r}\,
   \omega{}_{;}{}_{s}{}_{t}{}_{u} +
  60\,\omega{}_{;}{}_{p}\,\omega{}_{;}{}_{q}{}_{r}\,
   \omega{}_{;}{}_{s}{}_{t}{}_{u} +
  10\,\omega{}_{;}{}_{p}{}_{q}{}_{r}\,\omega{}_{;}{}_{s}{}_{t}{}_{u}
 \right. \cr && \hspace{10mm} \left. \left.
 +15\,\omega{}_{;}{}_{p}\,\omega{}_{;}{}_{q}\,
   \omega{}_{;}{}_{r}{}_{s}{}_{t}{}_{u} +
  15\,\omega{}_{;}{}_{p}{}_{q}\,\omega{}_{;}{}_{r}{}_{s}{}_{t}{}_{u} +
  6\,\omega{}_{;}{}_{p}\,\omega{}_{;}{}_{q}{}_{r}{}_{s}{}_{t}{}_{u} +
  \omega{}_{;}{}_{p}{}_{q}{}_{r}{}_{s}{}_{t}{}_{u} \right)
%
\right) + O\left( \sigma^\frac{7}{2} \right)
\end{eqnarray}
we can get the expansion of $\Sigma$.
Multiplying together the above series and (\ref{Apx-bsigma2}) and
subtracting
$\sigma$, we determine the expansion
\begin{eqnarray}
\Sigma &=&
 \Sigma^{(4)}_{pqrs}  \sigma^p \sigma^q \sigma^r \sigma^s
 + \Sigma^{(5)}_{pqrst} \sigma^p \sigma^q \sigma^r \sigma^s \sigma^t
 + \Sigma^{(6)}_{pqrstu}\sigma^p \sigma^q \sigma^r \sigma^s \sigma^t
\sigma^u\cr
&&
 + \Sigma^{(7)}_{pqrstuv}
   \sigma^p \sigma^q \sigma^r \sigma^s \sigma^t \sigma^u \sigma^v
 + \Sigma^{(8)}_{pqrstuvw}
   \sigma^p \sigma^q \sigma^r \sigma^s \sigma^t \sigma^u \sigma^v \sigma^w
\label{Apx-Sigma1}
\end{eqnarray}
where the expansion tensors are
\begin{subequations}
\begin{eqnarray}
%
\Sigma^{(4)}_{abcd} &\symmeq& {\frac{g{}_{a}{}_{b}}{3}} \left(
\omega{}_{;}{}_{c}\,\omega{}_{;}{}_{d} + \omega{}_{;}{}_{c}{}_{d}
\right)
  - {\frac{g{}_{a}{}_{b}\,g{}_{c}{}_{d}}{6}}
\omega{}_{;}{}_{p}\,\omega{}^{;}{}^{p}
\label{Apx-Sigma4}
\\ \cr
%
%
\Sigma^{(5)}_{abcde} &\symmeq& -
{\frac{g{}_{a}{}_{b}}{3\,{\sqrt{2}}}} \left(
2\,\omega{}_{;}{}_{c}\,\omega{}_{;}{}_{d}{}_{e} +
  \omega{}_{;}{}_{c}{}_{d}{}_{e} \right)
+{\frac{g{}_{a}{}_{b}\,g{}_{c}{}_{d}}{3\,{\sqrt{2}}}}
\omega{}_{;}{}_{p}\,\omega{}_{;}{}_{e}{}^{p}
\\ \cr
%
%
\Sigma^{(6)}_{abcdef} &\symmeq& {\frac{g{}_{a}{}_{b}}{30}}\left(
   \omega{}_{;}{}_{c}\,\omega{}_{;}{}_{d}\,\omega{}_{;}{}_{e}\,
   \omega{}_{;}{}_{f} + 2\,\omega{}_{;}{}_{c}\,\omega{}_{;}{}_{d}\,
   \omega{}_{;}{}_{e}{}_{f} + 7\,\omega{}_{;}{}_{c}{}_{d}\,
   \omega{}_{;}{}_{e}{}_{f}
\right. \cr && \hspace{6mm} \left.
   + 6\,\omega{}_{;}{}_{c}\,\omega{}_{;}{}_{d}{}_{e}{}_{f} +
  3\,\omega{}_{;}{}_{c}{}_{d}{}_{e}{}_{f}
\right)\cr
&&-
  {\frac{g{}_{a}{}_{b}\,g{}_{c}{}_{d}}{90}}\left(
   3\,\omega{}_{;}{}_{r}\,\omega{}_{;}{}_{e}\,\omega{}_{;}{}_{f}\,
   \omega{}^{;}{}^{r} + 7\,\omega{}_{;}{}_{p}\,\omega{}^{;}{}^{p}\,
   \omega{}_{;}{}_{e}{}_{f}
\right. \cr && \hspace{10mm} \left.
   + 9\,\omega{}_{;}{}_{p}\,\omega{}_{;}{}_{e}{}_{f}{}^{p} +
  12\,\omega{}_{;}{}_{p}\,\omega{}_{;}{}_{q}\,R{}_{e}{}^{q}{}_{f}{}^{p}
\right)\cr
&&+ \frac{1}{720}
  8\,g{}_{a}{}_{b}\,g{}_{c}{}_{d}\,g{}_{e}{}_{f} \,\,
    \omega{}_{;}{}_{p}\,\omega{}_{;}{}_{q}\,
  \left( \omega{}^{;}{}^{p}\,\omega{}^{;}{}^{q} -
    3\,\omega{}^{;}{}^{p}{}^{q} \right)
\\ \cr
%
%
\Sigma^{(7)}_{abcdefg} &\symmeq&
-{\frac{g{}_{a}{}_{b}}{45\,{\sqrt{2}}}} \left(
6\,\omega{}_{;}{}_{c}\,\omega{}_{;}{}_{d}\,\omega{}_{;}{}_{e}\,
   \omega{}_{;}{}_{f}{}_{g} + 6\,\omega{}_{;}{}_{c}\,
   \omega{}_{;}{}_{d}{}_{e}\,\omega{}_{;}{}_{f}{}_{g} +
  3\,\omega{}_{;}{}_{c}\,\omega{}_{;}{}_{d}\,\omega{}_{;}{}_{e}{}_{f}{}_{g}
\right. \cr && \hspace{10mm} \left.
     + 15\,\omega{}_{;}{}_{c}{}_{d}\,\omega{}_{;}{}_{e}{}_{f}{}_{g} +
  4\,\omega{}_{;}{}_{c}\,\omega{}_{;}{}_{d}{}_{e}{}_{f}{}_{g} +
  2\,\omega{}_{;}{}_{c}{}_{d}{}_{e}{}_{f}{}_{g}
\right)\cr
&& {\frac{g{}_{a}{}_{b}\,g{}_{c}{}_{d}}{270\,{\sqrt{2}}}} \left(
18\,\omega{}_{;}{}_{p}\,\omega{}_{;}{}_{e}\,\omega{}^{;}{}^{p}\,
   \omega{}_{;}{}_{f}{}_{g} - 50\,\omega{}_{;}{}_{p}\,
   \omega{}_{;}{}_{e}{}^{p}\,\omega{}_{;}{}_{f}{}_{g} +
  18\,\omega{}_{;}{}_{p}\,\omega{}_{;}{}_{e}\,\omega{}_{;}{}_{f}\,
   \omega{}_{;}{}_{g}{}^{p} \right.
\cr && \hspace{10mm}
 -24\,\omega{}_{;}{}_{e}\,\omega{}_{;}{}_{p}{}_{f}\,\omega{}_{;}{}_{g}{}^{p}
+
  68\,\omega{}_{;}{}_{p}\,\omega{}_{;}{}_{e}{}_{f}\,
   \omega{}_{;}{}_{g}{}^{p} + 21\,\omega{}_{;}{}_{p}\,\omega{}^{;}{}^{p}\,
   \omega{}_{;}{}_{e}{}_{f}{}_{g}
\cr && \hspace{10mm}
 -24\,\omega{}_{;}{}_{p}\,\omega{}_{;}{}_{e}\,\omega{}_{;}{}_{f}{}_{g}{}^{p}
+
  24\,\omega{}_{;}{}_{p}{}_{e}\,\omega{}_{;}{}_{f}{}_{g}{}^{p} +
  6\,\omega{}_{;}{}_{p}{}_{e}\,\omega{}_{;}{}_{f}{}^{p}{}_{g}
\cr && \hspace{10mm}
 +12\,\omega{}_{;}{}_{p}\,\omega{}_{;}{}_{e}{}_{f}{}_{g}{}^{p} +
  6\,\omega{}_{;}{}_{p}\,\omega{}_{;}{}_{e}{}_{f}{}^{p}{}_{g} -
  6\,\omega{}_{;}{}_{p}\,\omega{}_{;}{}_{e}{}^{p}{}_{f}{}_{g}
\cr && \hspace{10mm} \left.
+27\,\omega{}_{;}{}_{p}\,\omega{}_{;}{}_{q}\,
  R{}_{e}{}^{p}{}_{f}{}^{q}{}_{;}{}_{g} - 12\,\omega{}_{;}{}_{p}\,\left(
2\,\omega{}_{;}{}_{q}\,\omega{}_{;}{}_{e} -
    7\,\omega{}_{;}{}_{q}{}_{e} \right) \,R{}_{f}{}^{p}{}_{g}{}^{q}
\right) \cr
&&
-{\frac{g{}_{a}{}_{b}\,g{}_{c}{}_{d}\,g{}_{e}{}_{f}}{180\,{\sqrt{2}}}}
\,\, \omega{}_{;}{}_{p} \left(
 8\,\omega{}_{;}{}_{q}\,\omega{}^{;}{}^{p}\,\omega{}_{;}{}_{g}{}^{q} -
  3\,\omega{}_{;}{}_{q}{}^{p}\,\omega{}_{;}{}_{g}{}^{q} -
  9\,\omega{}_{;}{}_{q}{}_{g}\,\omega{}^{;}{}^{p}{}^{q} \right.
\cr && \hspace{27mm} \left.
 -4\,\omega{}_{;}{}_{q}\,\omega{}_{;}{}_{g}{}^{p}{}^{q} -
  2\,\omega{}_{;}{}_{q}\,\omega{}^{;}{}^{p}{}^{q}{}_{g} +
  2\,\omega{}_{;}{}_{q}\,\omega{}_{;}{}_{r}\,R{}_{g}{}^{p}{}^{q}{}^{r}
\right) \end{eqnarray} \begin{eqnarray}
%
%
\Sigma^{(8)}_{abcdefgh} &\symmeq& {\frac{g{}_{a}{}_{b}}{630}}
\left\{
\omega{}_{;}{}_{c}\,\omega{}_{;}{}_{d}\,\omega{}_{;}{}_{e}\,
   \omega{}_{;}{}_{f}\,\omega{}_{;}{}_{g}\,\omega{}_{;}{}_{h} +
  3\,\omega{}_{;}{}_{c}\,\omega{}_{;}{}_{d}\,\omega{}_{;}{}_{e}\,
   \omega{}_{;}{}_{f}\,\omega{}_{;}{}_{g}{}_{h}
\right. \cr && \hspace{6mm} \left.
 +69\,\omega{}_{;}{}_{c}\,\omega{}_{;}{}_{d}\,\omega{}_{;}{}_{e}{}_{f}\,
   \omega{}_{;}{}_{g}{}_{h} + 27\,\omega{}_{;}{}_{c}{}_{d}\,
   \omega{}_{;}{}_{e}{}_{f}\,\omega{}_{;}{}_{g}{}_{h}
\right. \cr && \hspace{6mm} \left.
 +26\,\omega{}_{;}{}_{c}\,\omega{}_{;}{}_{d}\,\omega{}_{;}{}_{e}\,
   \omega{}_{;}{}_{f}{}_{g}{}_{h} +
  66\,\omega{}_{;}{}_{c}\,\omega{}_{;}{}_{d}{}_{e}\,
   \omega{}_{;}{}_{f}{}_{g}{}_{h}
\right. \cr && \hspace{6mm} \left.
 +40\,\omega{}_{;}{}_{c}{}_{d}{}_{e}\,\omega{}_{;}{}_{f}{}_{g}{}_{h} +
  13\,\omega{}_{;}{}_{c}\,\omega{}_{;}{}_{d}\,
   \omega{}_{;}{}_{e}{}_{f}{}_{g}{}_{h} +
  53\,\omega{}_{;}{}_{c}{}_{d}\,\omega{}_{;}{}_{e}{}_{f}{}_{g}{}_{h}
\right. \cr && \hspace{6mm} \left.
 +10\,\omega{}_{;}{}_{c}\,\omega{}_{;}{}_{d}{}_{e}{}_{f}{}_{g}{}_{h} +
  5\,\omega{}_{;}{}_{c}{}_{d}{}_{e}{}_{f}{}_{g}{}_{h}
\right\}\cr
%
%
&&-{\frac{g{}_{a}{}_{b}\,g{}_{c}{}_{d}}{7560}} \left( 12\,\left(
8\,\omega{}_{;}{}_{p}{}_{e}\,
     \omega{}_{;}{}_{f}{}_{g}{}^{p}{}_{h} -
    4\,\omega{}_{;}{}_{p}{}_{e}\,\omega{}_{;}{}_{f}{}^{p}{}_{g}{}_{h} +
    5\,\omega{}_{;}{}_{p}\,\omega{}_{;}{}_{e}{}_{f}{}_{g}{}_{h}{}^{p}
\right)
\right. \cr && \hspace{10mm} + 12\,\omega{}_{;}{}_{p}\,\left( 5\,
     \omega{}_{;}{}_{e}{}_{f}{}_{g}{}_{h}{}^{p} +
    3\,\omega{}_{;}{}_{e}{}_{f}{}_{g}{}^{p}{}_{h} +
    \omega{}_{;}{}_{e}{}_{f}{}^{p}{}_{g}{}_{h} -
    4\,\omega{}_{;}{}_{e}{}^{p}{}_{f}{}_{g}{}_{h} \right)
\cr && \hspace{10mm}
 -84\,\omega{}_{;}{}_{p}{}_{e}{}_{f}\,
  \left( \omega{}_{;}{}_{g}{}_{h}{}^{p} - \omega{}_{;}{}_{g}{}^{p}{}_{h}
     \right)
\cr && \hspace{10mm} +9\,\left(
15\,\omega{}_{;}{}_{e}{}_{f}{}_{p}\,
     \omega{}_{;}{}_{g}{}_{h}{}^{p} +
    16\,\omega{}_{;}{}_{p}{}_{e}\,\omega{}_{;}{}_{f}{}_{g}{}_{h}{}^{p}
\right)
\cr && \hspace{10mm}
+6\,\omega{}_{;}{}_{p} \left(
 66\,\omega{}_{;}{}_{e}{}^{p}\,\omega{}_{;}{}_{f}{}_{g}{}_{h} -
  12\,\omega{}_{;}{}_{e}{}_{f}\,\omega{}_{;}{}_{g}{}_{h}{}^{p} +
  12\,\omega{}_{;}{}_{e}{}_{f}\,\omega{}_{;}{}_{g}{}^{p}{}_{h} \right.
\cr && \hspace{15mm} \left. +
29\,\omega{}^{;}{}^{p}\,\omega{}_{;}{}_{e}{}_{f}{}_{g}{}_{h} -
  28\,\omega{}_{;}{}_{e}\,\omega{}_{;}{}_{f}{}_{g}{}_{h}{}^{p} -
  20\,\omega{}_{;}{}_{e}\,\omega{}_{;}{}_{f}{}_{g}{}^{p}{}_{h} \right.
\cr && \hspace{15mm} \left.
 + 16\,\omega{}_{;}{}_{e}\,\omega{}_{;}{}_{f}{}^{p}{}_{g}{}_{h} \right)
 -24\,\omega{}_{;}{}_{e}\,\omega{}_{;}{}_{p}{}_{f}\,
  \left( 19\,\omega{}_{;}{}_{g}{}_{h}{}^{p} +
    4\,\omega{}_{;}{}_{g}{}^{p}{}_{h} \right)
\cr && \hspace{10mm}
 -8\,\omega{}_{;}{}_{p}{}_{e}\,\left( 25\,\omega{}_{;}{}_{f}{}^{p}\,
     \omega{}_{;}{}_{g}{}_{h} -
    21\,\omega{}_{;}{}_{f}{}_{g}\,\omega{}_{;}{}_{h}{}^{p} \right)
\cr && \hspace{10mm}
+4\,\omega{}_{;}{}_{p}\,\omega{}_{;}{}_{e} \left(
134\,\omega{}_{;}{}_{f}{}^{p}\,\omega{}_{;}{}_{g}{}_{h} -
  60\,\omega{}_{;}{}_{f}{}_{g}\,\omega{}_{;}{}_{h}{}^{p} +
  33\,\omega{}^{;}{}^{p}\,\omega{}_{;}{}_{f}{}_{g}{}_{h} \right.
\cr && \hspace{15mm} \left.
+39\,\omega{}_{;}{}_{f}\,\omega{}_{;}{}_{g}{}_{h}{}^{p} +
  12\,\omega{}_{;}{}_{f}\,\omega{}_{;}{}_{g}{}^{p}{}_{h} \right)
+266\,\omega{}_{;}{}_{p}\,\omega{}^{;}{}^{p}\,\omega{}_{;}{}_{e}{}_{f}\,
  \omega{}_{;}{}_{g}{}_{h}
\cr && \hspace{10mm} +
208\,\omega{}_{;}{}_{e}\,\omega{}_{;}{}_{f}\,\omega{}_{;}{}_{p}{}_{g}\,
  \omega{}_{;}{}_{h}{}^{p}
+ 18\,\omega{}_{;}{}_{p}\,\omega{}_{;}{}_{e}\,\omega{}_{;}{}_{f}\,
  \omega{}_{;}{}_{g}\,\omega{}_{;}{}_{h}\,\omega{}^{;}{}^{p}
\cr && \hspace{10mm} +
4\,\omega{}_{;}{}_{p}\,\omega{}_{;}{}_{e}\,\omega{}_{;}{}_{f}\,
  \left( 17\,\omega{}^{;}{}^{p}\,\omega{}_{;}{}_{g}{}_{h} -
    16\,\omega{}_{;}{}_{g}\,\omega{}_{;}{}_{h}{}^{p} \right)
\cr && \hspace{10mm}
+ 16 \left(
11\,\omega{}_{;}{}_{p}\,\omega{}_{;}{}_{q}\,\omega{}_{;}{}_{e}\,
   \omega{}_{;}{}_{f} - 74\,\omega{}_{;}{}_{p}\,\omega{}_{;}{}_{e}\,
   \omega{}_{;}{}_{q}{}_{f} + 41\,\omega{}_{;}{}_{p}{}_{e}\,
   \omega{}_{;}{}_{q}{}_{f} \right.
\cr && \hspace{15mm} \left.
+3\,\omega{}_{;}{}_{p}\,\omega{}_{;}{}_{q}\,\omega{}_{;}{}_{e}{}_{f}
+
  21\,\omega{}_{;}{}_{p}\,\omega{}_{;}{}_{q}{}_{e}{}_{f} +
  30\,\omega{}_{;}{}_{p}\,\omega{}_{;}{}_{e}{}_{f}{}_{q} \right)
R{}_{g}{}^{p}{}_{h}{}^{q} \cr && \hspace{10mm}
 -312\,\omega{}_{;}{}_{p}\,\omega{}_{;}{}_{q}\,\omega{}_{;}{}_{e}\,
   R{}_{f}{}^{p}{}_{g}{}^{q}{}_{;}{}_{h} -
  912\,\omega{}_{;}{}_{p}\,R{}_{q}{}_{e}{}_{f}{}^{p}{}_{;}{}_{g}\,
   \omega{}_{;}{}_{h}{}^{q}
\cr && \hspace{10mm}  \left. +
180\,\omega{}_{;}{}_{p}\,\omega{}_{;}{}_{q}\,
   R{}_{e}{}^{p}{}_{f}{}^{q}{}_{;}{}_{g}{}_{h} -
  336\,\omega{}_{;}{}_{p}\,\omega{}_{;}{}_{q}\,R{}_{r}{}_{e}{}_{f}{}^{p}\,
   R{}_{g}{}^{r}{}_{h}{}^{q}
\right\} \cr
%
&& {\frac{g{}_{a}{}_{b}\,g{}_{c}{}_{d}\,g{}_{e}{}_{f}}{15120}}
\left\{ -256\,\omega{}_{;}{}_{p}{}_{q}\,\omega{}_{;}{}_{g}{}^{p}\,
   \omega{}_{;}{}_{h}{}^{q} - 300\,\omega{}_{;}{}_{p}\,
   \omega{}_{;}{}_{q}{}^{p}\,\omega{}_{;}{}_{g}{}_{h}{}^{q}  \right.
\cr && \hspace{10mm}
-24\,\omega{}_{;}{}_{p}\,\omega{}_{;}{}_{q}{}_{g}\,
  \left( 4\,\omega{}_{;}{}_{h}{}^{p}{}^{q} +
    11\,\omega{}_{;}{}_{h}{}^{q}{}^{p} + 7\,\omega{}^{;}{}^{p}{}^{q}{}_{h}
     \right)
\cr && \hspace{10mm} -24\,\omega{}_{;}{}_{p}\,\omega{}_{;}{}_{q}\,
  \left( 6\,\omega{}_{;}{}_{g}{}_{h}{}^{p}{}^{q} -
    3\,\omega{}_{;}{}_{g}{}^{p}{}_{h}{}^{q} +
    \omega{}_{;}{}_{g}{}^{p}{}^{q}{}_{h} +
    2\,\omega{}^{;}{}^{p}{}^{q}{}_{g}{}_{h} \right)
\cr && \hspace{10mm} +8\,\omega{}_{;}{}_{p}\,\omega{}^{;}{}^{p}\,
  \left( 20\,\omega{}_{;}{}_{q}{}_{g}\,\omega{}_{;}{}_{h}{}^{q} +
    3\,\omega{}_{;}{}_{q}\,\omega{}_{;}{}_{g}{}_{h}{}^{q} +
    18\,\omega{}_{;}{}_{q}\,\omega{}_{;}{}_{g}{}^{q}{}_{h} \right)
\cr && \hspace{10mm}
+24\,\omega{}_{;}{}_{p}\,\omega{}_{;}{}_{q}\,\omega{}_{;}{}_{g}\,
  \left( 2\,\omega{}_{;}{}_{h}{}^{p}{}^{q} +
    3\,\omega{}^{;}{}^{p}{}^{q}{}_{h} \right)
+256\,\omega{}_{;}{}_{p}\,\omega{}_{;}{}_{g}\,\omega{}_{;}{}_{q}{}^{p}\,
  \omega{}_{;}{}_{h}{}^{q}
\cr && \hspace{10mm} +16\,\omega{}_{;}{}_{p}\,\omega{}_{;}{}_{q}\,
  \left( 29\,\omega{}_{;}{}_{g}{}^{p}\,\omega{}_{;}{}_{h}{}^{q} -
    18\,\omega{}_{;}{}_{g}{}_{h}\,\omega{}^{;}{}^{p}{}^{q} \right)
\cr && \hspace{10mm} +8\,\omega{}_{;}{}_{p}\,\omega{}_{;}{}_{q}\,
  \left( 9\,\omega{}^{;}{}^{p}\,\omega{}^{;}{}^{q}\,
     \omega{}_{;}{}_{g}{}_{h} -
    12\,\omega{}_{;}{}_{g}\,\omega{}^{;}{}^{p}\,\omega{}_{;}{}_{h}{}^{q} -
    5\,\omega{}_{;}{}_{g}\,\omega{}_{;}{}_{h}\,\omega{}^{;}{}^{p}{}^{q}
    \right)
\cr && \hspace{10mm}
+24\,\omega{}_{;}{}_{p}\,\omega{}_{;}{}_{q}\,\omega{}_{;}{}_{g}\,
  \omega{}_{;}{}_{h}\,\omega{}^{;}{}^{p}\,\omega{}^{;}{}^{q}
\cr && \hspace{10mm}
-3\,\omega{}_{;}{}_{p}\,\omega{}_{;}{}_{q}\,\omega{}_{;}{}_{r}\,
  \left( 15\,R{}_{g}{}^{p}{}_{h}{}^{q}{}^{;}{}^{r} +
    45\,R{}_{g}{}^{p}{}_{h}{}^{r}{}^{;}{}^{q} -
    26\,R{}_{g}{}^{p}{}^{q}{}^{r}{}_{;}{}_{h} \right)
\cr && \hspace{10mm} +8\,\omega{}_{;}{}_{p}\,\omega{}_{;}{}_{q}\,
  \left( 9\,\omega{}_{;}{}_{r}\,\omega{}_{;}{}_{g}\,
     R{}_{h}{}^{p}{}^{q}{}^{r} +
    18\,\omega{}_{;}{}_{r}{}_{g}\,R{}_{h}{}^{p}{}^{q}{}^{r} +
    16\,\omega{}_{;}{}_{r}{}_{g}\,R{}_{h}{}^{r}{}^{p}{}^{q} \right)
\cr && \hspace{10mm} \left.
+48\,\omega{}_{;}{}_{p}\,\omega{}_{;}{}_{q}\,
  \left( 2\,\omega{}_{;}{}_{r}\,\omega{}^{;}{}^{p} -
    15\,\omega{}_{;}{}_{r}{}^{p} \right) \,R{}_{g}{}^{q}{}_{h}{}^{r}
\right\} \cr
%
%
&&-{\frac{g{}_{a}{}_{b}\,g{}_{c}{}_{d}\,g{}_{e}{}_{f}\,g{}_{g}{}_{h}}{2520}}
\,\, \omega{}_{;}{}_{p}\,\omega{}_{;}{}_{q} \left(
\omega{}_{;}{}_{r}\,\omega{}^{;}{}^{p}\,\omega{}^{;}{}^{q}\,
   \omega{}^{;}{}^{r} - 6\,\omega{}_{;}{}_{r}\,\omega{}^{;}{}^{p}\,
   \omega{}^{;}{}^{q}{}^{r} \right.
\cr && \hspace{40mm} \left.
+17\,\omega{}_{;}{}_{r}{}^{p}\,\omega{}^{;}{}^{q}{}^{r} +
  6\,\omega{}_{;}{}_{r}\,\omega{}^{;}{}^{p}{}^{q}{}^{r} \right)
\end{eqnarray}
\end{subequations}

Turning to the last expansion we need, we recall the
Van Vleck-Morette determinant on the optical metric, $\bar\VanD$,
satisfies the differential equation
\begin{equation}
\bar\VanD\left(4-{\bar \square}{\bar\sigma}\right)
    - 2 \bar\VanD_{\,\,|p} \bar\sigma^{|p} = 0
\end{equation}
Using the conformal transformation property
\begin{equation}
\bar\square\bar\sigma = e^{2\w}
  \left(\square\bar\sigma  -2 \w^{;p}\bar\sigma_{;p}\right)
\end{equation}
we determine $\bar\VanD$ satisfies the equation
\begin{equation}
\bar\VanD\left(4- e^{2\w}\left(\square\bar\sigma
  -2 \w^{;p}\bar\sigma_{;p}\right)    \right)
    - 2 e^{2\w}\bar\VanD_{\,\,;p} \bar\sigma^{;p} = 0
\label{Apx-bVanD1}
\end{equation}
on the physical metric. Since we have the expansion for
$\bar\sigma$, we need only assume the expansion
\begin{eqnarray}
\bar\VanD &=& 1
 + \bar\Delta^{(2)}_{pq}    \sigma^p \sigma^q
 + \bar\Delta^{(3)}_{pqr}   \sigma^p \sigma^q \sigma^r
 + \bar\Delta^{(4)}_{pqrs}  \sigma^p \sigma^q \sigma^r \sigma^s
 + \bar\Delta^{(5)}_{pqrst} \sigma^p \sigma^q \sigma^r \sigma^s \sigma^t \cr
&&\hspace{2cm}
 + \bar\Delta^{(6)}_{pqrstu}
      \sigma^p \sigma^q \sigma^r \sigma^s \sigma^t \sigma^u
\label{Apx-bVanD2}
\end{eqnarray}
and substitute this and (\ref{Apx-bsigma2}) into (\ref{Apx-bVanD1})
and solve for the expansion tensors.
Following the now well defined method outlined above, they are found to be
%
%
\begin{subequations}
\begin{eqnarray} \bar\Delta^{(2)}_{ab} &\symmeq & \Delta^{(2)}_{ab} +
{\frac{1}{6}} \left( \omega{}_{;}{}_{a}\,\omega{}_{;}{}_{b} +
\omega{}_{;}{}_{a}{}_{b} \right) -{\frac{g{}_{a}{}_{b}}{12}}
\left( 2\,\omega{}_{;}{}_{p}\,\omega{}^{;}{}^{p} -
\omega{}_{;}{}_{p}{}^{p} \right)
\label{Apx-bVan2}
\\ \cr
%
%
\bar\Delta^{(3)}_{abc} &\symmeq & \Delta^{(3)}_{abc} -
{\frac{1}{12}} \left(
2\,\omega{}_{;}{}_{a}\,\omega{}_{;}{}_{b}{}_{c} +
  \omega{}_{;}{}_{a}{}_{b}{}_{c} \right)
+{\frac{g{}_{a}{}_{b}}{24}} \left(
4\,\omega{}_{;}{}_{p}\,\omega{}_{;}{}_{c}{}^{p} -
  \omega{}_{;}{}_{p}{}^{p}{}_{c} \right)
\\ \cr
%
%
\bar\Delta^{(4)}_{abcd} &\symmeq & \Delta^{(4)}_{abcd}
+\frac{1}{72} \left( \omega{}_{;}{}_{c}\,\omega{}_{;}{}_{d} +
\omega{}_{;}{}_{c}{}_{d}
     \right) \,R{}_{a}{}_{b}
+\frac{1}{120}\left(
7\,\omega{}_{;}{}_{a}{}_{b}\,\omega{}_{;}{}_{c}{}_{d} +
  6\,\omega{}_{;}{}_{a}\,\omega{}_{;}{}_{b}{}_{c}{}_{d} \right)
\cr && +\frac{1}{120} \left(
  \omega{}_{;}{}_{a}\,\omega{}_{;}{}_{b}\,\omega{}_{;}{}_{c}\,
   \omega{}_{;}{}_{d} + 2\,\omega{}_{;}{}_{a}\,\omega{}_{;}{}_{b}\,
   \omega{}_{;}{}_{c}{}_{d} + 3\,\omega{}_{;}{}_{a}{}_{b}{}_{c}{}_{d}
\right) \cr
&&+ {\frac{g{}_{a}{}_{b}}{720}} \left\{
 6\,\omega{}_{;}{}_{c}\,\left( \omega{}_{;}{}_{d}\,\omega{}_{;}{}_{p}{}^{p}
-
    \omega{}_{;}{}_{p}{}^{p}{}_{d} \right)  \right.
\cr && \hspace{10mm} +3\,\left(
4\,\omega{}_{;}{}_{p}{}^{p}\,\omega{}_{;}{}_{c}{}_{d} -
    12\,\omega{}_{;}{}_{p}{}_{c}\,\omega{}_{;}{}_{d}{}^{p} +
    3\,\omega{}_{;}{}_{p}{}^{p}{}_{c}{}_{d} \right)
\cr && \hspace{10mm}
  -12\,\omega{}_{;}{}_{p}\,\left( \omega{}_{;}{}_{c}\,\omega{}_{;}{}_{d}\,
     \omega{}^{;}{}^{p} + 2\,\omega{}^{;}{}^{p}\,\omega{}_{;}{}_{c}{}_{d} -
    2\,\omega{}_{;}{}_{c}\,\omega{}_{;}{}_{d}{}^{p} -
    \omega{}_{;}{}_{c}{}_{d}{}^{p} + 4\,\omega{}_{;}{}_{c}{}^{p}{}_{d}
\right)
\cr && \hspace{10mm} +6\,\omega{}_{;}{}_{p}\,\left(
R{}_{c}{}_{d}{}^{;}{}^{p} -
    R{}_{c}{}^{p}{}_{;}{}_{d} \right)  -5\,\left(
2\,\omega{}_{;}{}_{p}\,\omega{}^{;}{}^{p} -
    \omega{}_{;}{}_{p}{}^{p} \right) \,R{}_{c}{}_{d}
\cr && \hspace{10mm} \left. -2\,\left(
2\,\omega{}_{;}{}_{p}\,\omega{}_{;}{}_{d} -
    \omega{}_{;}{}_{p}{}_{d} \right) \,R{}_{c}{}^{p} +4\,\left(
\omega{}_{;}{}_{p}\,\omega{}_{;}{}_{q} +
    \omega{}_{;}{}_{p}{}_{q} \right) \,R{}_{c}{}^{p}{}_{d}{}^{q}
\right\} \cr
&&+ {\frac{g{}_{a}{}_{b}\,g{}_{c}{}_{d}}{1440}} \left\{
2\,\omega{}_{;}{}_{p}\,\omega{}_{;}{}_{q}\,
  \left( 6\,\omega{}^{;}{}^{p}\,\omega{}^{;}{}^{q} -
    14\,\omega{}^{;}{}^{p}{}^{q} + 3\,R{}^{p}{}^{q} \right)
\right. \cr && \hspace{10mm} \left.
-14\,\omega{}_{;}{}_{p}\,\omega{}^{;}{}^{p}\,\omega{}_{;}{}_{q}{}^{q}
+
  5\,\omega{}_{;}{}_{p}{}^{p}\,\omega{}_{;}{}_{q}{}^{q} +
  4\,\omega{}_{;}{}_{p}{}_{q}\,\omega{}^{;}{}^{p}{}^{q} +
  12\,\omega{}_{;}{}_{p}\,\omega{}_{;}{}_{q}{}^{q}{}^{p}
\right\} \\ \cr
%
%
\bar\Delta^{(5)}_{abcde} &\symmeq & \Delta^{(5)}_{abcde}
-\frac{1}{360} \left(
  6\,\omega{}_{;}{}_{a}\,\omega{}_{;}{}_{b}\,\omega{}_{;}{}_{c}\,
   \omega{}_{;}{}_{d}{}_{e} + 6\,\omega{}_{;}{}_{a}\,
   \omega{}_{;}{}_{b}{}_{c}\,\omega{}_{;}{}_{d}{}_{e} +
  3\,\omega{}_{;}{}_{a}\,\omega{}_{;}{}_{b}\,\omega{}_{;}{}_{c}{}_{d}{}_{e}
\right. \cr && \hspace{5mm} \left. +
15\,\omega{}_{;}{}_{a}{}_{b}\,\omega{}_{;}{}_{c}{}_{d}{}_{e} +
  4\,\omega{}_{;}{}_{a}\,\omega{}_{;}{}_{b}{}_{c}{}_{d}{}_{e} +
  2\,\omega{}_{;}{}_{a}{}_{b}{}_{c}{}_{d}{}_{e} \right)
\cr && -\frac{1}{144} \left( R{}_{c}{}_{d}{}_{;}{}_{e}\,\left(
\omega{}_{;}{}_{a}\,\omega{}_{;}{}_{b} +
     \omega{}_{;}{}_{a}{}_{b} \right)  +
  \left( 2\,\omega{}_{;}{}_{a}\,\omega{}_{;}{}_{b}{}_{c} +
     \omega{}_{;}{}_{a}{}_{b}{}_{c} \right) \,R{}_{d}{}_{e}
\right) \cr &&
-{\frac{g{}_{a}{}_{b}}{1440}} \left\{
6\,\omega{}_{;}{}_{c}\,\left( 2\,\omega{}_{;}{}_{p}{}^{p}\,
     \omega{}_{;}{}_{d}{}_{e} +
    \omega{}_{;}{}_{d}\,\omega{}_{;}{}_{p}{}^{p}{}_{e} \right)  \right.
\cr && \hspace{10mm} +2\,\left(
3\,\omega{}_{;}{}_{c}{}_{d}\,\omega{}_{;}{}_{p}{}^{p}{}_{e} +
    6\,\omega{}_{;}{}_{p}{}^{p}\,\omega{}_{;}{}_{c}{}_{d}{}_{e} -
    3\,\omega{}_{;}{}_{c}\,\omega{}_{;}{}_{p}{}^{p}{}_{d}{}_{e} +
    2\,\omega{}_{;}{}_{p}{}^{p}{}_{c}{}_{d}{}_{e} \right)
\cr && \hspace{10mm} +24\,\left(
\omega{}_{;}{}_{c}\,\omega{}_{;}{}_{p}{}_{d}\,
     \omega{}_{;}{}_{e}{}^{p} +
    3\,\omega{}_{;}{}_{p}{}_{c}\,\omega{}_{;}{}_{d}{}_{e}{}^{p} -
    5\,\omega{}_{;}{}_{p}{}_{c}\,\omega{}_{;}{}_{d}{}^{p}{}_{e} \right)
\cr && \hspace{10mm} +2\,\omega{}_{;}{}_{p}\,\left(
5\,\omega{}_{;}{}_{c}{}_{d}{}_{e}{}^{p} +
    \omega{}_{;}{}_{c}{}_{d}{}^{p}{}_{e} -
    14\,\omega{}_{;}{}_{c}{}^{p}{}_{d}{}_{e} \right)
\cr && \hspace{10mm} -24\,\omega{}_{;}{}_{p}\,\omega{}_{;}{}_{c}\,
  \left( \omega{}^{;}{}^{p}\,\omega{}_{;}{}_{d}{}_{e} +
    \omega{}_{;}{}_{d}\,\omega{}_{;}{}_{e}{}^{p} \right)
\cr && \hspace{10mm} -8\,\omega{}_{;}{}_{p}\,\left(
3\,\omega{}_{;}{}_{c}{}_{d}\,
     \omega{}_{;}{}_{e}{}^{p} +
    3\,\omega{}^{;}{}^{p}\,\omega{}_{;}{}_{c}{}_{d}{}_{e} -
    10\,\omega{}_{;}{}_{c}\,\omega{}_{;}{}_{d}{}_{e}{}^{p} +
    7\,\omega{}_{;}{}_{c}\,\omega{}_{;}{}_{d}{}^{p}{}_{e} \right)
\cr && \hspace{10mm}
%
+ 4\,\left(
20\,\omega{}_{;}{}_{p}\,\omega{}_{;}{}_{q}\,\omega{}_{;}{}_{c} +
    22\,\omega{}_{;}{}_{q}\,\omega{}_{;}{}_{p}{}_{c} +
    \omega{}_{;}{}_{p}{}_{q}{}_{c} + 10\,\omega{}_{;}{}_{p}\,R{}_{q}{}_{c}
     \right) \,R{}_{d}{}^{p}{}_{e}{}^{q}
\cr && \hspace{10mm} + 4\,R{}_{c}{}^{p}{}_{d}{}^{q}{}_{;}{}_{e}\,
  \left( \omega{}_{;}{}_{p}\,\omega{}_{;}{}_{q} +
    \omega{}_{;}{}_{p}{}_{q} \right)
+ \omega{}_{;}{}_{p}\,\left( 5\,R{}_{c}{}_{d}{}_{;}{}_{e}{}^{p} +
    R{}_{c}{}_{d}{}^{;}{}^{p}{}_{e} - 6\,R{}_{c}{}^{p}{}_{;}{}_{d}{}_{e}
     \right)
\cr && \hspace{10mm} -2\,\left(
2\,\omega{}_{;}{}_{c}\,\omega{}_{;}{}_{p}{}_{d} +
    2\,\omega{}_{;}{}_{p}\,\omega{}_{;}{}_{c}{}_{d} +
    14\,\omega{}_{;}{}_{p}{}_{c}{}_{d} - 15\,\omega{}_{;}{}_{c}{}_{d}{}_{p}
     \right) \,R{}_{e}{}^{p}
\cr && \hspace{10mm} -5\,R{}_{c}{}_{d}{}_{;}{}_{e}\,\left(
2\,\omega{}_{;}{}_{p}\,
      \omega{}^{;}{}^{p} - \omega{}_{;}{}_{p}{}^{p} \right)  -
  5\,\left( 4\,\omega{}_{;}{}_{p}\,\omega{}_{;}{}_{c}{}^{p} -
     \omega{}_{;}{}_{p}{}^{p}{}_{c} \right) \,R{}_{d}{}_{e}
\cr && \hspace{10mm} \left. -4\,R{}_{d}{}^{p}{}_{;}{}_{e}\,\left(
\omega{}_{;}{}_{p}\,
     \omega{}_{;}{}_{c} + \omega{}_{;}{}_{p}{}_{c} \right)  +
6\,R{}_{d}{}_{e}{}^{;}{}^{p}\,\omega{}_{;}{}_{p}{}_{c} \right\}
\cr
&&-{\frac{g{}_{a}{}_{b}\,g{}_{c}{}_{d}}{1440}} \left\{
 2\,\omega{}_{;}{}_{p}\,\omega{}_{;}{}_{q}\,
  \left( 12\,\omega{}^{;}{}^{p}\,\omega{}_{;}{}_{e}{}^{q} -
    10\,\omega{}_{;}{}_{e}{}^{p}{}^{q} + 3\,\omega{}^{;}{}^{p}{}^{q}{}_{e}
     \right)  \right.
\cr && \hspace{10mm}
-14\,\omega{}_{;}{}_{p}\,\omega{}_{;}{}_{q}{}^{q}\,\omega{}_{;}{}_{e}{}^{p}
-
  28\,\omega{}_{;}{}_{p}\,\omega{}_{;}{}_{q}{}_{e}\,
   \omega{}^{;}{}^{p}{}^{q} - 7\,\omega{}_{;}{}_{p}\,\omega{}^{;}{}^{p}\,
   \omega{}_{;}{}_{q}{}^{q}{}_{e}
\cr && \hspace{10mm}
5\,\omega{}_{;}{}_{p}{}^{p}\,\omega{}_{;}{}_{q}{}^{q}{}_{e} +
  6\,\omega{}_{;}{}_{p}{}_{e}\,\omega{}_{;}{}_{q}{}^{q}{}^{p} +
  4\,\omega{}_{;}{}_{p}{}_{q}\,\omega{}^{;}{}^{p}{}^{q}{}_{e} +
  5\,\omega{}_{;}{}_{p}\,\omega{}_{;}{}_{q}{}^{q}{}_{e}{}^{p}
\cr && \hspace{10mm} \left.
+\omega{}_{;}{}_{p}\,\omega{}_{;}{}_{q}{}^{q}{}^{p}{}_{e} +
3\,\omega{}_{;}{}_{p}\,\left(
\omega{}_{;}{}_{q}\,R{}^{p}{}^{q}{}_{;}{}_{e} +
    2\,\omega{}_{;}{}_{q}{}_{e}\,R{}^{p}{}^{q} \right)
\right\}
\\ \cr
%
%
%
\bar\Delta^{(6)}_{abcdef} &\symmeq & \Delta^{(6)}_{abcdef} +
\bar\Delta^{(6')}_{abcdef} + g_{ab} \bar\Delta^{(6')}_{cdef} +
g_{ab} g_{cd} \bar\Delta^{(6')}_{ef} + g_{ab} g_{cd} g_{ef}
\bar\Delta^{(6')} \end{eqnarray}
\end{subequations}
\begin{subequations}
\begin{eqnarray}
%
181440 \bar\Delta^{(6')}_{abcdef} &=&
 -22556\,\omega{}_{;}{}_{a}\,\omega{}_{;}{}_{b}\,\omega{}_{;}{}_{c}\,
  \omega{}_{;}{}_{d}\,\omega{}_{;}{}_{e}\,\omega{}_{;}{}_{f}
-67668\,\omega{}_{;}{}_{a}\,\omega{}_{;}{}_{b}\,\omega{}_{;}{}_{c}\,
  \omega{}_{;}{}_{d}\,\omega{}_{;}{}_{e}{}_{f} \cr &&
 -48444\,\omega{}_{;}{}_{a}\,\omega{}_{;}{}_{b}\,\omega{}_{;}{}_{c}{}_{d}\,
   \omega{}_{;}{}_{e}{}_{f} -
10296\,\omega{}_{;}{}_{a}\,\omega{}_{;}{}_{b}\,
   \omega{}_{;}{}_{c}\,\omega{}_{;}{}_{d}{}_{e}{}_{f} \cr &&
 -2612\,\omega{}_{;}{}_{a}{}_{b}\,\omega{}_{;}{}_{c}{}_{d}\,
   \omega{}_{;}{}_{e}{}_{f} - 11016\,\omega{}_{;}{}_{a}\,
   \omega{}_{;}{}_{b}{}_{c}\,\omega{}_{;}{}_{d}{}_{e}{}_{f} -
  108\,\omega{}_{;}{}_{a}\,\omega{}_{;}{}_{b}\,
   \omega{}_{;}{}_{c}{}_{d}{}_{e}{}_{f} \cr &&
+900\,\omega{}_{;}{}_{a}{}_{b}{}_{c}\,\omega{}_{;}{}_{d}{}_{e}{}_{f}
+
  1332\,\omega{}_{;}{}_{a}{}_{b}\,\omega{}_{;}{}_{c}{}_{d}{}_{e}{}_{f} +
  360\,\omega{}_{;}{}_{a}\,\omega{}_{;}{}_{b}{}_{c}{}_{d}{}_{e}{}_{f}  \cr
&&
+180\,\omega{}_{;}{}_{a}{}_{b}{}_{c}{}_{d}{}_{e}{}_{f} \cr &&
%
+90\,\left( \omega{}_{;}{}_{a}\,\omega{}_{;}{}_{b} +
     \omega{}_{;}{}_{a}{}_{b} \right) \,R{}_{c}{}_{d}{}_{;}{}_{e}{}_{f} -
  5040\,\omega{}_{;}{}_{a}\,R{}_{d}{}^{p}{}_{e}{}^{q}{}_{;}{}_{f}\,
   R{}_{p}{}_{b}{}_{q}{}_{c} \cr &&
 -90\,R{}_{d}{}_{e}{}_{;}{}_{f}\,
  \left( 56\,\omega{}_{;}{}_{a}\,\omega{}_{;}{}_{b}\,\omega{}_{;}{}_{c} +
    54\,\omega{}_{;}{}_{a}\,\omega{}_{;}{}_{b}{}_{c} -
    \omega{}_{;}{}_{a}{}_{b}{}_{c} \right)  \cr &&
-30 \left(
751\,\omega{}_{;}{}_{a}\,\omega{}_{;}{}_{b}\,\omega{}_{;}{}_{c}\,
   \omega{}_{;}{}_{d} + 1166\,\omega{}_{;}{}_{a}\,\omega{}_{;}{}_{b}\,
   \omega{}_{;}{}_{c}{}_{d} + 73\,\omega{}_{;}{}_{a}{}_{b}\,
   \omega{}_{;}{}_{c}{}_{d} \right.
\cr && \quad \left. +
162\,\omega{}_{;}{}_{a}\,\omega{}_{;}{}_{b}{}_{c}{}_{d} -
  3\,\omega{}_{;}{}_{a}{}_{b}{}_{c}{}_{d} \right)
                 R{}_{e}{}_{f} \cr &&
+105\,\left( \omega{}_{;}{}_{a}\,\omega{}_{;}{}_{b} +
    \omega{}_{;}{}_{a}{}_{b} \right) \,R{}_{c}{}_{d}\,R{}_{e}{}_{f} \cr &&
-1260\,\left( 9\,\omega{}_{;}{}_{a}\,\omega{}_{;}{}_{b} +
    \omega{}_{;}{}_{a}{}_{b} \right) \,R{}_{p}{}_{c}{}_{q}{}_{d}\,
  R{}_{e}{}^{p}{}_{f}{}^{q}
\end{eqnarray} \begin{eqnarray}
%
120960 \bar\Delta^{(6')}_{abcd} &=&
 31672\,\omega{}_{;}{}_{p}\,\omega{}_{;}{}_{a}\,\omega{}_{;}{}_{b}\,
  \omega{}_{;}{}_{c}\,\omega{}_{;}{}_{d}\,\omega{}^{;}{}^{p}
-9116\,\omega{}_{;}{}_{a}\,\omega{}_{;}{}_{b}\,\omega{}_{;}{}_{c}\,
  \omega{}_{;}{}_{d}\,\omega{}_{;}{}_{p}{}^{p} \cr &&
+41376\,\omega{}_{;}{}_{p}\,\omega{}_{;}{}_{a}\,\omega{}_{;}{}_{b}\,
   \omega{}^{;}{}^{p}\,\omega{}_{;}{}_{c}{}_{d} +
  43936\,\omega{}_{;}{}_{p}\,\omega{}_{;}{}_{a}\,\omega{}_{;}{}_{b}\,
   \omega{}_{;}{}_{c}\,\omega{}_{;}{}_{d}{}^{p} \cr &&
 -15568\,\omega{}_{;}{}_{a}\,\omega{}_{;}{}_{b}\,\omega{}_{;}{}_{p}{}^{p}\,
   \omega{}_{;}{}_{c}{}_{d} + 4088\,\omega{}_{;}{}_{p}\,\omega{}^{;}{}^{p}\,
   \omega{}_{;}{}_{a}{}_{b}\,\omega{}_{;}{}_{c}{}_{d} \cr &&
+15152\,\omega{}_{;}{}_{a}\,\omega{}_{;}{}_{b}\,\omega{}_{;}{}_{p}{}_{c}\,
   \omega{}_{;}{}_{d}{}^{p} +
32416\,\omega{}_{;}{}_{p}\,\omega{}_{;}{}_{a}\,
   \omega{}_{;}{}_{b}{}_{c}\,\omega{}_{;}{}_{d}{}^{p} \cr &&
 -2664\,\omega{}_{;}{}_{a}\,\omega{}_{;}{}_{b}\,\omega{}_{;}{}_{c}\,
   \omega{}_{;}{}_{p}{}^{p}{}_{d} +
  4320\,\omega{}_{;}{}_{p}\,\omega{}_{;}{}_{a}\,\omega{}^{;}{}^{p}\,
   \omega{}_{;}{}_{b}{}_{c}{}_{d} \cr &&
+2256\,\omega{}_{;}{}_{p}\,\omega{}_{;}{}_{a}\,\omega{}_{;}{}_{b}\,
   \omega{}_{;}{}_{c}{}_{d}{}^{p} +
  4896\,\omega{}_{;}{}_{p}\,\omega{}_{;}{}_{a}\,\omega{}_{;}{}_{b}\,
   \omega{}_{;}{}_{c}{}^{p}{}_{d} \cr &&
 -1196\,\omega{}_{;}{}_{p}{}^{p}\,\omega{}_{;}{}_{a}{}_{b}\,
   \omega{}_{;}{}_{c}{}_{d} + 3056\,\omega{}_{;}{}_{p}{}_{a}\,
   \omega{}_{;}{}_{b}{}_{c}\,\omega{}_{;}{}_{d}{}^{p} -
  2952\,\omega{}_{;}{}_{a}\,\omega{}_{;}{}_{b}{}_{c}\,
   \omega{}_{;}{}_{p}{}^{p}{}_{d} \cr &&
 -2736\,\omega{}_{;}{}_{a}\,\omega{}_{;}{}_{p}{}^{p}\,
   \omega{}_{;}{}_{b}{}_{c}{}_{d} +
  1152\,\omega{}_{;}{}_{p}\,\omega{}_{;}{}_{a}{}^{p}\,
   \omega{}_{;}{}_{b}{}_{c}{}_{d} +
  3264\,\omega{}_{;}{}_{a}\,\omega{}_{;}{}_{p}{}_{b}\,
   \omega{}_{;}{}_{c}{}_{d}{}^{p} \cr &&
+8256\,\omega{}_{;}{}_{p}\,\omega{}_{;}{}_{a}{}_{b}\,
   \omega{}_{;}{}_{c}{}_{d}{}^{p} +
  4800\,\omega{}_{;}{}_{a}\,\omega{}_{;}{}_{p}{}_{b}\,
   \omega{}_{;}{}_{c}{}^{p}{}_{d} -
  5424\,\omega{}_{;}{}_{p}\,\omega{}_{;}{}_{a}{}_{b}\,
   \omega{}_{;}{}_{c}{}^{p}{}_{d} \cr &&
 -12\,\omega{}_{;}{}_{a}\,\omega{}_{;}{}_{b}\,
   \omega{}_{;}{}_{p}{}^{p}{}_{c}{}_{d} -
  408\,\omega{}_{;}{}_{p}\,\omega{}^{;}{}^{p}\,
   \omega{}_{;}{}_{a}{}_{b}{}_{c}{}_{d} +
  1848\,\omega{}_{;}{}_{p}\,\omega{}_{;}{}_{a}\,
   \omega{}_{;}{}_{b}{}_{c}{}_{d}{}^{p} \cr &&
 -264\,\omega{}_{;}{}_{p}\,\omega{}_{;}{}_{a}\,
   \omega{}_{;}{}_{b}{}_{c}{}^{p}{}_{d} -
  624\,\omega{}_{;}{}_{p}\,\omega{}_{;}{}_{a}\,
   \omega{}_{;}{}_{b}{}^{p}{}_{c}{}_{d}
+1560\,\omega{}_{;}{}_{p}{}_{a}{}_{b}\,\omega{}_{;}{}_{c}{}_{d}{}^{p}
\cr &&
+1260\,\omega{}_{;}{}_{a}{}_{b}{}_{p}\,\omega{}_{;}{}_{c}{}_{d}{}^{p}
-
  3180\,\omega{}_{;}{}_{p}{}_{a}{}_{b}\,\omega{}_{;}{}_{c}{}^{p}{}_{d} -
  156\,\omega{}_{;}{}_{a}{}_{b}\,\omega{}_{;}{}_{p}{}^{p}{}_{c}{}_{d} \cr &&
+108\,\omega{}_{;}{}_{p}{}^{p}\,\omega{}_{;}{}_{a}{}_{b}{}_{c}{}_{d}
+
  1728\,\omega{}_{;}{}_{p}{}_{a}\,\omega{}_{;}{}_{b}{}_{c}{}_{d}{}^{p} -
  192\,\omega{}_{;}{}_{p}{}_{a}\,\omega{}_{;}{}_{b}{}_{c}{}^{p}{}_{d} \cr &&
 -2112\,\omega{}_{;}{}_{p}{}_{a}\,\omega{}_{;}{}_{b}{}^{p}{}_{c}{}_{d} -
  144\,\omega{}_{;}{}_{a}\,\omega{}_{;}{}_{p}{}^{p}{}_{b}{}_{c}{}_{d} +
  216\,\omega{}_{;}{}_{p}\,\omega{}_{;}{}_{a}{}_{b}{}_{c}{}_{d}{}^{p} \cr &&
+96\,\omega{}_{;}{}_{p}\,\omega{}_{;}{}_{a}{}_{b}{}_{c}{}^{p}{}_{d}
-
  24\,\omega{}_{;}{}_{p}\,\omega{}_{;}{}_{a}{}_{b}{}^{p}{}_{c}{}_{d} -
  528\,\omega{}_{;}{}_{p}\,\omega{}_{;}{}_{a}{}^{p}{}_{b}{}_{c}{}_{d} \cr &&
+60\,\omega{}_{;}{}_{p}{}^{p}{}_{a}{}_{b}{}_{c}{}_{d}  \cr
%
&&+12\,\omega{}_{;}{}_{p}\,\left(
9\,R{}_{a}{}_{b}{}_{;}{}_{c}{}_{d}{}^{p} +
    4\,R{}_{a}{}_{b}{}_{;}{}_{c}{}^{p}{}_{d} -
    R{}_{a}{}_{b}{}^{;}{}^{p}{}_{c}{}_{d} -
    12\,R{}_{a}{}^{p}{}_{;}{}_{b}{}_{c}{}_{d} \right)  \cr &&
-6\,\left( 10\,\omega{}_{;}{}_{p}\,\omega{}^{;}{}^{p} -
     21\,\omega{}_{;}{}_{p}{}^{p} \right) \,R{}_{a}{}_{b}{}_{;}{}_{c}{}_{d}
+
  12\,\left( 3\,\omega{}_{;}{}_{p}\,\omega{}_{;}{}_{a} +
     16\,\omega{}_{;}{}_{p}{}_{a} \right) \,R{}_{b}{}_{c}{}_{;}{}_{d}{}^{p}
\cr && -12\,\left( 5\,\omega{}_{;}{}_{p}\,\omega{}_{;}{}_{a} -
     6\,\omega{}_{;}{}_{p}{}_{a} \right) \,R{}_{b}{}_{c}{}^{;}{}^{p}{}_{d} -
  72\,\left( \omega{}_{;}{}_{p}\,\omega{}_{;}{}_{a} +
     3\,\omega{}_{;}{}_{p}{}_{a} \right) \,R{}_{b}{}^{p}{}_{;}{}_{c}{}_{d}
\cr
&&
+3360\,\omega{}_{;}{}_{p}\,\omega{}_{;}{}_{a}\,\omega{}^{;}{}^{p}\,
   R{}_{b}{}_{c}{}_{;}{}_{d} + 288\,\omega{}_{;}{}_{p}\,\omega{}_{;}{}_{a}\,
   \omega{}_{;}{}_{b}\,R{}_{c}{}_{d}{}^{;}{}^{p} \cr &&
+696\,\omega{}_{;}{}_{p}\,\omega{}_{;}{}_{a}\,\omega{}_{;}{}_{b}\,
   R{}_{c}{}^{p}{}_{;}{}_{d} - 96\,\omega{}_{;}{}_{a}\,
   R{}_{c}{}_{d}{}^{;}{}^{p}\,\omega{}_{;}{}_{p}{}_{b} -
96\,\omega{}_{;}{}_{a}\,R{}_{c}{}^{p}{}_{;}{}_{d}\,\omega{}_{;}{}_{p}{}_{b}
\cr && +396\,\omega{}_{;}{}_{p}\,R{}_{c}{}_{d}{}^{;}{}^{p}\,
   \omega{}_{;}{}_{a}{}_{b} + 384\,\omega{}_{;}{}_{p}\,
   R{}_{c}{}^{p}{}_{;}{}_{d}\,\omega{}_{;}{}_{a}{}_{b} +
600\,\omega{}_{;}{}_{p}\,R{}_{b}{}_{c}{}_{;}{}_{d}\,\omega{}_{;}{}_{a}{}^{p}
\cr &&
 -480\,R{}_{c}{}_{d}{}^{;}{}^{p}\,\omega{}_{;}{}_{p}{}_{a}{}_{b} -
  1320\,R{}_{c}{}^{p}{}_{;}{}_{d}\,\omega{}_{;}{}_{p}{}_{a}{}_{b} +
  210\,R{}_{b}{}_{c}{}_{;}{}_{d}\,\omega{}_{;}{}_{p}{}^{p}{}_{a} \cr &&
+630\,R{}_{c}{}_{d}{}^{;}{}^{p}\,\omega{}_{;}{}_{a}{}_{b}{}_{p} +
  1260\,R{}_{c}{}^{p}{}_{;}{}_{d}\,\omega{}_{;}{}_{a}{}_{b}{}_{p} +
84\,\omega{}_{;}{}_{p}\,\left( R{}_{c}{}_{d}{}^{;}{}^{p} -
    R{}_{c}{}^{p}{}_{;}{}_{d} \right) \,R{}_{a}{}_{b} \cr &&
 -840\,\omega{}_{;}{}_{p}\,\left( R{}_{c}{}_{d}{}^{;}{}^{q} +
    2\,R{}_{c}{}^{q}{}_{;}{}_{d} \right) \,R{}_{q}{}_{a}{}_{b}{}^{p}
 -96\,\left( \omega{}_{;}{}_{p}\,\omega{}_{;}{}_{q} +
    \omega{}_{;}{}_{p}{}_{q} \right) \,
  R{}_{a}{}^{p}{}_{b}{}^{q}{}_{;}{}_{c}{}_{d} \cr &&
-12\,R{}_{b}{}^{p}{}_{c}{}^{q}{}_{;}{}_{d} \left(
80\,\omega{}_{;}{}_{p}\,\omega{}_{;}{}_{q}\,\omega{}_{;}{}_{a} +
  222\,\omega{}_{;}{}_{a}\,\omega{}_{;}{}_{p}{}_{q} -
  112\,\omega{}_{;}{}_{q}\,\omega{}_{;}{}_{p}{}_{a} \right. \cr && \quad
  \left. + 15\,\omega{}_{;}{}_{p}{}_{q}{}_{a} -
70\,\omega{}_{;}{}_{q}\,R{}_{p}{}_{a} \right)
+348\,\omega{}_{;}{}_{p}\,\left(
R{}_{b}{}^{q}{}_{c}{}^{r}{}_{;}{}_{d} +
    R{}_{b}{}^{r}{}_{c}{}^{q}{}_{;}{}_{d} \right)
\,R{}_{q}{}_{a}{}_{r}{}^{p}
\cr && +24\,\omega{}_{;}{}_{p}\,\left(
11\,R{}_{c}{}^{q}{}_{d}{}^{r}{}^{;}{}^{p} +
    15\,R{}_{c}{}^{q}{}^{p}{}^{r}{}_{;}{}_{d} +
    15\,R{}_{c}{}^{r}{}^{p}{}^{q}{}_{;}{}_{d} \right) \,
  R{}_{q}{}_{a}{}_{r}{}_{b} \cr &&
+2 \left(
8300\,\omega{}_{;}{}_{p}\,\omega{}_{;}{}_{a}\,\omega{}_{;}{}_{b}\,
   \omega{}^{;}{}^{p} + 42\,\omega{}_{;}{}_{a}\,\omega{}_{;}{}_{b}\,
   \omega{}_{;}{}_{p}{}^{p} + 1688\,\omega{}_{;}{}_{p}\,\omega{}^{;}{}^{p}\,
   \omega{}_{;}{}_{a}{}_{b} \right. \cr && \quad
 +84\,\omega{}_{;}{}_{p}{}^{p}\,\omega{}_{;}{}_{a}{}_{b} +
  6504\,\omega{}_{;}{}_{p}\,\omega{}_{;}{}_{a}\,\omega{}_{;}{}_{b}{}^{p} +
  132\,\omega{}_{;}{}_{p}{}_{a}\,\omega{}_{;}{}_{b}{}^{p}  \cr && \quad
\left.
  -42\,\omega{}_{;}{}_{a}\,\omega{}_{;}{}_{p}{}^{p}{}_{b} +
  1188\,\omega{}_{;}{}_{p}\,\omega{}_{;}{}_{a}{}_{b}{}^{p} -
  768\,\omega{}_{;}{}_{p}\,\omega{}_{;}{}_{a}{}^{p}{}_{b} +
  63\,\omega{}_{;}{}_{p}{}^{p}{}_{a}{}_{b}  \right) R{}_{c}{}_{d} \cr &&
+24 \left(
246\,\omega{}_{;}{}_{p}\,\omega{}_{;}{}_{a}\,\omega{}_{;}{}_{b}\,
   \omega{}_{;}{}_{c} + 19\,\omega{}_{;}{}_{a}\,\omega{}_{;}{}_{b}\,
   \omega{}_{;}{}_{p}{}_{c} + 304\,\omega{}_{;}{}_{p}\,\omega{}_{;}{}_{a}\,
   \omega{}_{;}{}_{b}{}_{c} \right. \cr && \quad
 +11\,\omega{}_{;}{}_{p}{}_{a}\,\omega{}_{;}{}_{b}{}_{c} +
  44\,\omega{}_{;}{}_{a}\,\omega{}_{;}{}_{p}{}_{b}{}_{c} +
  26\,\omega{}_{;}{}_{p}\,\omega{}_{;}{}_{a}{}_{b}{}_{c}  \cr && \quad
\left.
  -49\,\omega{}_{;}{}_{a}\,\omega{}_{;}{}_{b}{}_{c}{}_{p} -
  19\,\omega{}_{;}{}_{p}{}_{a}{}_{b}{}_{c} -
  7\,\omega{}_{;}{}_{a}{}_{b}{}_{p}{}_{c} +
  28\,\omega{}_{;}{}_{a}{}_{b}{}_{c}{}_{p}  \right) R{}_{d}{}^{p} \cr &&
+56\,\left( 37\,\omega{}_{;}{}_{p}\,\omega{}_{;}{}_{q} +
    \omega{}_{;}{}_{p}{}_{q} \right) \,R{}_{a}{}_{b}\,
  R{}_{c}{}^{p}{}_{d}{}^{q} \cr &&
 -1120\,\left( \omega{}_{;}{}_{p}\,\omega{}_{;}{}_{a} -
    2\,\omega{}_{;}{}_{p}{}_{a} \right) \,R{}_{q}{}_{b}\,
  R{}_{c}{}^{q}{}_{d}{}^{p} \cr
%
&& -4\left\{
 1844\,\omega{}_{;}{}_{p}\,\omega{}_{;}{}_{q}\,\omega{}_{;}{}_{a}\,
   \omega{}_{;}{}_{b} + 2308\,\omega{}_{;}{}_{a}\,\omega{}_{;}{}_{b}\,
   \omega{}_{;}{}_{p}{}_{q} + 256\,\omega{}_{;}{}_{q}\,\omega{}_{;}{}_{a}\,
   \omega{}_{;}{}_{p}{}_{b} \right. \cr && \quad
 -1144\,\omega{}_{;}{}_{p}{}_{b}\,\omega{}_{;}{}_{q}{}_{a} -
  1208\,\omega{}_{;}{}_{p}\,\omega{}_{;}{}_{a}\,\omega{}_{;}{}_{q}{}_{b} +
  56\,\omega{}_{;}{}_{p}{}_{a}\,\omega{}_{;}{}_{q}{}_{b} \cr && \quad
 -1520\,\omega{}_{;}{}_{p}\,\omega{}_{;}{}_{q}\,\omega{}_{;}{}_{a}{}_{b} +
  316\,\omega{}_{;}{}_{p}{}_{q}\,\omega{}_{;}{}_{a}{}_{b} +
  660\,\omega{}_{;}{}_{a}\,\omega{}_{;}{}_{p}{}_{q}{}_{b} \cr && \quad
 -768\,\omega{}_{;}{}_{q}\,\omega{}_{;}{}_{p}{}_{a}{}_{b} -
  18\,\omega{}_{;}{}_{p}\,\omega{}_{;}{}_{q}{}_{a}{}_{b} -
  420\,\omega{}_{;}{}_{p}\,\omega{}_{;}{}_{a}{}_{b}{}_{q} +
  24\,\omega{}_{;}{}_{p}{}_{q}{}_{a}{}_{b} \cr && \quad
 -7\,\left( 30\,\omega{}_{;}{}_{r}\,\omega{}^{;}{}^{r} +
    \omega{}_{;}{}_{r}{}^{r} \right) \,R{}_{p}{}_{a}{}_{q}{}_{b} \cr &&
\quad
 -12\,\left( 59\,\omega{}_{;}{}_{r}\,\omega{}_{;}{}_{a} +
    2\,\omega{}_{;}{}_{r}{}_{a} \right) \,R{}_{p}{}_{b}{}_{q}{}^{r} \cr &&
\quad
 -4\,\left( 191\,\omega{}_{;}{}_{r}\,\omega{}_{;}{}_{a} +
    20\,\omega{}_{;}{}_{r}{}_{a} \right) \,R{}_{p}{}^{r}{}_{q}{}_{b} \cr &&
\quad \left.
 -\left( \left( 19\,\omega{}_{;}{}_{p}\,\omega{}_{;}{}_{r} +
      16\,\omega{}_{;}{}_{p}{}_{r} \right) \,R{}_{q}{}_{a}{}_{b}{}^{r}
\right)
      \right\} R{}_{c}{}^{p}{}_{d}{}^{q}
\end{eqnarray} \begin{eqnarray}
%
120960 \bar\Delta^{(6')}_{ab} &=&
 -18232\,\omega{}_{;}{}_{p}\,\omega{}_{;}{}_{q}\,\omega{}_{;}{}_{a}\,
  \omega{}_{;}{}_{b}\,\omega{}^{;}{}^{p}\,\omega{}^{;}{}^{q}+
10364\,\omega{}_{;}{}_{p}\,\omega{}_{;}{}_{a}\,\omega{}_{;}{}_{b}\,
  \omega{}^{;}{}^{p}\,\omega{}_{;}{}_{q}{}^{q} \cr &&
-8\,\omega{}_{;}{}_{p}\,\omega{}_{;}{}_{q}\,
  \left( 520\,\omega{}^{;}{}^{p}\,\omega{}^{;}{}^{q}\,
     \omega{}_{;}{}_{a}{}_{b} +
    3518\,\omega{}_{;}{}_{a}\,\omega{}^{;}{}^{p}\,\omega{}_{;}{}_{b}{}^{q} +
    1353\,\omega{}_{;}{}_{a}\,\omega{}_{;}{}_{b}\,\omega{}^{;}{}^{p}{}^{q}
     \right)  \cr &&
+14\,\omega{}_{;}{}_{a}\,\omega{}_{;}{}_{b}\,\omega{}_{;}{}_{p}{}^{p}\,
   \omega{}_{;}{}_{q}{}^{q}
+ 2720\,\omega{}_{;}{}_{p}\,\omega{}^{;}{}^{p}\,
   \omega{}_{;}{}_{q}{}^{q}\,\omega{}_{;}{}_{a}{}_{b}
\cr &&
+
9888\,\omega{}_{;}{}_{p}\,\omega{}_{;}{}_{a}\,\omega{}_{;}{}_{q}{}^{q}\,
   \omega{}_{;}{}_{b}{}^{p}
 -2632\,\omega{}_{;}{}_{p}\,\omega{}^{;}{}^{p}\,\omega{}_{;}{}_{q}{}_{a}\,
   \omega{}_{;}{}_{b}{}^{q}
\cr &&
- 4576\,\omega{}_{;}{}_{p}\,\omega{}_{;}{}_{q}\,
   \omega{}_{;}{}_{a}{}^{p}\,\omega{}_{;}{}_{b}{}^{q}
-
2368\,\omega{}_{;}{}_{a}\,\omega{}_{;}{}_{b}\,\omega{}_{;}{}_{p}{}_{q}\,
   \omega{}^{;}{}^{p}{}^{q}
\cr &&
 -9152\,\omega{}_{;}{}_{p}\,\omega{}_{;}{}_{a}\,\omega{}_{;}{}_{q}{}_{b}\,
   \omega{}^{;}{}^{p}{}^{q}
 - 3440\,\omega{}_{;}{}_{p}\,\omega{}_{;}{}_{q}\,
   \omega{}_{;}{}_{a}{}_{b}\,\omega{}^{;}{}^{p}{}^{q}
\cr &&
+2700\,\omega{}_{;}{}_{p}\,\omega{}_{;}{}_{a}\,\omega{}^{;}{}^{p}\,
   \omega{}_{;}{}_{q}{}^{q}{}_{b}
+240\,\omega{}_{;}{}_{p}\,\omega{}_{;}{}_{a}\,\omega{}_{;}{}_{b}\,
   \omega{}_{;}{}_{q}{}^{q}{}^{p}
\cr &&
-
3024\,\omega{}_{;}{}_{p}\,\omega{}_{;}{}_{q}\,\omega{}^{;}{}^{p}\,
   \omega{}_{;}{}_{a}{}_{b}{}^{q}
+
1104\,\omega{}_{;}{}_{p}\,\omega{}_{;}{}_{q}\,\omega{}^{;}{}^{p}\,
   \omega{}_{;}{}_{a}{}^{q}{}_{b}
\cr &&
 -96\,\omega{}_{;}{}_{p}\,\omega{}_{;}{}_{q}\,\omega{}_{;}{}_{a}\,
   \omega{}_{;}{}_{b}{}^{p}{}^{q} -
  2712\,\omega{}_{;}{}_{p}\,\omega{}_{;}{}_{q}\,\omega{}_{;}{}_{a}\,
   \omega{}^{;}{}^{p}{}^{q}{}_{b} \cr &&
+98\,\omega{}_{;}{}_{p}{}^{p}\,\omega{}_{;}{}_{q}{}^{q}\,
   \omega{}_{;}{}_{a}{}_{b} + 408\,\omega{}_{;}{}_{p}{}_{a}\,
   \omega{}_{;}{}_{q}{}^{q}\,\omega{}_{;}{}_{b}{}^{p} -
  2096\,\omega{}_{;}{}_{p}{}_{q}\,\omega{}_{;}{}_{a}{}^{p}\,
   \omega{}_{;}{}_{b}{}^{q} \cr &&
+784\,\omega{}_{;}{}_{p}{}_{a}\,\omega{}_{;}{}_{q}{}_{b}\,
   \omega{}^{;}{}^{p}{}^{q} - 424\,\omega{}_{;}{}_{p}{}_{q}\,
   \omega{}_{;}{}_{a}{}_{b}\,\omega{}^{;}{}^{p}{}^{q} -
  84\,\omega{}_{;}{}_{a}\,\omega{}_{;}{}_{p}{}^{p}\,
   \omega{}_{;}{}_{q}{}^{q}{}_{b} \cr &&
+756\,\omega{}_{;}{}_{p}\,\omega{}_{;}{}_{a}{}^{p}\,
   \omega{}_{;}{}_{q}{}^{q}{}_{b} -
  120\,\omega{}_{;}{}_{a}\,\omega{}_{;}{}_{p}{}_{b}\,
   \omega{}_{;}{}_{q}{}^{q}{}^{p} +
  444\,\omega{}_{;}{}_{p}\,\omega{}_{;}{}_{a}{}_{b}\,
   \omega{}_{;}{}_{q}{}^{q}{}^{p} \cr &&
+1656\,\omega{}_{;}{}_{p}\,\omega{}_{;}{}_{q}{}^{q}\,
   \omega{}_{;}{}_{a}{}_{b}{}^{p} -
  2316\,\omega{}_{;}{}_{p}\,\omega{}_{;}{}_{q}{}^{p}\,
   \omega{}_{;}{}_{a}{}_{b}{}^{q} -
  648\,\omega{}_{;}{}_{p}\,\omega{}_{;}{}_{q}{}^{q}\,
   \omega{}_{;}{}_{a}{}^{p}{}_{b} \cr &&
+504\,\omega{}_{;}{}_{p}\,\omega{}_{;}{}_{q}{}^{p}\,
   \omega{}_{;}{}_{a}{}^{q}{}_{b} +
  864\,\omega{}_{;}{}_{p}\,\omega{}_{;}{}_{q}{}_{a}\,
   \omega{}_{;}{}_{b}{}^{p}{}^{q} -
  2928\,\omega{}_{;}{}_{p}\,\omega{}_{;}{}_{q}{}_{a}\,
   \omega{}_{;}{}_{b}{}^{q}{}^{p} \cr &&
 -1944\,\omega{}_{;}{}_{a}\,\omega{}_{;}{}_{p}{}_{q}\,
   \omega{}^{;}{}^{p}{}^{q}{}_{b} -
  1104\,\omega{}_{;}{}_{p}\,\omega{}_{;}{}_{q}{}_{a}\,
   \omega{}^{;}{}^{p}{}^{q}{}_{b} +
  30\,\omega{}_{;}{}_{p}\,\omega{}^{;}{}^{p}\,
   \omega{}_{;}{}_{q}{}^{q}{}_{a}{}_{b} \cr &&
 -72\,\omega{}_{;}{}_{p}\,\omega{}_{;}{}_{a}\,
   \omega{}_{;}{}_{q}{}^{q}{}_{b}{}^{p} -
  24\,\omega{}_{;}{}_{p}\,\omega{}_{;}{}_{a}\,
   \omega{}_{;}{}_{q}{}^{q}{}^{p}{}_{b} +
  108\,\omega{}_{;}{}_{p}\,\omega{}_{;}{}_{q}\,
   \omega{}_{;}{}_{a}{}_{b}{}^{p}{}^{q} \cr &&
 -648\,\omega{}_{;}{}_{p}\,\omega{}_{;}{}_{q}\,
   \omega{}_{;}{}_{a}{}^{p}{}_{b}{}^{q} -
  120\,\omega{}_{;}{}_{p}\,\omega{}_{;}{}_{q}\,
   \omega{}_{;}{}_{a}{}^{p}{}^{q}{}_{b} +
  132\,\omega{}_{;}{}_{p}\,\omega{}_{;}{}_{q}\,
   \omega{}^{;}{}^{p}{}^{q}{}_{a}{}_{b} \cr &&
+105\,\omega{}_{;}{}_{p}{}^{p}{}_{a}\,\omega{}_{;}{}_{q}{}^{q}{}_{b}
-
  480\,\omega{}_{;}{}_{p}{}_{a}{}_{b}\,\omega{}_{;}{}_{q}{}^{q}{}^{p} +
  630\,\omega{}_{;}{}_{p}{}^{p}{}^{q}\,\omega{}_{;}{}_{a}{}_{b}{}_{q} -
  90\,\omega{}_{;}{}_{p}{}_{q}{}_{a}\,\omega{}^{;}{}^{p}{}^{q}{}_{b} \cr &&
+126\,\omega{}_{;}{}_{p}{}^{p}\,\omega{}_{;}{}_{q}{}^{q}{}_{a}{}_{b}
+
  192\,\omega{}_{;}{}_{p}{}_{a}\,\omega{}_{;}{}_{q}{}^{q}{}_{b}{}^{p} +
  72\,\omega{}_{;}{}_{p}{}_{a}\,\omega{}_{;}{}_{q}{}^{q}{}^{p}{}_{b} -
  96\,\omega{}_{;}{}_{p}{}_{q}\,\omega{}^{;}{}^{p}{}^{q}{}_{a}{}_{b} \cr &&
+108\,\omega{}_{;}{}_{p}\,\omega{}_{;}{}_{q}{}^{q}{}_{a}{}_{b}{}^{p}
+
  48\,\omega{}_{;}{}_{p}\,\omega{}_{;}{}_{q}{}^{q}{}_{a}{}^{p}{}_{b} -
  12\,\omega{}_{;}{}_{p}\,\omega{}_{;}{}_{q}{}^{q}{}^{p}{}_{a}{}_{b} \cr &&
%
%
+18\,\omega{}_{;}{}_{p}\,\omega{}_{;}{}_{q}\,
  \left( 3\,R{}_{a}{}_{b}{}^{;}{}^{p}{}^{q} -
    6\,R{}_{a}{}^{p}{}_{;}{}_{b}{}^{q} + 2\,R{}_{a}{}^{p}{}^{;}{}^{q}{}_{b}
+
    5\,R{}^{p}{}^{q}{}_{;}{}_{a}{}_{b} \right)  \cr &&
-12\,\omega{}_{;}{}_{p}\,\omega{}_{;}{}_{q}\,
  \left( \omega{}^{;}{}^{p}\,\left( 26\,R{}_{a}{}_{b}{}^{;}{}^{q} +
       53\,R{}_{a}{}^{q}{}_{;}{}_{b} \right)  +
    2\,\omega{}_{;}{}_{a}\,\left( 2\,R{}_{b}{}^{p}{}^{;}{}^{q} +
       R{}^{p}{}^{q}{}_{;}{}_{b} \right)  \right)  \cr &&
-24\,\omega{}_{;}{}_{p}\,\left( R{}_{b}{}^{p}{}^{;}{}^{q} -
    2\,R{}_{b}{}^{q}{}^{;}{}^{p} - 11\,R{}^{p}{}^{q}{}_{;}{}_{b} \right) \,
  \omega{}_{;}{}_{q}{}_{a} \cr &&
+6\,\omega{}_{;}{}_{p}\,\left( \left(
17\,R{}_{a}{}_{b}{}^{;}{}^{q} -
       18\,R{}_{a}{}^{q}{}_{;}{}_{b} \right) \,\omega{}_{;}{}_{q}{}^{p} +
    14\,\left( R{}_{a}{}_{b}{}^{;}{}^{p} - R{}_{a}{}^{p}{}_{;}{}_{b} \right)
       \,\omega{}_{;}{}_{q}{}^{q} \right)  \cr &&
+24\,\omega{}_{;}{}_{p}\,\omega{}_{;}{}_{q}\,\omega{}_{;}{}_{r}\,
  \left( 11\,R{}_{a}{}^{p}{}_{b}{}^{q}{}^{;}{}^{r} -
    30\,R{}_{a}{}^{p}{}^{q}{}^{r}{}_{;}{}_{b} \right)  \cr &&
+24\,\omega{}_{;}{}_{p}\,\left(
11\,R{}_{a}{}^{q}{}_{b}{}^{r}{}^{;}{}^{p} +
    30\,R{}_{a}{}^{q}{}^{p}{}^{r}{}_{;}{}_{b} \right) \,
  \omega{}_{;}{}_{q}{}_{r}  +
42\,\omega{}_{;}{}_{p}\,\omega{}_{;}{}_{q}\,R{}_{a}{}_{b}\,R{}^{p}{}^{q}\cr
&&
%
-\left(
1580\,\omega{}_{;}{}_{p}\,\omega{}_{;}{}_{q}\,\omega{}^{;}{}^{p}\,
   \omega{}^{;}{}^{q} + 98\,\omega{}_{;}{}_{p}\,\omega{}^{;}{}^{p}\,
   \omega{}_{;}{}_{q}{}^{q} - 35\,\omega{}_{;}{}_{p}{}^{p}\,
   \omega{}_{;}{}_{q}{}^{q} \right. \cr && \quad \left.
  +964\,\omega{}_{;}{}_{p}\,\omega{}_{;}{}_{q}\,\omega{}^{;}{}^{p}{}^{q} -
  28\,\omega{}_{;}{}_{p}{}_{q}\,\omega{}^{;}{}^{p}{}^{q} -
  84\,\omega{}_{;}{}_{p}\,\omega{}_{;}{}_{q}{}^{q}{}^{p}
\right) R{}_{a}{}_{b} \cr &&
-4\left(
1470\,\omega{}_{;}{}_{p}\,\omega{}_{;}{}_{q}\,\omega{}_{;}{}_{a}\,
   \omega{}^{;}{}^{q} + 101\,\omega{}_{;}{}_{q}\,\omega{}^{;}{}^{q}\,
   \omega{}_{;}{}_{p}{}_{a} + 16\,\omega{}_{;}{}_{q}\,\omega{}_{;}{}_{a}\,
   \omega{}_{;}{}_{p}{}^{q} \right. \cr && \quad
 +14\,\omega{}_{;}{}_{p}\,\omega{}_{;}{}_{a}\,\omega{}_{;}{}_{q}{}^{q} -
  7\,\omega{}_{;}{}_{p}{}_{a}\,\omega{}_{;}{}_{q}{}^{q} +
  736\,\omega{}_{;}{}_{p}\,\omega{}_{;}{}_{q}\,\omega{}_{;}{}_{a}{}^{q} -
  2\,\omega{}_{;}{}_{p}{}_{q}\,\omega{}_{;}{}_{a}{}^{q} \cr && \quad \left.
 +180\,\omega{}_{;}{}_{q}\,\omega{}_{;}{}_{p}{}_{a}{}^{q} +
  24\,\omega{}_{;}{}_{q}\,\omega{}_{;}{}_{p}{}^{q}{}_{a} -
  210\,\omega{}_{;}{}_{q}\,\omega{}_{;}{}_{a}{}^{q}{}_{p}
\right) R{}_{b}{}^{p} \cr &&
-4\left(
89\,\omega{}_{;}{}_{p}\,\omega{}_{;}{}_{q}\,\omega{}_{;}{}_{a}\,
   \omega{}_{;}{}_{b} + 28\,\omega{}_{;}{}_{p}\,\omega{}_{;}{}_{a}\,
   \omega{}_{;}{}_{q}{}_{b} - 34\,\omega{}_{;}{}_{p}{}_{a}\,
   \omega{}_{;}{}_{q}{}_{b} \right. \cr && \quad \left.
  +63\,\omega{}_{;}{}_{p}\,\omega{}_{;}{}_{q}\,\omega{}_{;}{}_{a}{}_{b} -
  36\,\omega{}_{;}{}_{p}\,\omega{}_{;}{}_{q}{}_{a}{}_{b}
\right) R{}^{p}{}^{q} +
672\,\omega{}_{;}{}_{p}\,\omega{}_{;}{}_{q}\,R{}_{r}{}_{a}\,
  R{}_{b}{}^{p}{}^{q}{}^{r} \cr &&
%
-8\left(
179\,\omega{}_{;}{}_{p}\,\omega{}_{;}{}_{q}\,\omega{}_{;}{}_{r}\,
   \omega{}^{;}{}^{r} - 165\,\omega{}_{;}{}_{r}\,\omega{}^{;}{}^{r}\,
   \omega{}_{;}{}_{p}{}_{q} + 378\,\omega{}_{;}{}_{q}\,\omega{}_{;}{}_{r}\,
   \omega{}_{;}{}_{p}{}^{r} \right. \cr && \quad
  -8\,\omega{}_{;}{}_{p}\,\omega{}_{;}{}_{r}\,\omega{}_{;}{}_{q}{}^{r} +
  8\,\omega{}_{;}{}_{p}{}_{r}\,\omega{}_{;}{}_{q}{}^{r} -
  175\,\omega{}_{;}{}_{p}\,\omega{}_{;}{}_{q}\,\omega{}_{;}{}_{r}{}^{r} -
  7\,\omega{}_{;}{}_{p}{}_{q}\,\omega{}_{;}{}_{r}{}^{r} \cr && \quad \left.
  -33\,\omega{}_{;}{}_{r}\,\omega{}_{;}{}_{p}{}_{q}{}^{r} -
  105\,\omega{}_{;}{}_{p}\,\omega{}_{;}{}_{r}{}^{r}{}_{q} +
  45\,\omega{}_{;}{}_{r}\,\omega{}_{;}{}_{s}\,R{}_{p}{}^{r}{}_{q}{}^{s}
\right) R{}_{a}{}^{p}{}_{b}{}^{q} \cr &&
4\left(
1124\,\omega{}_{;}{}_{q}\,\omega{}_{;}{}_{a}\,\omega{}_{;}{}_{p}{}_{r}
+
  400\,\omega{}_{;}{}_{p}\,\omega{}_{;}{}_{q}\,\omega{}_{;}{}_{r}{}_{a} -
  104\,\omega{}_{;}{}_{p}{}_{q}\,\omega{}_{;}{}_{r}{}_{a} \right. \cr &&
\quad
\left.
  +174\,\omega{}_{;}{}_{q}\,\omega{}_{;}{}_{p}{}_{r}{}_{a} -
  87\,\omega{}_{;}{}_{q}\,\omega{}_{;}{}_{s}\,R{}_{p}{}_{a}{}_{r}{}^{s} -
  87\,\omega{}_{;}{}_{q}\,\omega{}_{;}{}_{s}\,R{}_{p}{}^{s}{}_{r}{}_{a}
\right) R{}_{b}{}^{p}{}^{q}{}^{r} \end{eqnarray} \begin{eqnarray}
%
362880 \bar\Delta^{(6')} &=&
 4792\,\omega{}_{;}{}_{p}\,\omega{}_{;}{}_{q}\,\omega{}_{;}{}_{r}\,
  \omega{}^{;}{}^{p}\,\omega{}^{;}{}^{q}\,\omega{}^{;}{}^{r}
-3732\,\omega{}_{;}{}_{p}\,\omega{}_{;}{}_{q}\,\omega{}^{;}{}^{p}\,
  \omega{}^{;}{}^{q}\,\omega{}_{;}{}_{r}{}^{r} \cr &&
+8808\,\omega{}_{;}{}_{p}\,\omega{}_{;}{}_{q}\,\omega{}_{;}{}_{r}\,
  \omega{}^{;}{}^{p}\,\omega{}^{;}{}^{q}{}^{r}
-84\,\omega{}_{;}{}_{p}\,\omega{}^{;}{}^{p}\,\omega{}_{;}{}_{q}{}^{q}\,
  \omega{}_{;}{}_{r}{}^{r} \cr &&
-2568\,\omega{}_{;}{}_{p}\,\omega{}_{;}{}_{q}\,\omega{}_{;}{}_{r}{}^{r}\,
   \omega{}^{;}{}^{p}{}^{q} + 1440\,\omega{}_{;}{}_{p}\,\omega{}^{;}{}^{p}\,
   \omega{}_{;}{}_{q}{}_{r}\,\omega{}^{;}{}^{q}{}^{r} \cr &&
+1248\,\omega{}_{;}{}_{p}\,\omega{}_{;}{}_{q}\,\omega{}_{;}{}_{r}{}^{p}\,
   \omega{}^{;}{}^{q}{}^{r} - 828\,\omega{}_{;}{}_{p}\,\omega{}_{;}{}_{q}\,
   \omega{}^{;}{}^{p}\,\omega{}_{;}{}_{r}{}^{r}{}^{q} \cr &&
+360\,\omega{}_{;}{}_{p}\,\omega{}_{;}{}_{q}\,\omega{}_{;}{}_{r}\,
  \omega{}^{;}{}^{p}{}^{q}{}^{r}
+35\,\omega{}_{;}{}_{p}{}^{p}\,\omega{}_{;}{}_{q}{}^{q}\,
   \omega{}_{;}{}_{r}{}^{r} + 84\,\omega{}_{;}{}_{p}{}_{q}\,
   \omega{}_{;}{}_{r}{}^{r}\,\omega{}^{;}{}^{p}{}^{q} \cr &&
+-64\,\omega{}_{;}{}_{p}{}_{q}\,\omega{}_{;}{}_{r}{}^{p}\,
   \omega{}^{;}{}^{q}{}^{r} + 252\,\omega{}_{;}{}_{p}\,
   \omega{}_{;}{}_{q}{}^{q}\,\omega{}_{;}{}_{r}{}^{r}{}^{p} +
  306\,\omega{}_{;}{}_{p}\,\omega{}_{;}{}_{q}{}^{p}\,
   \omega{}_{;}{}_{r}{}^{r}{}^{q} \cr &&
+792\,\omega{}_{;}{}_{p}\,\omega{}_{;}{}_{q}{}_{r}\,
   \omega{}^{;}{}^{q}{}^{r}{}^{p} +
  162\,\omega{}_{;}{}_{p}\,\omega{}_{;}{}_{q}\,
   \omega{}_{;}{}_{r}{}^{r}{}^{p}{}^{q} \cr &&
+18\,\omega{}_{;}{}_{p}\,\left(
56\,\omega{}_{;}{}_{q}\,\omega{}_{;}{}_{r}\,
     \omega{}^{;}{}^{r} + 18\,\omega{}_{;}{}_{r}\,\omega{}_{;}{}_{q}{}^{r} +
    7\,\omega{}_{;}{}_{q}\,\omega{}_{;}{}_{r}{}^{r} \right) \,R{}^{p}{}^{q}
\cr && +108\,\omega{}_{;}{}_{p}\,\omega{}_{;}{}_{q}\,
  \left( \omega{}_{;}{}_{r}\,R{}^{p}{}^{q}{}^{;}{}^{r} -
    10\,\omega{}_{;}{}_{r}{}_{s}\,R{}^{p}{}^{r}{}^{q}{}^{s} \right)
\end{eqnarray}
\end{subequations}

\section{Expansion tensors for $G_{\rm div,ren}$}
\label{apx-Gdiv}
We present the explicit expressions for the expansion tensors
for $G_{\rm div,ren}$. The expansion scalars in the body are
related to these tensors via
$G^{(n)}_{{\rm div,ren}} = 2^\frac{n}{2} p^{a_1}\cdots p^{a_n}
G^{(n)}_{{\rm div,ren},a_1\cdots a_n}$.

\begin{subequations}
\begin{eqnarray} G^{(0)}_{\rm div,ren} &=& \frac{1}{6} \left(
           \w{}_{;}{}_{p}{}^{p} - \w{}_{;}{}_{p}\,\w{}^{;}{}^{p} \right)
\\
G^{(1)}_{{\rm div,ren},a} &=&
   {\frac{1}{12}} \left(
        2\,\w{}_{;}{}_{p}\,\w{}_{;}{}_{a}{}^{p} - \w{}_{;}{}_{p}{}^{p}{}_{a}
\right)
\\
G^{(2)}_{{\rm div,ren},ab} &=&
\left(
4\,\w{}_{;}{}_{p}\,\w{}_{;}{}_{a}\,\w{}_{;}{}_{b}\,\w{}^{;}{}^{p}
+
  15\,\w{}_{;}{}_{p}\,R{}_{a}{}_{b}{}^{;}{}^{p} -
  15\,\w{}_{;}{}_{p}\,R{}_{a}{}^{p}{}_{;}{}_{b} -
  4\,\w{}_{;}{}_{a}\,\w{}_{;}{}_{b}\,\w{}_{;}{}_{p}{}^{p} \right. \cr &&
+2\,\w{}_{;}{}_{p}{}^{p}\,\w{}_{;}{}_{a}{}_{b} +
  8\,\w{}_{;}{}_{p}\,\w{}_{;}{}_{a}\,\w{}_{;}{}_{b}{}^{p} -
  20\,\w{}_{;}{}_{p}{}_{a}\,\w{}_{;}{}_{b}{}^{p} +
  6\,\w{}_{;}{}_{a}\,\w{}_{;}{}_{p}{}_{b}{}^{p} \cr &&
 -12\,\w{}_{;}{}_{a}\,\w{}_{;}{}_{p}{}^{p}{}_{b} +
  36\,\w{}_{;}{}_{p}\,\w{}_{;}{}_{a}{}_{b}{}^{p} -
  54\,\w{}_{;}{}_{p}\,\w{}_{;}{}_{a}{}^{p}{}_{b} -
  12\,\w{}_{;}{}_{p}{}_{a}{}_{b}{}^{p} \cr &&
+3\,\w{}_{;}{}_{p}{}_{a}{}^{p}{}_{b} +
9\,\w{}_{;}{}_{p}{}^{p}{}_{a}{}_{b} +
  9\,\w{}_{;}{}_{a}{}_{b}{}_{p}{}^{p} -
  5\,\w{}_{;}{}_{p}\,\w{}^{;}{}^{p}\,R{}_{a}{}_{b} \cr &&
+5\,\w{}_{;}{}_{p}{}^{p}\,R{}_{a}{}_{b} -
  10\,\w{}_{;}{}_{p}\,\w{}_{;}{}_{a}\,R{}_{b}{}^{p} +
  5\,\w{}_{;}{}_{p}{}_{a}\,R{}_{b}{}^{p} +
  34\,\w{}_{;}{}_{p}\,\w{}_{;}{}_{q}\,R{}_{a}{}^{q}{}_{b}{}^{p} \cr &&
\left.
+ 10\,\w{}_{;}{}_{p}{}_{q}\,R{}_{a}{}^{q}{}_{b}{}^{p} \right) /
360 \cr
%
&& -g_{ab} \left(
2\,\w{}_{;}{}_{p}\,\w{}_{;}{}_{q}\,\w{}^{;}{}^{p}\,\w{}^{;}{}^{q}
+
  4\,\w{}_{;}{}_{p}\,\w{}^{;}{}^{p}\,\w{}_{;}{}_{q}{}^{q} -
  5\,\w{}_{;}{}_{p}{}^{p}\,\w{}_{;}{}_{q}{}^{q} +
  16\,\w{}_{;}{}_{p}\,\w{}_{;}{}_{q}\,\w{}^{;}{}^{p}{}^{q} \right. \cr &&
\left.
 -4\,\w{}_{;}{}_{p}{}_{q}\,\w{}^{;}{}^{p}{}^{q} -
  12\,\w{}_{;}{}_{p}\,\w{}_{;}{}_{q}{}^{q}{}^{p} -
  6\,\w{}_{;}{}_{p}\,\w{}_{;}{}_{q}\,R{}^{p}{}^{q} \right) / 720
\\
\cr
G^{(3)}_{{\rm div,ren},abc} &=&
\left(
25\,\w{}_{;}{}_{p}\,\w{}^{;}{}^{p}\,R{}_{a}{}_{b}{}_{;}{}_{c} +
  50\,\w{}_{;}{}_{p}\,\w{}_{;}{}_{a}\,R{}_{b}{}_{c}{}^{;}{}^{p} -
  10\,\w{}_{;}{}_{p}\,\w{}_{;}{}_{a}\,R{}_{b}{}^{p}{}_{;}{}_{c}
\right. \cr &&
 -
130\,\w{}_{;}{}_{p}\,\w{}_{;}{}_{q}\,R{}_{a}{}^{q}{}_{b}{}^{p}{}_{;}{}_{c}
 -25\,R{}_{a}{}_{b}{}_{;}{}_{c}\,\w{}_{;}{}_{p}{}^{p} -
  40\,\w{}_{;}{}_{p}\,\w{}_{;}{}_{a}\,\w{}^{;}{}^{p}\,\w{}_{;}{}_{b}{}_{c}
\cr &&
 +  40\,\w{}_{;}{}_{a}\,\w{}_{;}{}_{p}{}^{p}\,\w{}_{;}{}_{b}{}_{c} -
  52\,\w{}_{;}{}_{p}\,\w{}_{;}{}_{a}{}^{p}\,\w{}_{;}{}_{b}{}_{c} \cr &&
 -40\,\w{}_{;}{}_{p}\,\w{}_{;}{}_{a}\,\w{}_{;}{}_{b}\,\w{}_{;}{}_{c}{}^{p} +
  80\,R{}_{p}{}_{a}{}_{;}{}_{b}\,\w{}_{;}{}_{c}{}^{p} -
  100\,R{}_{a}{}_{b}{}_{;}{}_{p}\,\w{}_{;}{}_{c}{}^{p} -
  40\,\w{}_{;}{}_{a}\,\w{}_{;}{}_{p}{}_{b}\,\w{}_{;}{}_{c}{}^{p} \cr &&
+12\,\w{}_{;}{}_{p}\,\w{}_{;}{}_{a}{}_{b}\,\w{}_{;}{}_{c}{}^{p} -
  40\,R{}_{p}{}_{a}{}_{q}{}_{b}{}_{;}{}_{c}\,\w{}^{;}{}^{p}{}^{q} -
  21\,\w{}_{;}{}_{p}\,R{}_{a}{}_{b}{}_{;}{}_{c}{}^{p} -
  39\,\w{}_{;}{}_{p}\,R{}_{a}{}_{b}{}^{;}{}^{p}{}_{c} \cr &&
+60\,\w{}_{;}{}_{p}\,R{}_{a}{}^{p}{}_{;}{}_{b}{}_{c} +
  20\,\w{}_{;}{}_{a}\,\w{}_{;}{}_{b}\,\w{}_{;}{}_{p}{}_{c}{}^{p} -
  40\,\w{}_{;}{}_{a}{}_{b}\,\w{}_{;}{}_{p}{}_{c}{}^{p} +
  60\,\w{}_{;}{}_{a}{}_{b}\,\w{}_{;}{}_{p}{}^{p}{}_{c} \cr &&
 -10\,\w{}_{;}{}_{p}{}^{p}\,\w{}_{;}{}_{a}{}_{b}{}_{c} +
  20\,\w{}_{;}{}_{p}\,\w{}_{;}{}_{a}\,\w{}_{;}{}_{b}{}_{c}{}^{p} -
  140\,\w{}_{;}{}_{p}{}_{a}\,\w{}_{;}{}_{b}{}_{c}{}^{p} -
  60\,\w{}_{;}{}_{p}\,\w{}_{;}{}_{a}\,\w{}_{;}{}_{b}{}^{p}{}_{c} \cr &&
+280\,\w{}_{;}{}_{p}{}_{a}\,\w{}_{;}{}_{b}{}^{p}{}_{c} -
  80\,\w{}_{;}{}_{a}\,\w{}_{;}{}_{p}{}_{b}{}_{c}{}^{p} +
  10\,\w{}_{;}{}_{a}\,\w{}_{;}{}_{p}{}_{b}{}^{p}{}_{c} +
  50\,\w{}_{;}{}_{a}\,\w{}_{;}{}_{p}{}^{p}{}_{b}{}_{c} \cr &&
 -110\,\w{}_{;}{}_{p}\,\w{}_{;}{}_{a}{}_{b}{}_{c}{}^{p} -
  30\,\w{}_{;}{}_{p}\,\w{}_{;}{}_{a}{}_{b}{}^{p}{}_{c} +
  180\,\w{}_{;}{}_{p}\,\w{}_{;}{}_{a}{}^{p}{}_{b}{}_{c} +
  50\,\w{}_{;}{}_{a}\,\w{}_{;}{}_{b}{}_{c}{}_{p}{}^{p} \cr &&
+20\,\w{}_{;}{}_{p}{}_{a}{}_{b}{}_{c}{}^{p} +
  20\,\w{}_{;}{}_{p}{}_{a}{}_{b}{}^{p}{}_{c} -
  10\,\w{}_{;}{}_{p}{}_{a}{}^{p}{}_{b}{}_{c} -
  20\,\w{}_{;}{}_{p}{}^{p}{}_{a}{}_{b}{}_{c} \cr &&
+10\,\w{}_{;}{}_{a}{}_{b}{}_{p}{}_{c}{}^{p} -
  20\,\w{}_{;}{}_{a}{}_{b}{}_{p}{}^{p}{}_{c} -
  20\,\w{}_{;}{}_{a}{}_{b}{}_{c}{}_{p}{}^{p} +
  50\,\w{}_{;}{}_{p}\,\w{}_{;}{}_{a}{}^{p}\,R{}_{b}{}_{c} \cr &&
 -25\,\w{}_{;}{}_{p}{}^{p}{}_{a}\,R{}_{b}{}_{c} +
  10\,\w{}_{;}{}_{p}\,R{}_{a}{}^{p}\,R{}_{b}{}_{c} -
  20\,\w{}_{;}{}_{p}\,\w{}_{;}{}_{a}\,\w{}_{;}{}_{b}\,R{}_{c}{}^{p} +
  70\,\w{}_{;}{}_{a}\,\w{}_{;}{}_{p}{}_{b}\,R{}_{c}{}^{p} \cr &&
+60\,\w{}_{;}{}_{p}\,\w{}_{;}{}_{a}{}_{b}\,R{}_{c}{}^{p} -
  50\,\w{}_{;}{}_{p}{}_{a}{}_{b}\,R{}_{c}{}^{p} +
  20\,\w{}_{;}{}_{a}{}_{b}{}_{p}\,R{}_{c}{}^{p} -
  10\,\w{}_{;}{}_{p}\,R{}_{a}{}_{b}\,R{}_{c}{}^{p} \cr &&
+20\,\w{}_{;}{}_{p}\,\w{}_{;}{}_{q}\,\w{}_{;}{}_{a}\,
   R{}_{b}{}^{q}{}_{c}{}^{p} + 20\,\w{}_{;}{}_{a}\,\w{}_{;}{}_{p}{}_{q}\,
   R{}_{b}{}^{q}{}_{c}{}^{p} \cr &&
  - 480\,\w{}_{;}{}_{p}\,\w{}_{;}{}_{q}{}_{a}\,
   R{}_{b}{}^{q}{}_{c}{}^{p} + 20\,\w{}_{;}{}_{p}{}_{q}{}_{a}\,
   R{}_{b}{}^{q}{}_{c}{}^{p} \cr &&
 -80\,\w{}_{;}{}_{p}{}_{a}{}_{q}\,R{}_{b}{}^{q}{}_{c}{}^{p} -
  32\,\w{}_{;}{}_{p}\,R{}_{q}{}_{a}\,R{}_{b}{}^{q}{}_{c}{}^{p}
 \cr &&
-
32\,\w{}_{;}{}_{p}\,R{}_{q}{}_{r}{}_{a}{}^{p}\,R{}_{b}{}^{r}{}_{c}{}^{q}
+
  24\,\w{}_{;}{}_{p}\,R{}_{q}{}_{a}{}_{r}{}^{p}\,R{}_{b}{}^{r}{}_{c}{}^{q}
\cr
&& \left.
+32\,\w{}_{;}{}_{p}\,R{}_{q}{}^{p}{}_{r}{}_{a}\,R{}_{b}{}^{r}{}_{c}{}^{q}
-
  16\,\w{}_{;}{}_{p}\,R{}_{q}{}_{a}{}_{r}{}_{b}\,R{}_{c}{}^{q}{}^{p}{}^{r}
\right) / 3600 \cr
%
&&-g_{ab}\left(
15\,\w{}_{;}{}_{p}\,\w{}_{;}{}_{q}\,R{}_{c}{}^{p}{}^{;}{}^{q} -
  100\,\w{}_{;}{}_{p}\,\w{}_{;}{}_{q}\,\w{}^{;}{}^{q}\,\w{}_{;}{}_{c}{}^{p}
\right. \cr && -
20\,\w{}_{;}{}_{p}\,\w{}_{;}{}_{q}{}^{q}\,\w{}_{;}{}_{c}{}^{p} +
  80\,\w{}_{;}{}_{p}\,\w{}_{;}{}_{q}\,\w{}^{;}{}^{p}\,\w{}_{;}{}_{c}{}^{q}
\cr &&
+132\,\w{}_{;}{}_{p}\,\w{}_{;}{}_{q}{}^{p}\,\w{}_{;}{}_{c}{}^{q} -
  212\,\w{}_{;}{}_{p}\,\w{}_{;}{}_{q}{}_{c}\,\w{}^{;}{}^{p}{}^{q}
\cr && -
10\,\w{}_{;}{}_{p}\,\w{}^{;}{}^{p}\,\w{}_{;}{}_{q}{}_{c}{}^{q} +
  10\,\w{}_{;}{}_{p}\,\w{}_{;}{}_{c}\,\w{}_{;}{}_{q}{}^{p}{}^{q} \cr &&
+25\,\w{}_{;}{}_{p}{}^{p}\,\w{}_{;}{}_{q}{}^{q}{}_{c} -
  10\,\w{}_{;}{}_{p}\,\w{}_{;}{}_{c}\,\w{}_{;}{}_{q}{}^{q}{}^{p} +
  30\,\w{}_{;}{}_{p}{}_{c}\,\w{}_{;}{}_{q}{}^{q}{}^{p} -
  70\,\w{}_{;}{}_{p}\,\w{}_{;}{}_{q}\,\w{}_{;}{}_{c}{}^{p}{}^{q} \cr &&
+10\,\w{}_{;}{}_{p}{}_{q}\,\w{}_{;}{}_{c}{}^{p}{}^{q} +
  30\,\w{}_{;}{}_{p}\,\w{}_{;}{}_{q}\,\w{}^{;}{}^{p}{}^{q}{}_{c} +
  10\,\w{}_{;}{}_{p}{}_{q}\,\w{}^{;}{}^{p}{}^{q}{}_{c} -
  20\,\w{}_{;}{}_{p}\,\w{}_{;}{}_{q}{}_{c}{}^{p}{}^{q} \cr &&
+5\,\w{}_{;}{}_{p}\,\w{}_{;}{}_{q}{}_{c}{}^{q}{}^{p} +
  10\,\w{}_{;}{}_{p}\,\w{}_{;}{}_{q}{}^{p}{}_{c}{}^{q} -
  5\,\w{}_{;}{}_{p}\,\w{}_{;}{}_{q}{}^{p}{}^{q}{}_{c} +
  25\,\w{}_{;}{}_{p}\,\w{}_{;}{}_{q}{}^{q}{}_{c}{}^{p} \cr &&
+5\,\w{}_{;}{}_{p}\,\w{}_{;}{}_{q}{}^{q}{}^{p}{}_{c} +
  10\,\w{}_{;}{}_{p}\,\w{}_{;}{}_{c}{}^{p}{}_{q}{}^{q} +
  10\,\w{}_{;}{}_{p}\,\w{}_{;}{}_{q}\,\w{}^{;}{}^{p}\,R{}_{c}{}^{q} -
  5\,\w{}_{;}{}_{p}\,\w{}_{;}{}_{q}{}^{p}\,R{}_{c}{}^{q} \cr && \left.
 -10\,\w{}_{;}{}_{p}\,\w{}_{;}{}_{q}\,\w{}_{;}{}_{c}\,R{}^{p}{}^{q} +
  35\,\w{}_{;}{}_{p}\,\w{}_{;}{}_{q}{}_{c}\,R{}^{p}{}^{q} \right) / 3600
\end{eqnarray} \begin{eqnarray} G^{(4)}_{{\rm div,ren},abcd} &=&
\left(
-960\,\w{}_{;}{}_{p}\,\w{}_{;}{}_{a}\,\w{}_{;}{}_{b}\,\w{}_{;}{}_{c}\,
   \w{}_{;}{}_{d}\,\w{}^{;}{}^{p} +
  2520\,\w{}_{;}{}_{p}\,\w{}_{;}{}_{a}\,\w{}_{;}{}_{b}\,
   R{}_{c}{}^{p}{}_{;}{}_{d}
\right. \cr &&
 - 4320\,\w{}_{;}{}_{p}\,\w{}_{;}{}_{q}\,
   \w{}_{;}{}_{a}\,R{}_{b}{}^{q}{}_{c}{}^{p}{}_{;}{}_{d}
+960\,\w{}_{;}{}_{a}\,\w{}_{;}{}_{b}\,\w{}_{;}{}_{c}\,\w{}_{;}{}_{d}\,
   \w{}_{;}{}_{p}{}^{p}
\cr &&
 - 1440\,\w{}_{;}{}_{p}\,\w{}_{;}{}_{a}\,
   \w{}_{;}{}_{b}\,\w{}^{;}{}^{p}\,\w{}_{;}{}_{c}{}_{d} -
  6300\,\w{}_{;}{}_{p}\,R{}_{a}{}_{b}{}^{;}{}^{p}\,\w{}_{;}{}_{c}{}_{d}
\cr &&
+480\,\w{}_{;}{}_{a}\,\w{}_{;}{}_{b}\,\w{}_{;}{}_{p}{}^{p}\,
   \w{}_{;}{}_{c}{}_{d}
 + 2760\,\w{}_{;}{}_{p}\,\w{}^{;}{}^{p}\,
   \w{}_{;}{}_{a}{}_{b}\,\w{}_{;}{}_{c}{}_{d}
\cr &&
 -   2640\,\w{}_{;}{}_{p}{}^{p}\,\w{}_{;}{}_{a}{}_{b}\,\w{}_{;}{}_{c}{}_{d}
 -48544\,\w{}_{;}{}_{p}\,\w{}_{;}{}_{a}\,\w{}_{;}{}_{b}{}^{p}\,
   \w{}_{;}{}_{c}{}_{d}
\cr &&
 + 7384\,\w{}_{;}{}_{p}{}_{a}\,\w{}_{;}{}_{b}{}^{p}\,
   \w{}_{;}{}_{c}{}_{d}
 - 960\,\w{}_{;}{}_{p}\,\w{}_{;}{}_{a}\,
   \w{}_{;}{}_{b}\,\w{}_{;}{}_{c}\,\w{}_{;}{}_{d}{}^{p}
\cr &&
+5040\,\w{}_{;}{}_{a}\,R{}_{p}{}_{b}{}_{;}{}_{c}\,\w{}_{;}{}_{d}{}^{p}
 -   6300\,\w{}_{;}{}_{p}\,R{}_{a}{}_{b}{}_{;}{}_{c}\,\w{}_{;}{}_{d}{}^{p}
\cr &&
 -   12600\,\w{}_{;}{}_{a}\,R{}_{b}{}_{c}{}_{;}{}_{p}\,\w{}_{;}{}_{d}{}^{p}
+3840\,\w{}_{;}{}_{a}\,\w{}_{;}{}_{b}\,\w{}_{;}{}_{p}{}_{c}\,
   \w{}_{;}{}_{d}{}^{p}
\cr &&
 + 57664\,\w{}_{;}{}_{p}\,\w{}_{;}{}_{a}\,
   \w{}_{;}{}_{b}{}_{c}\,\w{}_{;}{}_{d}{}^{p}
 -   2104\,\w{}_{;}{}_{p}{}_{a}\,\w{}_{;}{}_{b}{}_{c}\,\w{}_{;}{}_{d}{}^{p}
\cr &&
 -47520\,\w{}_{;}{}_{p}\,R{}_{q}{}_{a}{}_{b}{}^{p}{}_{;}{}_{c}\,
   \w{}_{;}{}_{d}{}^{q}
 - 2520\,\w{}_{;}{}_{a}\,
   R{}_{p}{}_{b}{}_{q}{}_{c}{}_{;}{}_{d}\,\w{}^{;}{}^{p}{}^{q}
\cr &&
 -   1890\,\w{}_{;}{}_{p}\,\w{}^{;}{}^{p}\,R{}_{a}{}_{b}{}_{;}{}_{c}{}_{d}
+1890\,\w{}_{;}{}_{p}{}^{p}\,R{}_{a}{}_{b}{}_{;}{}_{c}{}_{d} \cr
&&
 + 588\,\w{}_{;}{}_{p}\,\w{}_{;}{}_{a}\,R{}_{b}{}_{c}{}_{;}{}_{d}{}^{p}
 +
  2352\,\w{}_{;}{}_{p}{}_{a}\,R{}_{b}{}_{c}{}_{;}{}_{d}{}^{p} \cr &&
 -5628\,\w{}_{;}{}_{p}\,\w{}_{;}{}_{a}\,R{}_{b}{}_{c}{}^{;}{}^{p}{}_{d} +
  7728\,\w{}_{;}{}_{p}{}_{a}\,R{}_{b}{}_{c}{}^{;}{}^{p}{}_{d} +
  2520\,\w{}_{;}{}_{p}\,\w{}_{;}{}_{a}\,R{}_{b}{}^{p}{}_{;}{}_{c}{}_{d} \cr
&&
 -8820\,\w{}_{;}{}_{p}{}_{a}\,R{}_{b}{}^{p}{}_{;}{}_{c}{}_{d} +
  7920\,\w{}_{;}{}_{p}\,\w{}_{;}{}_{q}\,
   R{}_{a}{}^{q}{}_{b}{}^{p}{}_{;}{}_{c}{}_{d} +
  2520\,\w{}_{;}{}_{p}{}_{q}\,R{}_{a}{}^{q}{}_{b}{}^{p}{}_{;}{}_{c}{}_{d}
\cr &&
 -720\,\w{}_{;}{}_{a}\,\w{}_{;}{}_{b}\,\w{}_{;}{}_{c}\,
   \w{}_{;}{}_{p}{}_{d}{}^{p} -
  6480\,\w{}_{;}{}_{a}\,\w{}_{;}{}_{b}{}_{c}\,\w{}_{;}{}_{p}{}_{d}{}^{p} +
  2160\,\w{}_{;}{}_{a}\,\w{}_{;}{}_{b}\,\w{}_{;}{}_{c}\,
   \w{}_{;}{}_{p}{}^{p}{}_{d} \cr &&
+3150\,R{}_{a}{}_{b}{}_{;}{}_{c}\,\w{}_{;}{}_{p}{}^{p}{}_{d} +
  1800\,\w{}_{;}{}_{a}\,\w{}_{;}{}_{b}{}_{c}\,\w{}_{;}{}_{p}{}^{p}{}_{d} +
  2880\,\w{}_{;}{}_{p}\,\w{}_{;}{}_{a}\,\w{}^{;}{}^{p}\,
   \w{}_{;}{}_{b}{}_{c}{}_{d} \cr &&
 -3240\,\w{}_{;}{}_{a}\,\w{}_{;}{}_{p}{}^{p}\,\w{}_{;}{}_{b}{}_{c}{}_{d} +
  2880\,\w{}_{;}{}_{p}\,\w{}_{;}{}_{a}{}^{p}\,\w{}_{;}{}_{b}{}_{c}{}_{d} +
  3600\,\w{}_{;}{}_{p}{}_{a}{}^{p}\,\w{}_{;}{}_{b}{}_{c}{}_{d} \cr &&
 -4500\,\w{}_{;}{}_{p}{}^{p}{}_{a}\,\w{}_{;}{}_{b}{}_{c}{}_{d} -
  6480\,\w{}_{;}{}_{p}\,\w{}_{;}{}_{a}\,\w{}_{;}{}_{b}\,
   \w{}_{;}{}_{c}{}_{d}{}^{p} -
  5040\,R{}_{p}{}_{a}{}_{;}{}_{b}\,\w{}_{;}{}_{c}{}_{d}{}^{p} \cr &&
+3150\,R{}_{a}{}_{b}{}_{;}{}_{p}\,\w{}_{;}{}_{c}{}_{d}{}^{p} -
  10080\,\w{}_{;}{}_{a}\,\w{}_{;}{}_{p}{}_{b}\,\w{}_{;}{}_{c}{}_{d}{}^{p} -
  360\,\w{}_{;}{}_{p}\,\w{}_{;}{}_{a}{}_{b}\,\w{}_{;}{}_{c}{}_{d}{}^{p} \cr
&&
 -9000\,\w{}_{;}{}_{p}{}_{a}{}_{b}\,\w{}_{;}{}_{c}{}_{d}{}^{p} +
  8730\,\w{}_{;}{}_{a}{}_{b}{}_{p}\,\w{}_{;}{}_{c}{}_{d}{}^{p} +
  9360\,\w{}_{;}{}_{p}\,\w{}_{;}{}_{a}\,\w{}_{;}{}_{b}\,
   \w{}_{;}{}_{c}{}^{p}{}_{d} \cr &&
 -1260\,R{}_{p}{}_{a}{}_{;}{}_{b}\,\w{}_{;}{}_{c}{}^{p}{}_{d} +
  6300\,R{}_{a}{}_{b}{}_{;}{}_{p}\,\w{}_{;}{}_{c}{}^{p}{}_{d} +
  19440\,\w{}_{;}{}_{a}\,\w{}_{;}{}_{p}{}_{b}\,\w{}_{;}{}_{c}{}^{p}{}_{d}
\cr &&
+5760\,\w{}_{;}{}_{p}\,\w{}_{;}{}_{a}{}_{b}\,\w{}_{;}{}_{c}{}^{p}{}_{d}
-
  6480\,\w{}_{;}{}_{p}{}_{a}{}_{b}\,\w{}_{;}{}_{c}{}^{p}{}_{d} +
  3780\,R{}_{p}{}_{a}{}_{q}{}_{b}{}_{;}{}_{c}\,\w{}_{;}{}_{d}{}^{p}{}^{q}
\cr &&
+3780\,R{}_{p}{}_{a}{}_{q}{}_{b}{}_{;}{}_{c}\,\w{}_{;}{}_{d}{}^{q}{}^{p}
-
  1260\,R{}_{p}{}_{a}{}_{q}{}_{b}{}_{;}{}_{c}\,\w{}^{;}{}^{p}{}^{q}{}_{d} +
  1470\,\w{}_{;}{}_{p}\,R{}_{a}{}_{b}{}_{;}{}_{c}{}_{d}{}^{p} \cr &&
+462\,\w{}_{;}{}_{p}\,R{}_{a}{}_{b}{}_{;}{}_{c}{}^{p}{}_{d} +
  798\,\w{}_{;}{}_{p}\,R{}_{a}{}_{b}{}^{;}{}^{p}{}_{c}{}_{d} -
  2730\,\w{}_{;}{}_{p}\,R{}_{a}{}^{p}{}_{;}{}_{b}{}_{c}{}_{d} \cr &&
 -1050\,\w{}_{;}{}_{p}\,R{}_{q}{}_{a}{}_{b}{}^{p}{}_{;}{}_{c}{}_{d}{}^{q} -
  2880\,\w{}_{;}{}_{a}\,\w{}_{;}{}_{b}\,\w{}_{;}{}_{p}{}_{c}{}_{d}{}^{p} +
  11520\,\w{}_{;}{}_{a}{}_{b}\,\w{}_{;}{}_{p}{}_{c}{}_{d}{}^{p} \cr &&
 -360\,\w{}_{;}{}_{a}\,\w{}_{;}{}_{b}\,\w{}_{;}{}_{p}{}_{c}{}^{p}{}_{d} -
  1080\,\w{}_{;}{}_{a}{}_{b}\,\w{}_{;}{}_{p}{}_{c}{}^{p}{}_{d} +
  1080\,\w{}_{;}{}_{a}\,\w{}_{;}{}_{b}\,\w{}_{;}{}_{p}{}^{p}{}_{c}{}_{d} \cr
&&
 -6840\,\w{}_{;}{}_{a}{}_{b}\,\w{}_{;}{}_{p}{}^{p}{}_{c}{}_{d} +
  720\,\w{}_{;}{}_{p}{}^{p}\,\w{}_{;}{}_{a}{}_{b}{}_{c}{}_{d} -
  360\,\w{}_{;}{}_{p}\,\w{}_{;}{}_{a}\,\w{}_{;}{}_{b}{}_{c}{}_{d}{}^{p} \cr
&&
+10800\,\w{}_{;}{}_{p}{}_{a}\,\w{}_{;}{}_{b}{}_{c}{}_{d}{}^{p} -
  3960\,\w{}_{;}{}_{p}\,\w{}_{;}{}_{a}\,\w{}_{;}{}_{b}{}_{c}{}^{p}{}_{d} +
  4320\,\w{}_{;}{}_{p}{}_{a}\,\w{}_{;}{}_{b}{}_{c}{}^{p}{}_{d} \cr &&
+7200\,\w{}_{;}{}_{p}\,\w{}_{;}{}_{a}\,\w{}_{;}{}_{b}{}^{p}{}_{c}{}_{d}
-
  23760\,\w{}_{;}{}_{p}{}_{a}\,\w{}_{;}{}_{b}{}^{p}{}_{c}{}_{d} +
  1080\,\w{}_{;}{}_{a}\,\w{}_{;}{}_{b}\,\w{}_{;}{}_{c}{}_{d}{}_{p}{}^{p} \cr
&&
 -6840\,\w{}_{;}{}_{a}{}_{b}\,\w{}_{;}{}_{c}{}_{d}{}_{p}{}^{p} +
  3600\,\w{}_{;}{}_{a}\,\w{}_{;}{}_{p}{}_{b}{}_{c}{}_{d}{}^{p} +
  3600\,\w{}_{;}{}_{a}\,\w{}_{;}{}_{p}{}_{b}{}_{c}{}^{p}{}_{d} \cr &&
 -1440\,\w{}_{;}{}_{a}\,\w{}_{;}{}_{p}{}_{b}{}^{p}{}_{c}{}_{d} -
  3240\,\w{}_{;}{}_{a}\,\w{}_{;}{}_{p}{}^{p}{}_{b}{}_{c}{}_{d} +
  6120\,\w{}_{;}{}_{p}\,\w{}_{;}{}_{a}{}_{b}{}_{c}{}_{d}{}^{p} \cr &&
+2880\,\w{}_{;}{}_{p}\,\w{}_{;}{}_{a}{}_{b}{}_{c}{}^{p}{}_{d} -
  360\,\w{}_{;}{}_{p}\,\w{}_{;}{}_{a}{}_{b}{}^{p}{}_{c}{}_{d} -
  10440\,\w{}_{;}{}_{p}\,\w{}_{;}{}_{a}{}^{p}{}_{b}{}_{c}{}_{d} \cr &&
+1800\,\w{}_{;}{}_{a}\,\w{}_{;}{}_{b}{}_{c}{}_{p}{}_{d}{}^{p} -
  3240\,\w{}_{;}{}_{a}\,\w{}_{;}{}_{b}{}_{c}{}_{p}{}^{p}{}_{d} -
  3240\,\w{}_{;}{}_{a}\,\w{}_{;}{}_{b}{}_{c}{}_{d}{}_{p}{}^{p} \cr &&
 -720\,\w{}_{;}{}_{p}{}_{a}{}_{b}{}_{c}{}_{d}{}^{p} -
  720\,\w{}_{;}{}_{p}{}_{a}{}_{b}{}_{c}{}^{p}{}_{d} -
  720\,\w{}_{;}{}_{p}{}_{a}{}_{b}{}^{p}{}_{c}{}_{d} \cr &&
+540\,\w{}_{;}{}_{p}{}_{a}{}^{p}{}_{b}{}_{c}{}_{d} +
  900\,\w{}_{;}{}_{p}{}^{p}{}_{a}{}_{b}{}_{c}{}_{d} -
  360\,\w{}_{;}{}_{a}{}_{b}{}_{p}{}_{c}{}_{d}{}^{p} \cr &&
 -360\,\w{}_{;}{}_{a}{}_{b}{}_{p}{}_{c}{}^{p}{}_{d} +
  900\,\w{}_{;}{}_{a}{}_{b}{}_{p}{}^{p}{}_{c}{}_{d} -
  360\,\w{}_{;}{}_{a}{}_{b}{}_{c}{}_{p}{}_{d}{}^{p} \cr &&
+900\,\w{}_{;}{}_{a}{}_{b}{}_{c}{}_{p}{}^{p}{}_{d} +
  900\,\w{}_{;}{}_{a}{}_{b}{}_{c}{}_{d}{}_{p}{}^{p} +
  840\,\w{}_{;}{}_{p}\,\w{}_{;}{}_{a}\,\w{}_{;}{}_{b}\,\w{}^{;}{}^{p}\,
   R{}_{c}{}_{d} \cr &&
+3150\,\w{}_{;}{}_{p}\,R{}_{a}{}_{b}{}^{;}{}^{p}\,R{}_{c}{}_{d} -
  3150\,\w{}_{;}{}_{p}\,R{}_{a}{}^{p}{}_{;}{}_{b}\,R{}_{c}{}_{d} -
  840\,\w{}_{;}{}_{a}\,\w{}_{;}{}_{b}\,\w{}_{;}{}_{p}{}^{p}\,R{}_{c}{}_{d}
\cr
&&
+420\,\w{}_{;}{}_{p}{}^{p}\,\w{}_{;}{}_{a}{}_{b}\,R{}_{c}{}_{d} +
  1680\,\w{}_{;}{}_{p}\,\w{}_{;}{}_{a}\,\w{}_{;}{}_{b}{}^{p}\,R{}_{c}{}_{d}
-
  4200\,\w{}_{;}{}_{p}{}_{a}\,\w{}_{;}{}_{b}{}^{p}\,R{}_{c}{}_{d} \cr &&
+1260\,\w{}_{;}{}_{a}\,\w{}_{;}{}_{p}{}_{b}{}^{p}\,R{}_{c}{}_{d} -
  2520\,\w{}_{;}{}_{a}\,\w{}_{;}{}_{p}{}^{p}{}_{b}\,R{}_{c}{}_{d} +
  7560\,\w{}_{;}{}_{p}\,\w{}_{;}{}_{a}{}_{b}{}^{p}\,R{}_{c}{}_{d} \cr &&
 -11340\,\w{}_{;}{}_{p}\,\w{}_{;}{}_{a}{}^{p}{}_{b}\,R{}_{c}{}_{d} -
  2520\,\w{}_{;}{}_{p}{}_{a}{}_{b}{}^{p}\,R{}_{c}{}_{d} +
  630\,\w{}_{;}{}_{p}{}_{a}{}^{p}{}_{b}\,R{}_{c}{}_{d} \cr &&
+1890\,\w{}_{;}{}_{p}{}^{p}{}_{a}{}_{b}\,R{}_{c}{}_{d} +
  1890\,\w{}_{;}{}_{a}{}_{b}{}_{p}{}^{p}\,R{}_{c}{}_{d} -
  525\,\w{}_{;}{}_{p}\,\w{}^{;}{}^{p}\,R{}_{a}{}_{b}\,R{}_{c}{}_{d} \cr &&
+525\,\w{}_{;}{}_{p}{}^{p}\,R{}_{a}{}_{b}\,R{}_{c}{}_{d} -
  980\,\w{}_{;}{}_{p}\,\w{}_{;}{}_{a}\,R{}_{b}{}^{p}\,R{}_{c}{}_{d} -
  770\,\w{}_{;}{}_{p}{}_{a}\,R{}_{b}{}^{p}\,R{}_{c}{}_{d} \cr &&
+1680\,\w{}_{;}{}_{p}\,\w{}_{;}{}_{a}\,\w{}_{;}{}_{b}\,\w{}_{;}{}_{c}\,
   R{}_{d}{}^{p} + 4200\,\w{}_{;}{}_{a}\,\w{}_{;}{}_{b}\,
   \w{}_{;}{}_{p}{}_{c}\,R{}_{d}{}^{p}
\cr &&
 +
  6720\,\w{}_{;}{}_{p}\,\w{}_{;}{}_{a}\,\w{}_{;}{}_{b}{}_{c}\,R{}_{d}{}^{p}
\cr &&
 -10920\,\w{}_{;}{}_{p}{}_{a}\,\w{}_{;}{}_{b}{}_{c}\,R{}_{d}{}^{p} -
  10080\,\w{}_{;}{}_{a}\,\w{}_{;}{}_{p}{}_{b}{}_{c}\,R{}_{d}{}^{p} -
  5040\,\w{}_{;}{}_{p}\,\w{}_{;}{}_{a}{}_{b}{}_{c}\,R{}_{d}{}^{p} \cr &&
+2520\,\w{}_{;}{}_{a}\,\w{}_{;}{}_{b}{}_{c}{}_{p}\,R{}_{d}{}^{p} +
  3780\,\w{}_{;}{}_{p}{}_{a}{}_{b}{}_{c}\,R{}_{d}{}^{p} -
  1260\,\w{}_{;}{}_{a}{}_{b}{}_{c}{}_{p}\,R{}_{d}{}^{p} \cr &&
 -1120\,\w{}_{;}{}_{p}\,\w{}_{;}{}_{a}\,R{}_{b}{}_{c}\,R{}_{d}{}^{p} +
  1820\,\w{}_{;}{}_{p}{}_{a}\,R{}_{b}{}_{c}\,R{}_{d}{}^{p} -
  1764\,\w{}_{;}{}_{p}\,R{}_{q}{}_{a}{}_{b}{}^{p}{}_{;}{}_{c}\,R{}_{d}{}^{q}
\cr &&
 -480\,\w{}_{;}{}_{p}\,\w{}_{;}{}_{q}\,R{}_{r}{}_{b}{}_{a}{}^{q}\,
   R{}_{c}{}^{p}{}_{d}{}^{r}
 - 6480\,\w{}_{;}{}_{p}\,\w{}_{;}{}_{q}\,
   \w{}_{;}{}_{a}\,\w{}_{;}{}_{b}\,R{}_{c}{}^{q}{}_{d}{}^{p}
\cr &&
 +
714\,\w{}_{;}{}_{p}\,R{}_{q}{}_{a}{}_{;}{}_{b}\,R{}_{c}{}^{q}{}_{d}{}^{p}
+5250\,\w{}_{;}{}_{p}\,R{}_{a}{}_{b}{}_{;}{}_{q}\,R{}_{c}{}^{q}{}_{d}{}^{p}
\cr &&
 -   1680\,\w{}_{;}{}_{a}\,\w{}_{;}{}_{b}\,\w{}_{;}{}_{p}{}_{q}\,
   R{}_{c}{}^{q}{}_{d}{}^{p}
 - 15360\,\w{}_{;}{}_{p}\,\w{}_{;}{}_{a}\,
   \w{}_{;}{}_{q}{}_{b}\,R{}_{c}{}^{q}{}_{d}{}^{p}
\cr && +34800\,\w{}_{;}{}_{p}{}_{a}\,\w{}_{;}{}_{q}{}_{b}\,
   R{}_{c}{}^{q}{}_{d}{}^{p}
 - 240\,\w{}_{;}{}_{p}\,\w{}_{;}{}_{q}\,
   \w{}_{;}{}_{a}{}_{b}\,R{}_{c}{}^{q}{}_{d}{}^{p}
\cr &&
 -
1680\,\w{}_{;}{}_{p}{}_{q}\,\w{}_{;}{}_{a}{}_{b}\,R{}_{c}{}^{q}{}_{d}{}^{p}
+5040\,\w{}_{;}{}_{a}\,\w{}_{;}{}_{p}{}_{q}{}_{b}\,R{}_{c}{}^{q}{}_{d}{}^{p}
\cr &&
 -   10080\,\w{}_{;}{}_{a}\,\w{}_{;}{}_{p}{}_{b}{}_{q}\,
   R{}_{c}{}^{q}{}_{d}{}^{p}
 + 20160\,\w{}_{;}{}_{p}\,
   \w{}_{;}{}_{q}{}_{a}{}_{b}\,R{}_{c}{}^{q}{}_{d}{}^{p}
\cr && +32040\,\w{}_{;}{}_{p}\,\w{}_{;}{}_{a}{}_{b}{}_{q}\,
   R{}_{c}{}^{q}{}_{d}{}^{p}
 - 2520\,\w{}_{;}{}_{p}{}_{q}{}_{a}{}_{b}\,
   R{}_{c}{}^{q}{}_{d}{}^{p}
\cr &&
 + 5040\,\w{}_{;}{}_{p}{}_{a}{}_{b}{}_{q}\,
   R{}_{c}{}^{q}{}_{d}{}^{p}
+2520\,\w{}_{;}{}_{a}{}_{b}{}_{p}{}_{q}\,R{}_{c}{}^{q}{}_{d}{}^{p}
\cr &&
 +   2016\,\w{}_{;}{}_{p}\,\w{}_{;}{}_{a}\,R{}_{q}{}_{b}\,
   R{}_{c}{}^{q}{}_{d}{}^{p}
 + 3024\,\w{}_{;}{}_{p}{}_{a}\,R{}_{q}{}_{b}\,
   R{}_{c}{}^{q}{}_{d}{}^{p}
\cr && +7140\,\w{}_{;}{}_{p}\,\w{}_{;}{}_{q}\,R{}_{a}{}_{b}\,
   R{}_{c}{}^{q}{}_{d}{}^{p}
+ 2100\,\w{}_{;}{}_{p}{}_{q}\,R{}_{a}{}_{b}\,
   R{}_{c}{}^{q}{}_{d}{}^{p}
\cr && - 3150\,\w{}_{;}{}_{p}\,
   R{}_{q}{}_{a}{}_{b}{}^{p}{}_{;}{}_{r}\,R{}_{c}{}^{q}{}_{d}{}^{r}
 -16320\,\w{}_{;}{}_{p}\,\w{}_{;}{}_{q}\,R{}_{r}{}_{b}{}_{a}{}^{q}\,
   R{}_{c}{}^{r}{}_{d}{}^{p}
\cr && - 6720\,\w{}_{;}{}_{p}{}_{q}\,
   R{}_{r}{}_{b}{}_{a}{}^{q}\,R{}_{c}{}^{r}{}_{d}{}^{p}
-   2016\,\w{}_{;}{}_{p}\,R{}_{q}{}_{r}{}_{a}{}^{p}{}_{;}{}_{b}\,
   R{}_{c}{}^{r}{}_{d}{}^{q}
\cr &&
+420\,\w{}_{;}{}_{p}\,R{}_{q}{}_{a}{}_{r}{}_{b}{}^{;}{}^{p}\,
   R{}_{c}{}^{r}{}_{d}{}^{q}
- 3990\,\w{}_{;}{}_{p}\,
   R{}_{q}{}^{p}{}_{r}{}_{a}{}_{;}{}_{b}\,R{}_{c}{}^{r}{}_{d}{}^{q}
\cr && -
3584\,\w{}_{;}{}_{p}\,\w{}_{;}{}_{a}\,R{}_{q}{}_{r}{}_{b}{}^{p}\,
   R{}_{c}{}^{r}{}_{d}{}^{q}
+5824\,\w{}_{;}{}_{p}{}_{a}\,R{}_{q}{}_{r}{}_{b}{}^{p}\,
   R{}_{c}{}^{r}{}_{d}{}^{q}
\cr && - 420\,\w{}_{;}{}_{p}\,\w{}^{;}{}^{p}\,
   R{}_{q}{}_{a}{}_{r}{}_{b}\,R{}_{c}{}^{r}{}_{d}{}^{q}
+   420\,\w{}_{;}{}_{p}{}^{p}\,R{}_{q}{}_{a}{}_{r}{}_{b}\,
   R{}_{c}{}^{r}{}_{d}{}^{q}
\cr &&
+2688\,\w{}_{;}{}_{p}\,\w{}_{;}{}_{a}\,R{}_{q}{}_{b}{}_{r}{}^{p}\,
   R{}_{c}{}^{r}{}_{d}{}^{q}
- 4368\,\w{}_{;}{}_{p}{}_{a}\,
   R{}_{q}{}_{b}{}_{r}{}^{p}\,R{}_{c}{}^{r}{}_{d}{}^{q}
\cr && +
3584\,\w{}_{;}{}_{p}\,\w{}_{;}{}_{a}\,R{}_{q}{}^{p}{}_{r}{}_{b}\,
   R{}_{c}{}^{r}{}_{d}{}^{q}
 -5824\,\w{}_{;}{}_{p}{}_{a}\,R{}_{q}{}^{p}{}_{r}{}_{b}\,
   R{}_{c}{}^{r}{}_{d}{}^{q}
\cr && - 3276\,\w{}_{;}{}_{p}\,
   R{}_{q}{}_{a}{}_{r}{}_{b}{}_{;}{}_{c}\,R{}_{d}{}^{p}{}^{q}{}^{r}
-   3318\,\w{}_{;}{}_{p}\,R{}_{q}{}_{a}{}_{r}{}_{b}{}_{;}{}_{c}\,
   R{}_{d}{}^{q}{}^{p}{}^{r}
\cr &&
  -2912\,\w{}_{;}{}_{p}\,\w{}_{;}{}_{a}\,R{}_{q}{}_{b}{}_{r}{}_{c}\,
   R{}_{d}{}^{q}{}^{p}{}^{r}
+ 3472\,\w{}_{;}{}_{p}{}_{a}\,
   R{}_{q}{}_{b}{}_{r}{}_{c}\,R{}_{d}{}^{q}{}^{p}{}^{r} \cr &&
\left. -
  672\,\w{}_{;}{}_{p}\,R{}_{q}{}_{a}{}_{r}{}_{b}{}_{;}{}_{c}\,
   R{}_{d}{}^{r}{}^{p}{}^{q} \right) / 907200 \cr\cr
%
&&+g_{ab}\left(
240\,\w{}_{;}{}_{p}\,\w{}_{;}{}_{q}\,\w{}_{;}{}_{c}\,\w{}_{;}{}_{d}\,
   \w{}^{;}{}^{p}\,\w{}^{;}{}^{q} +
  1050\,\w{}_{;}{}_{p}\,\w{}_{;}{}_{q}\,\w{}^{;}{}^{q}\,
   R{}_{c}{}_{d}{}^{;}{}^{p} \right. \cr &&
- 1050\,\w{}_{;}{}_{p}\,\w{}_{;}{}_{q}\,
   \w{}^{;}{}^{p}\,R{}_{c}{}_{d}{}^{;}{}^{q}
 -1050\,\w{}_{;}{}_{p}\,\w{}_{;}{}_{q}\,\w{}^{;}{}^{q}\,
   R{}_{c}{}^{p}{}_{;}{}_{d}
\cr && + 1470\,\w{}_{;}{}_{p}\,\w{}_{;}{}_{q}\,
   \w{}^{;}{}^{p}\,R{}_{c}{}^{q}{}_{;}{}_{d}
-   1440\,\w{}_{;}{}_{p}\,\w{}_{;}{}_{q}\,\w{}_{;}{}_{c}\,
   R{}_{d}{}^{p}{}^{;}{}^{q}
\cr && +480\,\w{}_{;}{}_{p}\,\w{}_{;}{}_{q}\,\w{}_{;}{}_{c}\,
   R{}^{p}{}^{q}{}_{;}{}_{d}
+ 225\,\w{}_{;}{}_{p}\,\w{}_{;}{}_{q}\,
   \w{}_{;}{}_{r}\,R{}_{c}{}^{p}{}_{d}{}^{q}{}^{;}{}^{r}
\cr && +   1995\,\w{}_{;}{}_{p}\,\w{}_{;}{}_{q}\,\w{}_{;}{}_{r}\,
   R{}_{c}{}^{r}{}_{d}{}^{p}{}^{;}{}^{q}
+80\,\w{}_{;}{}_{p}\,\w{}_{;}{}_{c}\,\w{}_{;}{}_{d}\,\w{}^{;}{}^{p}\,
   \w{}_{;}{}_{q}{}^{q}
\cr && + 1050\,\w{}_{;}{}_{p}\,R{}_{c}{}_{d}{}^{;}{}^{p}\,
   \w{}_{;}{}_{q}{}^{q}
- 1050\,\w{}_{;}{}_{p}\,R{}_{c}{}^{p}{}_{;}{}_{d}\,
   \w{}_{;}{}_{q}{}^{q}
\cr &&
 -280\,\w{}_{;}{}_{c}\,\w{}_{;}{}_{d}\,\w{}_{;}{}_{p}{}^{p}\,
   \w{}_{;}{}_{q}{}^{q}
- 240\,\w{}_{;}{}_{p}\,\w{}_{;}{}_{q}\,
   \w{}^{;}{}^{p}\,\w{}^{;}{}^{q}\,\w{}_{;}{}_{c}{}_{d}
\cr && +
280\,\w{}_{;}{}_{p}\,\w{}^{;}{}^{p}\,\w{}_{;}{}_{q}{}^{q}\,
   \w{}_{;}{}_{c}{}_{d}
+140\,\w{}_{;}{}_{p}{}^{p}\,\w{}_{;}{}_{q}{}^{q}\,\w{}_{;}{}_{c}{}_{d}
\cr && +
1680\,\w{}_{;}{}_{p}\,\w{}_{;}{}_{q}\,\w{}_{;}{}_{c}\,\w{}^{;}{}^{q}\,
   \w{}_{;}{}_{d}{}^{p}
- 640\,\w{}_{;}{}_{p}\,\w{}_{;}{}_{c}\,
   \w{}_{;}{}_{q}{}^{q}\,\w{}_{;}{}_{d}{}^{p} \cr &&
 -1800\,\w{}_{;}{}_{p}{}_{c}\,\w{}_{;}{}_{q}{}^{q}\,\w{}_{;}{}_{d}{}^{p} -
  760\,\w{}_{;}{}_{p}\,\w{}_{;}{}_{q}\,\w{}_{;}{}_{c}{}^{q}\,
   \w{}_{;}{}_{d}{}^{p} - 3360\,\w{}_{;}{}_{p}{}_{q}\,\w{}_{;}{}_{c}{}^{q}\,
   \w{}_{;}{}_{d}{}^{p} \cr &&
 -720\,\w{}_{;}{}_{p}\,\w{}_{;}{}_{q}\,\w{}_{;}{}_{c}\,\w{}^{;}{}^{p}\,
   \w{}_{;}{}_{d}{}^{q} + 1620\,\w{}_{;}{}_{p}\,R{}_{q}{}_{c}{}^{;}{}^{p}\,
   \w{}_{;}{}_{d}{}^{q} - 240\,\w{}_{;}{}_{p}\,R{}_{q}{}^{p}{}_{;}{}_{c}\,
   \w{}_{;}{}_{d}{}^{q} \cr &&
+360\,\w{}_{;}{}_{p}\,R{}_{c}{}^{p}{}_{;}{}_{q}\,\w{}_{;}{}_{d}{}^{q}
-
  320\,\w{}_{;}{}_{p}\,\w{}^{;}{}^{p}\,\w{}_{;}{}_{q}{}_{c}\,
   \w{}_{;}{}_{d}{}^{q} + 1120\,\w{}_{;}{}_{p}{}^{p}\,\w{}_{;}{}_{q}{}_{c}\,
   \w{}_{;}{}_{d}{}^{q} \cr &&
 -2936\,\w{}_{;}{}_{p}\,\w{}_{;}{}_{c}\,\w{}_{;}{}_{q}{}^{p}\,
   \w{}_{;}{}_{d}{}^{q} + 1280\,\w{}_{;}{}_{p}{}_{c}\,\w{}_{;}{}_{q}{}^{p}\,
   \w{}_{;}{}_{d}{}^{q}
\cr &&
 + 640\,\w{}_{;}{}_{p}\,\w{}_{;}{}_{q}\,
   \w{}_{;}{}_{c}\,\w{}_{;}{}_{d}\,\w{}^{;}{}^{p}{}^{q} \cr &&
 -1260\,\w{}_{;}{}_{p}\,R{}_{q}{}_{c}{}_{;}{}_{d}\,\w{}^{;}{}^{p}{}^{q} +
  1050\,\w{}_{;}{}_{p}\,R{}_{c}{}_{d}{}_{;}{}_{q}\,\w{}^{;}{}^{p}{}^{q} -
  200\,\w{}_{;}{}_{c}\,\w{}_{;}{}_{d}\,\w{}_{;}{}_{p}{}_{q}\,
   \w{}^{;}{}^{p}{}^{q} \cr &&
+3576\,\w{}_{;}{}_{p}\,\w{}_{;}{}_{c}\,\w{}_{;}{}_{q}{}_{d}\,
   \w{}^{;}{}^{p}{}^{q} + 80\,\w{}_{;}{}_{p}\,\w{}_{;}{}_{q}\,
   \w{}_{;}{}_{c}{}_{d}\,\w{}^{;}{}^{p}{}^{q} +
  160\,\w{}_{;}{}_{p}{}_{q}\,\w{}_{;}{}_{c}{}_{d}\,\w{}^{;}{}^{p}{}^{q} \cr
&&
+420\,\w{}_{;}{}_{p}\,R{}_{q}{}_{c}{}_{r}{}_{d}{}^{;}{}^{p}\,
   \w{}^{;}{}^{q}{}^{r} - 900\,\w{}_{;}{}_{p}\,
   R{}_{q}{}_{c}{}_{d}{}^{p}{}_{;}{}_{r}\,\w{}^{;}{}^{q}{}^{r} -
  60\,\w{}_{;}{}_{p}\,R{}_{q}{}^{p}{}_{r}{}_{c}{}_{;}{}_{d}\,
   \w{}^{;}{}^{q}{}^{r} \cr &&
+630\,\w{}_{;}{}_{p}\,\w{}_{;}{}_{q}\,R{}_{c}{}_{d}{}^{;}{}^{p}{}^{q}
+
  360\,\w{}_{;}{}_{p}\,\w{}_{;}{}_{q}\,R{}_{c}{}_{d}{}^{;}{}^{q}{}^{p} -
  924\,\w{}_{;}{}_{p}\,\w{}_{;}{}_{q}\,R{}_{c}{}^{p}{}_{;}{}_{d}{}^{q} \cr
&&
+384\,\w{}_{;}{}_{p}\,\w{}_{;}{}_{q}\,R{}_{c}{}^{p}{}^{;}{}^{q}{}_{d}
-
  360\,\w{}_{;}{}_{p}\,\w{}_{;}{}_{q}\,R{}_{c}{}^{q}{}_{;}{}_{d}{}^{p} +
  270\,\w{}_{;}{}_{p}\,\w{}_{;}{}_{q}\,R{}^{p}{}^{q}{}_{;}{}_{c}{}_{d} \cr
&&
+360\,\w{}_{;}{}_{p}\,\w{}_{;}{}_{c}\,\w{}^{;}{}^{p}\,
   \w{}_{;}{}_{q}{}_{d}{}^{q} +
  420\,\w{}_{;}{}_{c}\,\w{}_{;}{}_{p}{}^{p}\,\w{}_{;}{}_{q}{}_{d}{}^{q} -
  600\,\w{}_{;}{}_{p}\,\w{}_{;}{}_{c}{}^{p}\,\w{}_{;}{}_{q}{}_{d}{}^{q} \cr
&&
 -240\,\w{}_{;}{}_{p}\,\w{}_{;}{}_{c}\,\w{}_{;}{}_{d}\,
   \w{}_{;}{}_{q}{}^{p}{}^{q} +
  480\,\w{}_{;}{}_{c}\,\w{}_{;}{}_{p}{}_{d}\,\w{}_{;}{}_{q}{}^{p}{}^{q} +
  480\,\w{}_{;}{}_{p}\,\w{}_{;}{}_{c}{}_{d}\,\w{}_{;}{}_{q}{}^{p}{}^{q} \cr
&&
 -240\,\w{}_{;}{}_{p}\,\w{}_{;}{}_{c}\,\w{}^{;}{}^{p}\,
   \w{}_{;}{}_{q}{}^{q}{}_{d} -
  840\,\w{}_{;}{}_{c}\,\w{}_{;}{}_{p}{}^{p}\,\w{}_{;}{}_{q}{}^{q}{}_{d} -
  720\,\w{}_{;}{}_{p}\,\w{}_{;}{}_{c}{}^{p}\,\w{}_{;}{}_{q}{}^{q}{}_{d} \cr
&&
+525\,\w{}_{;}{}_{p}{}^{p}{}_{c}\,\w{}_{;}{}_{q}{}^{q}{}_{d} -
  360\,\w{}_{;}{}_{p}\,\w{}_{;}{}_{c}\,\w{}_{;}{}_{d}\,
   \w{}_{;}{}_{q}{}^{q}{}^{p} -
  1080\,\w{}_{;}{}_{c}\,\w{}_{;}{}_{p}{}_{d}\,\w{}_{;}{}_{q}{}^{q}{}^{p} \cr
&&
 -60\,\w{}_{;}{}_{p}\,\w{}_{;}{}_{c}{}_{d}\,\w{}_{;}{}_{q}{}^{q}{}^{p} +
  420\,\w{}_{;}{}_{p}{}_{c}{}_{d}\,\w{}_{;}{}_{q}{}^{q}{}^{p} +
  1260\,\w{}_{;}{}_{p}\,\w{}_{;}{}_{q}\,\w{}^{;}{}^{q}\,
   \w{}_{;}{}_{c}{}_{d}{}^{p} \cr &&
+3720\,\w{}_{;}{}_{p}\,\w{}_{;}{}_{q}{}^{q}\,\w{}_{;}{}_{c}{}_{d}{}^{p}
-
  2340\,\w{}_{;}{}_{p}\,\w{}_{;}{}_{q}\,\w{}^{;}{}^{p}\,
   \w{}_{;}{}_{c}{}_{d}{}^{q} +
  3360\,\w{}_{;}{}_{p}\,\w{}_{;}{}_{q}{}^{p}\,\w{}_{;}{}_{c}{}_{d}{}^{q} \cr
&&
+750\,\w{}_{;}{}_{p}{}^{p}{}_{q}\,\w{}_{;}{}_{c}{}_{d}{}^{q} -
  3780\,\w{}_{;}{}_{p}\,\w{}_{;}{}_{q}\,\w{}^{;}{}^{q}\,
   \w{}_{;}{}_{c}{}^{p}{}_{d} -
  4140\,\w{}_{;}{}_{p}\,\w{}_{;}{}_{q}{}^{q}\,\w{}_{;}{}_{c}{}^{p}{}_{d} \cr
&&
+4380\,\w{}_{;}{}_{p}\,\w{}_{;}{}_{q}\,\w{}^{;}{}^{p}\,
   \w{}_{;}{}_{c}{}^{q}{}_{d} -
  5160\,\w{}_{;}{}_{p}\,\w{}_{;}{}_{q}{}^{p}\,\w{}_{;}{}_{c}{}^{q}{}_{d} -
  420\,\w{}_{;}{}_{p}{}^{p}{}_{q}\,\w{}_{;}{}_{c}{}^{q}{}_{d} \cr &&
+480\,\w{}_{;}{}_{p}\,\w{}_{;}{}_{q}\,\w{}_{;}{}_{c}\,
   \w{}_{;}{}_{d}{}^{p}{}^{q} -
  4320\,\w{}_{;}{}_{p}\,\w{}_{;}{}_{q}{}_{c}\,\w{}_{;}{}_{d}{}^{p}{}^{q} +
  1500\,\w{}_{;}{}_{p}{}_{c}{}_{q}\,\w{}_{;}{}_{d}{}^{p}{}^{q} \cr &&
 -2040\,\w{}_{;}{}_{p}\,\w{}_{;}{}_{q}{}_{c}\,\w{}_{;}{}_{d}{}^{q}{}^{p} -
  1260\,\w{}_{;}{}_{p}{}_{c}{}_{q}\,\w{}_{;}{}_{d}{}^{q}{}^{p} -
  240\,\w{}_{;}{}_{p}\,\w{}_{;}{}_{q}\,\w{}_{;}{}_{c}\,
   \w{}^{;}{}^{p}{}^{q}{}_{d} \cr &&
 -360\,\w{}_{;}{}_{c}\,\w{}_{;}{}_{p}{}_{q}\,\w{}^{;}{}^{p}{}^{q}{}_{d} +
  2280\,\w{}_{;}{}_{p}\,\w{}_{;}{}_{q}{}_{c}\,\w{}^{;}{}^{p}{}^{q}{}_{d} +
  210\,\w{}_{;}{}_{p}{}_{q}{}_{c}\,\w{}^{;}{}^{p}{}^{q}{}_{d} \cr &&
 -480\,\w{}_{;}{}_{p}\,\w{}^{;}{}^{p}\,\w{}_{;}{}_{q}{}_{c}{}_{d}{}^{q} -
  840\,\w{}_{;}{}_{p}{}^{p}\,\w{}_{;}{}_{q}{}_{c}{}_{d}{}^{q} -
  60\,\w{}_{;}{}_{p}\,\w{}^{;}{}^{p}\,\w{}_{;}{}_{q}{}_{c}{}^{q}{}_{d} \cr
&&
+210\,\w{}_{;}{}_{p}{}^{p}\,\w{}_{;}{}_{q}{}_{c}{}^{q}{}_{d} +
  1200\,\w{}_{;}{}_{p}\,\w{}_{;}{}_{c}\,\w{}_{;}{}_{q}{}_{d}{}^{p}{}^{q} -
  960\,\w{}_{;}{}_{p}{}_{c}\,\w{}_{;}{}_{q}{}_{d}{}^{p}{}^{q} \cr &&
+240\,\w{}_{;}{}_{p}{}_{c}\,\w{}_{;}{}_{q}{}_{d}{}^{q}{}^{p} -
  480\,\w{}_{;}{}_{p}\,\w{}_{;}{}_{c}\,\w{}_{;}{}_{q}{}^{p}{}_{d}{}^{q} +
  240\,\w{}_{;}{}_{p}{}_{c}\,\w{}_{;}{}_{q}{}^{p}{}_{d}{}^{q} \cr &&
+360\,\w{}_{;}{}_{p}\,\w{}_{;}{}_{c}\,\w{}_{;}{}_{q}{}^{p}{}^{q}{}_{d}
-
  180\,\w{}_{;}{}_{p}{}_{c}\,\w{}_{;}{}_{q}{}^{p}{}^{q}{}_{d} +
  180\,\w{}_{;}{}_{p}\,\w{}^{;}{}^{p}\,\w{}_{;}{}_{q}{}^{q}{}_{c}{}_{d} \cr
&&
+630\,\w{}_{;}{}_{p}{}^{p}\,\w{}_{;}{}_{q}{}^{q}{}_{c}{}_{d} -
  840\,\w{}_{;}{}_{p}\,\w{}_{;}{}_{c}\,\w{}_{;}{}_{q}{}^{q}{}_{d}{}^{p} +
  960\,\w{}_{;}{}_{p}{}_{c}\,\w{}_{;}{}_{q}{}^{q}{}_{d}{}^{p} \cr &&
 -240\,\w{}_{;}{}_{p}\,\w{}_{;}{}_{c}\,\w{}_{;}{}_{q}{}^{q}{}^{p}{}_{d} +
  360\,\w{}_{;}{}_{p}{}_{c}\,\w{}_{;}{}_{q}{}^{q}{}^{p}{}_{d} +
  180\,\w{}_{;}{}_{p}\,\w{}^{;}{}^{p}\,\w{}_{;}{}_{c}{}_{d}{}_{q}{}^{q} \cr
&&
+630\,\w{}_{;}{}_{p}{}^{p}\,\w{}_{;}{}_{c}{}_{d}{}_{q}{}^{q} +
  2220\,\w{}_{;}{}_{p}\,\w{}_{;}{}_{q}\,\w{}_{;}{}_{c}{}_{d}{}^{p}{}^{q} +
  480\,\w{}_{;}{}_{p}{}_{q}\,\w{}_{;}{}_{c}{}_{d}{}^{p}{}^{q} \cr &&
 -4440\,\w{}_{;}{}_{p}\,\w{}_{;}{}_{q}\,\w{}_{;}{}_{c}{}^{p}{}_{d}{}^{q} -
  360\,\w{}_{;}{}_{p}{}_{q}\,\w{}_{;}{}_{c}{}^{p}{}_{d}{}^{q} -
  480\,\w{}_{;}{}_{p}\,\w{}_{;}{}_{q}\,\w{}_{;}{}_{c}{}^{p}{}^{q}{}_{d} \cr
&&
+120\,\w{}_{;}{}_{p}{}_{q}\,\w{}_{;}{}_{c}{}^{p}{}^{q}{}_{d} -
  480\,\w{}_{;}{}_{p}\,\w{}_{;}{}_{c}\,\w{}_{;}{}_{d}{}^{p}{}_{q}{}^{q} +
  660\,\w{}_{;}{}_{p}{}_{c}\,\w{}_{;}{}_{d}{}^{p}{}_{q}{}^{q} \cr &&
+1740\,\w{}_{;}{}_{p}\,\w{}_{;}{}_{q}\,\w{}^{;}{}^{p}{}^{q}{}_{c}{}_{d}
+
  240\,\w{}_{;}{}_{p}{}_{q}\,\w{}^{;}{}^{p}{}^{q}{}_{c}{}_{d} -
  360\,\w{}_{;}{}_{p}\,\w{}_{;}{}_{q}{}_{c}{}_{d}{}^{p}{}^{q} \cr &&
 -600\,\w{}_{;}{}_{p}\,\w{}_{;}{}_{q}{}_{c}{}_{d}{}^{q}{}^{p} -
  120\,\w{}_{;}{}_{p}\,\w{}_{;}{}_{q}{}_{c}{}^{p}{}_{d}{}^{q} -
  120\,\w{}_{;}{}_{p}\,\w{}_{;}{}_{q}{}_{c}{}^{p}{}^{q}{}_{d} \cr &&
+240\,\w{}_{;}{}_{p}\,\w{}_{;}{}_{q}{}_{c}{}^{q}{}_{d}{}^{p} +
  60\,\w{}_{;}{}_{p}\,\w{}_{;}{}_{q}{}_{c}{}^{q}{}^{p}{}_{d} +
  240\,\w{}_{;}{}_{p}\,\w{}_{;}{}_{q}{}^{p}{}_{c}{}_{d}{}^{q} \cr &&
+240\,\w{}_{;}{}_{p}\,\w{}_{;}{}_{q}{}^{p}{}_{c}{}^{q}{}_{d} -
  180\,\w{}_{;}{}_{p}\,\w{}_{;}{}_{q}{}^{p}{}^{q}{}_{c}{}_{d} +
  540\,\w{}_{;}{}_{p}\,\w{}_{;}{}_{q}{}^{q}{}_{c}{}_{d}{}^{p} \cr &&
+240\,\w{}_{;}{}_{p}\,\w{}_{;}{}_{q}{}^{q}{}_{c}{}^{p}{}_{d} -
  60\,\w{}_{;}{}_{p}\,\w{}_{;}{}_{q}{}^{q}{}^{p}{}_{c}{}_{d} -
  180\,\w{}_{;}{}_{p}\,\w{}_{;}{}_{c}{}_{d}{}_{q}{}^{p}{}^{q} \cr &&
+540\,\w{}_{;}{}_{p}\,\w{}_{;}{}_{c}{}_{d}{}_{q}{}^{q}{}^{p} +
  360\,\w{}_{;}{}_{p}\,\w{}_{;}{}_{c}{}_{d}{}^{p}{}_{q}{}^{q} +
  240\,\w{}_{;}{}_{p}\,\w{}_{;}{}_{c}{}^{p}{}_{q}{}_{d}{}^{q} \cr &&
 -180\,\w{}_{;}{}_{p}\,\w{}_{;}{}_{c}{}^{p}{}_{q}{}^{q}{}_{d} -
  180\,\w{}_{;}{}_{p}\,\w{}_{;}{}_{c}{}^{p}{}_{d}{}_{q}{}^{q} -
  70\,\w{}_{;}{}_{p}\,\w{}_{;}{}_{q}\,\w{}^{;}{}^{p}\,\w{}^{;}{}^{q}\,
   R{}_{c}{}_{d} \cr &&
 -140\,\w{}_{;}{}_{p}\,\w{}^{;}{}^{p}\,\w{}_{;}{}_{q}{}^{q}\,R{}_{c}{}_{d} +
  175\,\w{}_{;}{}_{p}{}^{p}\,\w{}_{;}{}_{q}{}^{q}\,R{}_{c}{}_{d} -
  560\,\w{}_{;}{}_{p}\,\w{}_{;}{}_{q}\,\w{}^{;}{}^{p}{}^{q}\,R{}_{c}{}_{d}
\cr
&&
+140\,\w{}_{;}{}_{p}{}_{q}\,\w{}^{;}{}^{p}{}^{q}\,R{}_{c}{}_{d} +
  420\,\w{}_{;}{}_{p}\,\w{}_{;}{}_{q}{}^{q}{}^{p}\,R{}_{c}{}_{d} -
  700\,\w{}_{;}{}_{p}\,\w{}_{;}{}_{q}\,\w{}_{;}{}_{c}\,\w{}^{;}{}^{q}\,
   R{}_{d}{}^{p} \cr &&
 -700\,\w{}_{;}{}_{p}\,\w{}_{;}{}_{c}\,\w{}_{;}{}_{q}{}^{q}\,R{}_{d}{}^{p} +
  350\,\w{}_{;}{}_{p}{}_{c}\,\w{}_{;}{}_{q}{}^{q}\,R{}_{d}{}^{p} +
  420\,\w{}_{;}{}_{p}\,\w{}_{;}{}_{q}\,\w{}_{;}{}_{c}\,\w{}^{;}{}^{p}\,
   R{}_{d}{}^{q} \cr &&
+700\,\w{}_{;}{}_{p}\,\w{}^{;}{}^{p}\,\w{}_{;}{}_{q}{}_{c}\,R{}_{d}{}^{q}
-
  560\,\w{}_{;}{}_{p}\,\w{}_{;}{}_{c}\,\w{}_{;}{}_{q}{}^{p}\,R{}_{d}{}^{q} +
  280\,\w{}_{;}{}_{p}\,\w{}_{;}{}_{q}\,\w{}_{;}{}_{c}{}^{p}\,R{}_{d}{}^{q}
\cr
&&
 -140\,\w{}_{;}{}_{p}{}_{q}\,\w{}_{;}{}_{c}{}^{p}\,R{}_{d}{}^{q} +
  840\,\w{}_{;}{}_{p}\,\w{}_{;}{}_{q}{}_{c}{}^{p}\,R{}_{d}{}^{q} -
  420\,\w{}_{;}{}_{p}\,\w{}_{;}{}_{q}{}^{p}{}_{c}\,R{}_{d}{}^{q} \cr &&
 -420\,\w{}_{;}{}_{p}\,\w{}_{;}{}_{c}{}^{p}{}_{q}\,R{}_{d}{}^{q} -
  40\,\w{}_{;}{}_{p}\,\w{}_{;}{}_{q}\,\w{}_{;}{}_{c}\,\w{}_{;}{}_{d}\,
   R{}^{p}{}^{q} - 1640\,\w{}_{;}{}_{p}\,\w{}_{;}{}_{c}\,
   \w{}_{;}{}_{q}{}_{d}\,R{}^{p}{}^{q} \cr &&
+920\,\w{}_{;}{}_{p}{}_{c}\,\w{}_{;}{}_{q}{}_{d}\,R{}^{p}{}^{q} -
  240\,\w{}_{;}{}_{p}\,\w{}_{;}{}_{q}\,\w{}_{;}{}_{c}{}_{d}\,R{}^{p}{}^{q} +
  420\,\w{}_{;}{}_{p}\,\w{}_{;}{}_{q}{}_{c}{}_{d}\,R{}^{p}{}^{q} \cr &&
+540\,\w{}_{;}{}_{p}\,\w{}_{;}{}_{c}{}_{d}{}_{q}\,R{}^{p}{}^{q} +
  210\,\w{}_{;}{}_{p}\,\w{}_{;}{}_{q}\,R{}_{c}{}_{d}\,R{}^{p}{}^{q} -
  840\,\w{}_{;}{}_{p}\,\w{}_{;}{}_{q}{}_{r}{}^{r}\,R{}_{c}{}^{p}{}_{d}{}^{q}
\cr &&
+1800\,\w{}_{;}{}_{p}\,\w{}_{;}{}_{q}{}^{q}{}_{r}\,R{}_{c}{}^{p}{}_{d}{}^{r}
+   3820\,\w{}_{;}{}_{p}\,\w{}_{;}{}_{q}\,\w{}_{;}{}_{r}{}^{r}\,
   R{}_{c}{}^{q}{}_{d}{}^{p}
\cr && + 700\,\w{}_{;}{}_{p}{}_{q}\,
   \w{}_{;}{}_{r}{}^{r}\,R{}_{c}{}^{q}{}_{d}{}^{p}
+840\,\w{}_{;}{}_{p}\,\w{}_{;}{}_{q}{}_{r}{}^{r}\,R{}_{c}{}^{q}{}_{d}{}^{p}
\cr && +
2680\,\w{}_{;}{}_{p}\,\w{}_{;}{}_{q}\,\w{}_{;}{}_{r}{}^{p}\,
   R{}_{c}{}^{q}{}_{d}{}^{r}
+ 920\,\w{}_{;}{}_{p}{}_{q}\,
   \w{}_{;}{}_{r}{}^{p}\,R{}_{c}{}^{q}{}_{d}{}^{r}
\cr &&
+840\,\w{}_{;}{}_{p}\,\w{}_{;}{}_{q}{}^{p}{}_{r}\,R{}_{c}{}^{q}{}_{d}{}^{r}
+   864\,\w{}_{;}{}_{p}\,\w{}_{;}{}_{q}\,R{}_{r}{}^{p}\,
   R{}_{c}{}^{q}{}_{d}{}^{r}
\cr && + 700\,\w{}_{;}{}_{p}\,\w{}_{;}{}_{q}\,
   \w{}_{;}{}_{r}\,\w{}^{;}{}^{q}\,R{}_{c}{}^{r}{}_{d}{}^{p}
 -168\,\w{}_{;}{}_{p}\,\w{}_{;}{}_{q}\,R{}_{r}{}^{q}\,
   R{}_{c}{}^{r}{}_{d}{}^{p}
\cr && - 2100\,\w{}_{;}{}_{p}\,\w{}_{;}{}_{q}\,
   \w{}_{;}{}_{r}\,\w{}^{;}{}^{p}\,R{}_{c}{}^{r}{}_{d}{}^{q}
-   280\,\w{}_{;}{}_{p}\,\w{}^{;}{}^{p}\,\w{}_{;}{}_{q}{}_{r}\,
   R{}_{c}{}^{r}{}_{d}{}^{q}
\cr &&
+960\,\w{}_{;}{}_{p}\,\w{}_{;}{}_{q}\,R{}_{r}{}^{q}{}_{s}{}^{p}\,
   R{}_{c}{}^{r}{}_{d}{}^{s}
- 1560\,\w{}_{;}{}_{p}\,\w{}_{;}{}_{q}\,
   R{}_{r}{}^{q}{}_{s}{}^{p}\,R{}_{c}{}^{s}{}_{d}{}^{r}
\cr && -
3280\,\w{}_{;}{}_{p}\,\w{}_{;}{}_{q}\,\w{}_{;}{}_{r}{}_{c}\,
   R{}_{d}{}^{p}{}^{q}{}^{r}
-280\,\w{}_{;}{}_{p}\,\w{}_{;}{}_{q}{}_{r}{}_{c}\,R{}_{d}{}^{p}{}^{q}{}^{r}
\cr && +   2480\,\w{}_{;}{}_{p}\,\w{}_{;}{}_{q}{}_{c}{}_{r}\,
   R{}_{d}{}^{p}{}^{q}{}^{r}
- 1176\,\w{}_{;}{}_{p}\,\w{}_{;}{}_{q}\,
   R{}_{r}{}_{c}\,R{}_{d}{}^{p}{}^{q}{}^{r}
\cr &&
+324\,\w{}_{;}{}_{p}\,\w{}_{;}{}_{q}\,R{}_{r}{}_{s}{}_{c}{}^{q}\,
   R{}_{d}{}^{p}{}^{r}{}^{s}
+ 1120\,\w{}_{;}{}_{p}\,
   \w{}_{;}{}_{q}{}_{c}{}_{r}\,R{}_{d}{}^{q}{}^{p}{}^{r}
\cr && +   432\,\w{}_{;}{}_{p}\,\w{}_{;}{}_{q}\,R{}_{r}{}_{c}\,
   R{}_{d}{}^{q}{}^{p}{}^{r}
 -320\,\w{}_{;}{}_{p}\,\w{}_{;}{}_{c}\,\w{}_{;}{}_{q}{}_{r}\,
   R{}_{d}{}^{r}{}^{p}{}^{q}
\cr && + 400\,\w{}_{;}{}_{p}{}_{c}\,
   \w{}_{;}{}_{q}{}_{r}\,R{}_{d}{}^{r}{}^{p}{}^{q}
-
1360\,\w{}_{;}{}_{p}\,\w{}_{;}{}_{q}{}_{c}{}_{r}\,R{}_{d}{}^{r}{}^{p}{}^{q}
\cr &&
 -80\,\w{}_{;}{}_{p}\,\w{}_{;}{}_{q}\,R{}_{r}{}_{s}{}_{c}{}^{q}\,
   R{}_{d}{}^{r}{}^{p}{}^{s}
+ 40\,\w{}_{;}{}_{p}\,\w{}_{;}{}_{q}\,
   R{}_{r}{}^{q}{}_{s}{}_{c}\,R{}_{d}{}^{r}{}^{p}{}^{s}
\cr && -
792\,\w{}_{;}{}_{p}\,\w{}_{;}{}_{q}\,R{}_{r}{}_{s}{}_{c}{}^{q}\,
   R{}_{d}{}^{s}{}^{p}{}^{r} \cr && \left.
+560\,\w{}_{;}{}_{p}\,\w{}_{;}{}_{q}\,R{}_{r}{}^{q}{}_{s}{}_{c}\,
  R{}_{d}{}^{s}{}^{p}{}^{r} \right) / 302400
\cr\cr
%
%
&&+g_{ab} g_{cd} \left(
-12\,\w{}_{;}{}_{p}\,\w{}_{;}{}_{q}\,\w{}_{;}{}_{r}\,\w{}^{;}{}^{p}\,
   \w{}^{;}{}^{q}\,\w{}^{;}{}^{r} \right. \cr && +
  108\,\w{}_{;}{}_{p}\,\w{}_{;}{}_{q}\,\w{}_{;}{}_{r}\,
   R{}^{p}{}^{r}{}^{;}{}^{q} - 60\,\w{}_{;}{}_{p}\,\w{}_{;}{}_{q}\,
   \w{}^{;}{}^{p}\,\w{}^{;}{}^{q}\,\w{}_{;}{}_{r}{}^{r}  \cr &&
+21\,\w{}_{;}{}_{p}\,\w{}^{;}{}^{p}\,\w{}_{;}{}_{q}{}^{q}\,
   \w{}_{;}{}_{r}{}^{r} + 35\,\w{}_{;}{}_{p}{}^{p}\,\w{}_{;}{}_{q}{}^{q}\,
   \w{}_{;}{}_{r}{}^{r} - 12\,\w{}_{;}{}_{p}\,\w{}_{;}{}_{q}\,
   \w{}_{;}{}_{r}{}^{r}\,\w{}^{;}{}^{p}{}^{q} \cr &&
+84\,\w{}_{;}{}_{p}{}_{q}\,\w{}_{;}{}_{r}{}^{r}\,\w{}^{;}{}^{p}{}^{q}
-
  588\,\w{}_{;}{}_{p}\,\w{}_{;}{}_{q}\,\w{}_{;}{}_{r}\,\w{}^{;}{}^{q}\,
   \w{}^{;}{}^{p}{}^{r} - 366\,\w{}_{;}{}_{p}\,\w{}_{;}{}_{q}\,
   \w{}_{;}{}_{r}{}^{q}\,\w{}^{;}{}^{p}{}^{r} \cr &&
+64\,\w{}_{;}{}_{p}{}_{q}\,\w{}_{;}{}_{r}{}^{q}\,\w{}^{;}{}^{p}{}^{r}
+
  252\,\w{}_{;}{}_{p}\,\w{}_{;}{}_{q}\,\w{}_{;}{}_{r}\,\w{}^{;}{}^{p}\,
   \w{}^{;}{}^{q}{}^{r} - 12\,\w{}_{;}{}_{p}\,\w{}^{;}{}^{p}\,
   \w{}_{;}{}_{q}{}_{r}\,\w{}^{;}{}^{q}{}^{r} \cr &&
 -36\,\w{}_{;}{}_{p}\,\w{}_{;}{}_{q}\,\w{}^{;}{}^{p}\,
   \w{}_{;}{}_{r}{}^{q}{}^{r} +
  252\,\w{}_{;}{}_{p}\,\w{}_{;}{}_{q}\,\w{}^{;}{}^{q}\,
   \w{}_{;}{}_{r}{}^{r}{}^{p} +
  252\,\w{}_{;}{}_{p}\,\w{}_{;}{}_{q}{}^{q}\,\w{}_{;}{}_{r}{}^{r}{}^{p} \cr
&&
 -216\,\w{}_{;}{}_{p}\,\w{}_{;}{}_{q}\,\w{}^{;}{}^{p}\,
   \w{}_{;}{}_{r}{}^{r}{}^{q} +
  306\,\w{}_{;}{}_{p}\,\w{}_{;}{}_{q}{}^{p}\,\w{}_{;}{}_{r}{}^{r}{}^{q} -
  108\,\w{}_{;}{}_{p}\,\w{}_{;}{}_{q}\,\w{}_{;}{}_{r}\,
   \w{}^{;}{}^{p}{}^{r}{}^{q} \cr &&
+108\,\w{}_{;}{}_{p}\,\w{}_{;}{}_{q}{}_{r}\,\w{}^{;}{}^{p}{}^{r}{}^{q}
+
  108\,\w{}_{;}{}_{p}\,\w{}_{;}{}_{q}{}_{r}\,\w{}^{;}{}^{q}{}^{r}{}^{p} -
  36\,\w{}_{;}{}_{p}\,\w{}_{;}{}_{q}\,\w{}_{;}{}_{r}{}^{p}{}^{q}{}^{r} \cr
&&
 -18\,\w{}_{;}{}_{p}\,\w{}_{;}{}_{q}\,\w{}_{;}{}_{r}{}^{p}{}^{r}{}^{q} +
  162\,\w{}_{;}{}_{p}\,\w{}_{;}{}_{q}\,\w{}_{;}{}_{r}{}^{r}{}^{p}{}^{q} +
  54\,\w{}_{;}{}_{p}\,\w{}_{;}{}_{q}\,\w{}^{;}{}^{p}{}^{q}{}_{r}{}^{r} \cr
&&
+126\,\w{}_{;}{}_{p}\,\w{}_{;}{}_{q}\,\w{}_{;}{}_{r}{}^{r}\,R{}^{p}{}^{q}
+
  126\,\w{}_{;}{}_{p}\,\w{}_{;}{}_{q}\,\w{}_{;}{}_{r}\,\w{}^{;}{}^{q}\,
   R{}^{p}{}^{r}
\cr &&
 - 108\,\w{}_{;}{}_{p}\,\w{}_{;}{}_{q}\,\w{}_{;}{}_{r}\,
   \w{}^{;}{}^{p}\,R{}^{q}{}^{r} \cr && \left.
+306\,\w{}_{;}{}_{p}\,\w{}_{;}{}_{q}\,\w{}_{;}{}_{r}{}^{p}\,R{}^{q}{}^{r}
+
  36\,\w{}_{;}{}_{p}\,\w{}_{;}{}_{q}\,\w{}_{;}{}_{r}{}_{s}\,
   R{}^{p}{}^{r}{}^{q}{}^{s} \right) / 181440
\end{eqnarray}
\end{subequations}

\section{Expansion tensors for $G_{\rm fin}$}
\label{apx-Gfin}
We present the explicit expressions for the expansion tensors
for $G_{\rm fin}$.
\begin{subequations}
\begin{eqnarray} G^{(0)}_{\rm fin} &=& {\frac{{{\kappa}^2}}{6\,{e^{2\,w}}}}
\\
G^{(1)}_{{\rm fin},a} &=&
{\frac{{{\kappa}^2}\,w{}_{;}{}_{a}}{6\,{e^{2\,w}}}}
\\
G^{(2)}_{{\rm fin},ab} &=&
 {\frac{{{\kappa}^2}}{72\,{e^{2\,w}}}} \left(
8\,w{}_{;}{}_{a}\,w{}_{;}{}_{b}
- 4\,w{}_{;}{}_{a}{}_{b} -
  2\,w{}_{;}{}_{p}\,w{}^{;}{}^{p}\,g{}_{a}{}_{b} +
  w{}_{;}{}_{p}{}^{p}\,g{}_{a}{}_{b} + R{}_{a}{}_{b} \right)
\cr && +{\frac{{{\kappa}^4}}{360\,{e^{4\,w}}}} \left(
4\,{e^{2\,w}}\,\delta{}_{a}^{\tau}\,\delta{}_{b}^{\tau} -
g{}_{a}{}_{b} \right)
\\
G^{(3)}_{{\rm fin},abc} &=&
 {\frac{{{\kappa}^2}}{144\,{e^{2\,w}}}}
   \left( 8\,w{}_{;}{}_{a}\,w{}_{;}{}_{b}\,w{}_{;}{}_{c} -
R{}_{a}{}_{b}{}_{;}{}_{c} -
  12\,w{}_{;}{}_{a}\,w{}_{;}{}_{b}{}_{c} + 2\,w{}_{;}{}_{a}{}_{b}{}_{c} -
  4\,w{}_{;}{}_{p}\,w{}_{;}{}_{a}\,w{}^{;}{}^{p}\,g{}_{b}{}_{c} \right.
\cr && \hspace{15mm} +
   \left. 2\,w{}_{;}{}_{a}\,w{}_{;}{}_{p}{}^{p}\,g{}_{b}{}_{c} +
  4\,w{}_{;}{}_{p}\,w{}_{;}{}_{a}{}^{p}\,g{}_{b}{}_{c} -
  w{}_{;}{}_{p}{}^{p}{}_{a}\,g{}_{b}{}_{c} + 2\,w{}_{;}{}_{a}\,R{}_{b}{}_{c}
\right) \cr && +{\frac{{{\kappa}^4}}{180\,{e^{4\,w}}}} \left(
2\,{e^{2\,w}}\,\Gamma^{\tau}{}_{a}{}_{b}\,\delta{}_{c}^{\tau} +
  2\,{e^{2\,w}}\,w{}_{;}{}_{a}\,\delta{}_{b}^{\tau}\,\delta{}_{c}^{\tau} -
  w{}_{;}{}_{a}\,g{}_{b}{}_{c} \right)
\\
G^{(3)}_{{\rm fin},abc} &=&
{\frac{{{\kappa}^2}}{8640\,{e^{2\,w}}}} \left(
  192\,w{}_{;}{}_{a}\,w{}_{;}{}_{b}\,w{}_{;}{}_{c}\,w{}_{;}{}_{d} -
  60\,w{}_{;}{}_{a}\,R{}_{b}{}_{c}{}_{;}{}_{d} -
  576\,w{}_{;}{}_{a}\,w{}_{;}{}_{b}\,w{}_{;}{}_{c}{}_{d} \right.\cr &&
\hspace{15mm}
+  144\,w{}_{;}{}_{a}{}_{b}\,w{}_{;}{}_{c}{}_{d} +
18\,R{}_{a}{}_{b}{}_{;}{}_{c}{}_{d} +
  192\,w{}_{;}{}_{a}\,w{}_{;}{}_{b}{}_{c}{}_{d} -
  24\,w{}_{;}{}_{a}{}_{b}{}_{c}{}_{d} \cr && \hspace{15mm}
+   80\,w{}_{;}{}_{a}\,w{}_{;}{}_{b}\,R{}_{c}{}_{d}
  -40\,w{}_{;}{}_{a}{}_{b}\,R{}_{c}{}_{d} + 5\,R{}_{a}{}_{b}\,R{}_{c}{}_{d}
+
  4\,R{}_{p}{}_{a}{}_{q}{}_{b}\,R{}_{c}{}^{p}{}_{d}{}^{q} \cr &&
\hspace{15mm}
%
%
-2g_{ab}\left\{
   72\,w{}_{;}{}_{p}\,w{}_{;}{}_{c}\,w{}_{;}{}_{d}\,w{}^{;}{}^{p}
  -15\,w{}_{;}{}_{p}\,R{}_{c}{}_{d}{}^{;}{}^{p}
  +15\,w{}_{;}{}_{p}\,R{}_{c}{}^{p}{}_{;}{}_{d} \right.
\cr && \hspace{25mm}
  -36\,w{}_{;}{}_{c}\,w{}_{;}{}_{d}\,w{}_{;}{}_{p}{}^{p}
  -36\,w{}_{;}{}_{p}\,w{}^{;}{}^{p}\,w{}_{;}{}_{c}{}_{d}
  +18\,w{}_{;}{}_{p}{}^{p}\,w{}_{;}{}_{c}{}_{d}
\cr && \hspace{25mm}
  -144\,w{}_{;}{}_{p}\,w{}_{;}{}_{c}\,w{}_{;}{}_{d}{}^{p}
  +36\,w{}_{;}{}_{p}{}_{c}\,w{}_{;}{}_{d}{}^{p}
  -6\,w{}_{;}{}_{c}\,w{}_{;}{}_{p}{}_{d}{}^{p}
\cr && \hspace{25mm}
  +42\,w{}_{;}{}_{c}\,w{}_{;}{}_{p}{}^{p}{}_{d}
  -18\,w{}_{;}{}_{p}\,w{}_{;}{}_{c}{}_{d}{}^{p}
  +54\,w{}_{;}{}_{p}\,w{}_{;}{}_{c}{}^{p}{}_{d}
\cr && \hspace{25mm}
  +12\,w{}_{;}{}_{p}{}_{c}{}_{d}{}^{p}
  -3\,w{}_{;}{}_{p}{}_{c}{}^{p}{}_{d}
  - 9\,w{}_{;}{}_{p}{}^{p}{}_{c}{}_{d}
\cr && \hspace{25mm}
  -9\,w{}_{;}{}_{c}{}_{d}{}_{p}{}^{p}
  +10\,w{}_{;}{}_{p}\,w{}^{;}{}^{p}\,R{}_{c}{}_{d}
  -5\,w{}_{;}{}_{p}{}^{p}\,R{}_{c}{}_{d}
\cr && \hspace{25mm}
 +10\,w{}_{;}{}_{p}\,w{}_{;}{}_{c}\,R{}_{d}{}^{p}
  -5\,w{}_{;}{}_{p}{}_{c}\,R{}_{d}{}^{p}
  -10\,w{}_{;}{}_{p}\,w{}_{;}{}_{q}\,R{}_{c}{}^{p}{}_{d}{}^{q}
\cr && \hspace{25mm} \left.
  -10\,w{}_{;}{}_{p}{}_{q}\,R{}_{c}{}^{q}{}_{d}{}^{p}
\right\} \cr && \hspace{15mm}
%
+g_{ab} g_{cd} \left\{
  12\,w{}_{;}{}_{p}\,w{}_{;}{}_{q}\,w{}^{;}{}^{p}\,w{}^{;}{}^{q} -
  14\,w{}_{;}{}_{p}\,w{}^{;}{}^{p}\,w{}_{;}{}_{q}{}^{q} +
  5\,w{}_{;}{}_{p}{}^{p}\,w{}_{;}{}_{q}{}^{q} \right.\cr && \hspace{27mm}
\left. \left.
  -28\,w{}_{;}{}_{p}\,w{}_{;}{}_{q}\,w{}^{;}{}^{p}{}^{q} +
  4\,w{}_{;}{}_{p}{}_{q}\,w{}^{;}{}^{p}{}^{q}
\right. \right. \cr && \hspace{27mm} \left. \left.
 +  12\,w{}_{;}{}_{p}\,w{}_{;}{}_{q}{}^{q}{}^{p} +
  6\,w{}_{;}{}_{p}\,w{}_{;}{}_{q}\,R{}^{p}{}^{q} \right\}
\right) \cr
&&+{\frac{{{\kappa}^4}}{4320\,{e^{4\,w}}}} \left(
  4\,{e^{2\,w}} \left\{
                3\,\Gamma^{\tau}{}_{a}{}_{b}\,\Gamma^{\tau}{}_{c}{}_{d} +
  12\,\Gamma^{\tau}{}_{a}{}_{b}\,w{}_{;}{}_{c}\,\delta{}_{d}^{\tau} -
  4\,\Gamma^{\tau}{}_{a}{}_{b}{}_{;}{}_{c}\,\delta{}_{d}^{\tau}
\right. \right. \cr && \hspace{32mm}
         \left.
+
8\,w{}_{;}{}_{a}\,w{}_{;}{}_{b}\,\delta{}_{c}^{\tau}\,\delta{}_{d}^{\tau}
-4\,w{}_{;}{}_{a}{}_{b}\,\delta{}_{c}^{\tau}\,\delta{}_{d}^{\tau}
+
  \delta{}_{c}^{\tau}\,\delta{}_{d}^{\tau}\,R{}_{a}{}_{b} \right\} \cr  &&
\hspace{20mm}
-g_{ab} \left\{
  28\,w{}_{;}{}_{c}\,w{}_{;}{}_{d} - 8\,w{}_{;}{}_{c}{}_{d} + R{}_{c}{}_{d}
\right. \cr && \hspace{32mm} \left.
  4\,{e^{2\,w}}\,\left( 2\,w{}_{;}{}_{p}\,w{}^{;}{}^{p} -
     w{}_{;}{}_{p}{}^{p} \right) \,\delta{}_{c}^{\tau}\,\delta{}_{d}^{\tau}
  R{}_{c}{}_{d} \right\} \cr && \hspace{20mm}
\left.+ g_{ab} g_{cd} \left\{
       3\,w{}_{;}{}_{p}\,w{}^{;}{}^{p} - w{}_{;}{}_{p}{}^{p}  \right\}
\right) \cr
&&+{\frac{{{\kappa}^6}}{15120\,{e^{6\,w}}}} \left(
  16\,{e^{4\,w}}\,\delta{}_{a}^{\tau}\,\delta{}_{b}^{\tau}\,
   \delta{}_{c}^{\tau}\,\delta{}_{d}^{\tau} -
  12\,{e^{2\,w}}\,\delta{}_{c}^{\tau}\,\delta{}_{d}^{\tau}\,g{}_{a}{}_{b} +
  g{}_{a}{}_{b}\,g{}_{c}{}_{d}
\right) \end{eqnarray}
\end{subequations}


\end{document}